\documentclass[runningheads,natbib]{svjour3}
\smartqed  
\usepackage{graphicx}
\usepackage{amsbsy}
\usepackage{amssymb}
\usepackage{amsmath}
\usepackage{bm}
\usepackage{eufrak}
\usepackage{upgreek}
\usepackage{revsymb}	
\usepackage{color}

\newcommand{\eq}[1] {Eq.\,(\ref{#1})}
\def\vec#1{\ensuremath{\mathchoice{\mbox{\boldmath$\displaystyle#1$}}
{\mbox{\boldmath$\textstyle#1$}}
{\mbox{\boldmath$\scriptstyle#1$}}
{\mbox{\boldmath$\scriptscriptstyle#1$}}}}

\def\sqr#1#2{{\vcenter{\vbox{\hrule height.#2pt
        \hbox{\vrule width.#2pt height#1pt \kern#1pt
                \vrule width.#2pt}
        \hrule height.#2pt}}}}
\def\square{\mathchoice\sqr34\sqr34\sqr{2.1}3\sqr{1.5}3}
\def\myf{\bm{\mathfrak{F}}}
\def\ob{\overline}
\def\rd{{\rm d}}
\journalname{Astronomy and Astrophysics Review}
\begin{document}

\title{The quest for the solar g modes
}
\subtitle{}


\author{T.Appourchaux        \and
        K.~Belkacem \and
        A.-M.~Broomhall   \and
        W.J. Chaplin \and
        D.O.~Gough           \and
        G.~Houdek            \and
        J. Provost\and
        F.~Baudin\and
        P.~Boumier\and
        Y.~Elsworth\and
        R.A.~Garc{\'{\i}}a\and
        B.Andersen\and
	W.Finsterle\and
	C.Fr\"ohlich\and
	A.Gabriel\and	
	G.Grec\and
	A.Jim\'enez\and
	A.Kosovichev\and
	T.Sekii\and
	T.Toutain	\and
	S.Turck-Chi\`eze
}


\institute{T.~Appourchaux \and F.~Baudin \and P.~Boumier \and A.~Gabriel \at
              Institut d'Astrophysique Spatiale, UMR8617, B\^atiment 121, 91045 Orsay Cedex, France\\
             \email{Thierry.Appourchaux@ias.u-psud.fr}           
	   \and
K.~Belkacem \at
            Institut d'Astrophysique et G\'eophysique, Universit\'e de Li\`ege, All\'ee du 6 Ao\^ut 17-B 4000 Li\`ege, Belgium
	  \and
A.-M.~Broomhall \and W.J.~Chaplin \and Y.~Elsworth \at 
            School of Physics and Astronomy, University of Birmingham, Edgbaston, Birmingham B15 2TT, UK
	\and
D.O.~Gough
              \at
             Institute of Astronomy and Department of Applied Mathematics and Theoretical Physics, University of Cambridge, Cambridge CB30HA, UK
             \and
G.~Houdek
             \at
             Institute of Astronomy, University of Vienna, T\"urkenschanzstra{\ss}e 17, A-1180 Vienna, Austria
              \at
             Institute of Astronomy, University of Cambridge, Cambridge CB30HA, UK
              \and
J.~Provost and G.~Grec\at
              Universit\'e de Nice Sophia-Antipolis, CNRS, Laboratoire Cassiop\'ee, Observatoire de la C\^ote d'Azur, BP4229, 06304 Nice Cedex4, France
	   \and
R.A.~Garc{\'{\i}}a \and S.~Turck-Chi\`eze\at
           Laboratoire AIM, CEA / DSM - CNRS, Universit\'e Paris Diderot, IRFU / SAp, Centre de Saclay, 91191 Gif-sur-Yvette, France
           \and
B.N.~Andersen\at
	Norwegian Space Centre, N-0212 Oslo, Norway
	\and
W.Finsterle \and C.Fr\"ohlich \at
           Physikalisch-Meteorologisches Observatorium Davos, World Radiation Center, 7260 Davos Dorf, Switzerland
           \and
A.~Jim\'enez \at
           Instituto de Astrofisica de Canarias, E-38205 La Laguna, Tenerife, Spain.
           \and
A.~Kosovichev \at
W.W. Hansen Experimental Physics Laboratory, Stanford University, Stanford, CA 94305, USA 
\and
T.~Sekii \at
National Astronomical Observatory of Japan, 2-21-1 Osawa, Mitaka, Tokyo 181-8588, Japan 
\and
T.~Toutain\at
	Center for Information Technology, University of Oslo, P.O. Box 1059 Blindern, N-0316 Oslo, Norway
}

\date{\today,~ $Revision: Final $}

\maketitle

\begin{quotation}
\noindent "A ce compte, toutes les sciences ne seraient que des applications inconscientes du calcul des probabilit\'es ; condamner ce calcul, ce serait condamner la science tout enti\`ere."

\noindent "From this point of view all the sciences would only be unconscious applications of the calculus of probabilities. And if this calculus be condemned, then the whole of the sciences must also be condemned."

Henri Poincar\'e, 1914, {\it La Science et l'hypoth\`ese}
\end{quotation}

\begin{abstract}

Solar gravity modes (or g modes) -- oscillations of the solar
interior for which buoyancy acts as the restoring force -- have
the potential to provide unprecedented inference on the
structure and dynamics of the solar core, inference that is not
possible with the well observed acoustic modes (or p modes). The high amplitude of the g-mode eigenfunctions in the core 
and the evanesence of the modes in the convection zone make the modes particularly sensitive to the physical and dynamical conditions in the core. Owing to the existence of the convection zone, the g modes have very low amplitudes at photospheric levels, which makes the modes
extremely hard to detect. In this paper, we review the current
state of play regarding attempts to detect g modes. We review
the theory of g modes, including theoretical estimation
of the g-mode frequencies, amplitudes and damping rates. Then
we go on to discuss the techniques that have been used to try
to detect g modes. We review results in the literature, and
finish by looking to the future, and the potential advances
that can be made -- from both data and data-analysis perspectives --
to give unambiguous detections of individual g modes.
The review ends by concluding that, at the time of writing, there is indeed a consensus
amongst the authors that {\it there is currently no undisputed detection of solar g  modes.}



\end{abstract}

\section{Introduction}
\label{intro}

The detection of solar oscillations by \citet{Leighton1962} marked the
start of a memorable era for solar physics, and the birth of a new
field, helioseismology.  The oscillatory motions detected in the solar
atmosphere had typical periods of about five minutes, and
\citet{Deubner75} subsequently verified observationally that the
oscillations were the visible manifestation of acoustic (p) modes
trapped in cavities in the solar interior. Later in the 1970s,
\citet{AC79} showed that the Sun supported truly global modes of
oscillation. The low-angular-degree (low-$l$) modes observed by
\citet{AC79} penetrated the solar core, and had therefore opened a
window on the structure of the deep interior of the Sun.  \citet{GG80} subsequently made long-duration observations of the Sun from the South Pole, providing the very first clean power spectrum free from day-night interruptions.  

While inference on the internal structure of the Sun made using
high-quality data on the p modes has gone from strength to strength
over the intervening 30 years, attempts to detect g modes have proven
altogether more elusive.  This is in stark contrast with the ubiquitous detection of stellar g modes across the Hertzsprung-Russell diagram such as in white dwarfs \citep{Winget2008}, in subdwarfs B stars \citep{Green2003}, in B stars \citep{Waelkens1991} and in $\gamma$ Dor stars \citep{Aerts1998}.  The detection, and subsequent exploitation,
of data on the solar g modes would provide much more precise inference on
the structure and dynamics of the solar core than is possible with p
modes. This is because the g modes are trapped in cavities in the
radiative interior, and they may have much higher displacement amplitude in the radiative zone / core
than their acoustic counterparts. However, this high sensitivity to
the core properties comes at a price. Because the g modes are trapped
beneath the convective envelope they have very low amplitudes at the
photospheric level where the perturbations due to the oscillations are
detected; hence their elusive nature.

The first attempt to detect long-period oscillations date back to the mid 1970s,
e.g., \citet{Severnyi1976}. Further claims, from ground-based Doppler
velocity observations, were made in the 1980s, e.g., \citet{PDPS83}.
It was realized very early on in the development of helioseismology as
a field that a space mission would be ideally suited not only for
observing the p modes but also for detecting g modes (in principle
providing the long-term stability needed to detect the low-frequency g
modes).  The DISCO\footnote{Dual Spectral Irradiance and Solar
Constant Orbiter} mission was proposed by \citet{Bonnet81} primarily
as an irradiance mission to which helioseismology was added later on. The initial design was subsequently
modified to accommodate several in-situ instruments and spectrographs,
and finally matured into the SOHO\footnote{SOlar and Heliospheric Observatory} mission \citep{Domingo1995}.

One of the goals of the helioseismology instruments on SOHO was to
detect g modes. Following its launch in 1995 December, initial
analyses of the helioseismology data failed to uncover evidence for
the modes. By 1997, it was realized that the searches would benefit
greatly from an internationally coordinated effort comprised of the
principal observational and theoretical scientists involved in g-mode
research. This led to the formation of the Phoebus Group, which has
been working since to use all available high-quality helioseismic 
data, and to develop the analysis techniques that are required, to
detect g modes.

In this review we collect and summarize the knowledge accumulated by
the Phoebus group over more than a decade of g-mode research.  Our
review is broken down as follows. In Section~2 we review the
theoretical properties of the solar oscillations. [A complete
presentation of the properties of solar and stellar oscillations may
be found for example in \citet{Unno89}, \citet{jcdgb91}, and
\citet{2002RvMP...74.1073C}.]  We also consider the accuracy of the
theoretical computations of the g mode frequencies, and detail the
various factors that contribute to the uncertainties (i.e., the
accuracy of the oscillation computations themselves, and the
sensitivity of the frequencies to the structure and dynamics of the
solar models).

In Section~3 we discuss the excitation and damping of the g modes,
again from a theoretical perspective. We consider the different
physical mechanisms that can potentially contribute to the stability
of g modes, and present theoretical estimates of the amplitudes of g
modes excited stochastically by convection in the convective envelope.

In Section~4, we turn to the observations. We begin by considering the
different detection techniques that are available to observers, and
discuss in detail the constituent elements required to ``build'' a
front-to-end analysis algorithm. Our presentation quite naturally
includes an in-depth discussion of frequency-domain analysis, and
hypothesis testing (frequentist versus Bayesian). We finish this
section by reviewing (from the literature) attempts to apply these
techniques to detect g modes.

Finally, in Section~5, we conclude with some remarks on the current
state of play. Our view is that, at the time of writing, there is no
unambiguous evidence for the detection of solar g modes, although some
of the most recent theoretical predictions suggest a positive result
may be attainable in the near future.  We therefore finish by looking
to the future prospects for finally detecting g modes.



\section{Eigenfrequencies for the solar model}
\label{theory}

The Sun can show oscillations whose restoring forces are due either to compressibility or
to buoyancy. Compressibility gives rise to acoustic waves: resonant
pressure (p) modes in the high-frequency range (typically about
$3000\,\rm \mu Hz$). The effects of buoyancy can give rise to internal gravity
(g) modes, which have very low frequencies, and are the focus of this
review. The so-called f modes (surface gravity waves) and mixed modes (having
dual p- and g-mode characteristics) occur at frequencies around the
transition between g modes and p modes.

\subsection{Properties of gravity modes }

\subsubsection{General properties (adiabatic)}

The Sun, as a mass of compressible self-gravitating gas, can oscillate
around its equilibrium state.  The small-amplitude oscillations
displayed by the Sun may be described by spherical harmonic functions,
$Y_l^m(\theta,\phi)$, of degree $l$ and order $m$ where ($\theta$,  $\phi$) are spherical polar co-ordinates, $\theta$ being colatitude (i.e., the angle from the polar axis) and $\phi$ longitude.  For example, for a singlet, the perturbation of a
thermodynamical quantity like the Eulerian pressure perturbation $p'$ can be
written:
\begin{equation}
p'=\tilde{p}(r)Y_l^m(\theta,\phi)e^{-2\pi {\rm i}\nu_{n, l,m}t},
\label{Eq1}
\end{equation}
\noindent where $\nu_{n, l, m}$ is the cyclic oscillation frequency
(which may be related to the angular frequency by $\omega=2\pi\nu$).
The displacement vector of the material from its equilibrium position
$\vec\xi$ may be written, in the frame of polar spherical coordinates, as:
\begin{equation}
\vec\xi=\left(\xi_{\rm r}(r) Y_l^m, L^{-1}\xi_{\rm h}(r) {{\rm d} \over {\rm d}\theta} Y_l^m, i m L^{-1}\xi_{\rm h}(r){ Y_l^m\over \sin \theta}\right)e^{-2\pi {\rm i}\nu_{n, l,m}t},
\label{xi}
\end{equation}
$\xi_{\rm r}$ and $\xi_{\rm h}$ being related to the radial and horizontal
components of the displacement and $L=\sqrt{l (l +1)}$.  Each mode is
characterized by three numbers, $n$, $l$, $m$: the radial order $n$ is
essentially the number of nodes\footnote{At the limit of low and high frequency,
the radial order $n$ is essentially the number of nodes of the radial displacement in a solar model. But  the classification of modes, i.e. the assignment of a radial order to them, may be more complicated depending on the frequency range (case of mixed modes)
and on the model \citep[e.g.][]{Eckart1960,Scufflaire1974,Osaki1975,Takata2006}.} of the radial displacement eigenfunction from centre to surface in the radial direction, the degree $l$ is related to the horizontal wavelength of a mode, and
corresponds to the number of nodal lines on the solar surface, and the magnitude of the
azimuthal order $m$, where $m$ satisfies $-l \le m \le l$, is twice the number
of nodes in longitude  \citep[e.g.][]{Unno89,jcdgb91,Gough1993}.  If the Sun were spherically
symmetric (no rotation and no magnetic fields) the frequency
$\nu_{n,l,m}$ of a mode would not depend on the azimuthal order $m$.


\begin{figure}
\begin{center}
\includegraphics[width=0.75\textwidth]{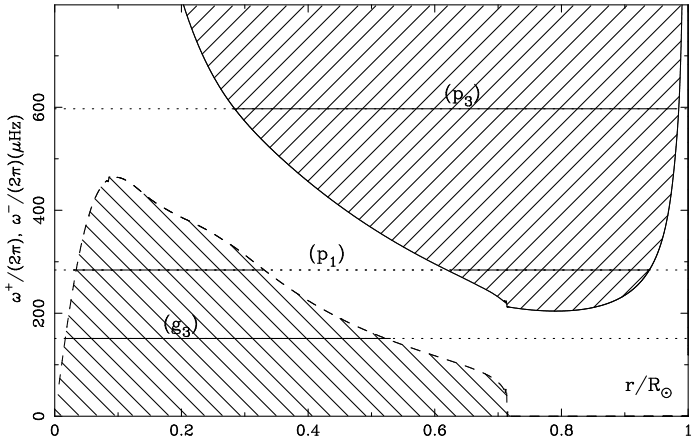}

\end{center}

\caption[]{Propagation diagram for a reference solar model. The solid
and dashed curves represent, respectively, the real values
$\omega^+/2\pi$ and $\omega^-/2\pi$ corresponding to zero values of
$K^2(r)$, and are plotted as a function of the radius, for $l$=1.  The
hatched areas indicate p-mode and g-mode propagation regions for this
degree.  The full horizontal lines indicate the propagation regions of
three modes of degree $l$=1: a p mode of frequency $\nu \simeq 600
\mu$Hz~(p$_3$), a g mode of frequency $\nu \simeq 150 \mu$Hz~(g$_3$),
and a mixed mode of frequency $\nu \simeq 280 \mu$Hz~(p$_1$). The
corresponding eigenfunctions of these modes are shown in
Figure~\ref{fig:mixed}.}

\label{fig:Vais}\end{figure}

The solar oscillations have two restoring
forces, compressibility and buoyancy, characterized respectively by
the sound speed $c$ and the buoyancy (Brunt-V\"ais\"al\"a) frequency $N$:
\begin{equation}
c^2={\gamma_1 p \over \rho}, \qquad
N^2 = g \left[{1\over
 \gamma_1 p}{{\rm d}p\over dr}-{1 \over \rho}{{\rm d}\rho\over {\rm d}r}\right],
\end{equation}
Here, $g$, $p$ and $\rho$ are, respectively, the gravity acceleration, the
pressure, and the density of the solar model, $r$ is the radius, and
$\gamma_1$ is the adiabatic exponent.  In order to have an insight
into the relation between the oscillations and the structure of the
model, one may consider a simplified wave equation, which provides an
approximate description the oscillations \citep{1984ARA&A..22..593D}.  If the wavelength is much less that the solar radius, the local effects of spherical geometry on the dynamics can be ignored.  This equation is then derived under the local wave-like approximation
JWKB\footnote{JWKB stands for Jeffreys-Wentzel-Kramers-Brillouin
\citep{Jeffreys1925}. See also \citet{2007AN....328..273G} in the asteroseismic context.}, and under the Cowling
approximation, i.e. where the perturbation to the gravitational
potential by the waves is neglected. The equation is
 \begin{equation}
 {{\rm d}^2 \Psi\over  {\rm d}r^2} + K(r)\Psi = 0
 \label{wave}
 \end{equation}
 \noindent
with 
 \begin{equation}
 K(r)\equiv {k_{\rm r}}^2 = \frac{1}{c^2}\left[\omega^2-
 \omega_{\rm c}^2-S_l^2\left(1-\frac{N^2}{\omega^2}\right)\right]
 \end{equation} 

The quantity $\Psi$ is related to the displacement of the wave, via
$\Psi = c^2\rho^{1/2} {\rm div} \vec\xi $.  $S_l$, $\omega_{\rm c}$ and $H$
are, respectively, the Lamb and the acoustic cut-off angular frequencies, and
the density scale height:
 \begin{equation}
 S_l = \sqrt{l(l+1)}{c\over r}\,\qquad 
 \omega_{\rm c}^2= {c^2\over 4 H^2 }\left(1 -2 {{\rm d}H\over {\rm d}r}\right) \,
 \qquad  H=- \left({1\over \rho}{{\rm d}\rho\over {\rm d}r}\right)^{-1}
 \end{equation}
In this plane-wave approximation, $k_{\rm r}$ represents a local radial wave
number, which may be associated with a local horizontal wave number,
$k_{\rm h}=L/r$.  In the wave equation, the term in $N^2 $ is the signature
of the buoyancy, which acts as restoring force only in the radiative
regions (where $N^2>0$).  The stratification in density close to the solar
surface induces an acoustic cut-off angular frequency, $\omega_{\rm c}$, which
corresponds to an upper limit to the frequencies of trapped, radial
modes ($l=0$).

Equation~(\ref{wave}) shows that a mode will oscillate as a function
of $r$ in regions where $K(r)> 0$. Such regions are called regions of
propagation. In contrast, a mode will be evanescent, i.e. its energy
will vary exponentially, in regions where $K(r)< 0$.  $K(r)$ is positive
either for $\omega> (\omega^+, \omega^-)$ or $\omega< (\omega^+, \omega^-)$ ,
the angular frequencies $\omega^+$ and  $\omega^-$ corresponding to the zeros of $K(r)$ .
$\omega^+$ and
$\omega^-$ are close to $S_l$ and $N$ respectively, except in
the external layers where $\omega^+$ is close to $\omega_{\rm c}$ and
$\omega^-$ is very small. The frequencies $\omega^+$ and $\omega^-$ are plotted, as functions of the radius,
in the so-called propagation diagram,  for a solar model (Figure~\ref{fig:Vais}). The horizontal lines represent modes; they are continuous in  the propagating regions and dotted in the evanescent regions.
There are two possible regions of propagation, according the frequency of the modes: a high frequency region ($\omega \gtrsim S_l,\omega_c$) corresponding to pressure modes (or p modes)resulting from the compressibility,
and a low frequency region ($\omega \lesssim N$) corresponding  to internal gravity modes caused by  the buoyancy.
In between, some
modes, called mixed modes, can experience a substantial restoring force from both the gradient of
pressure and buoyancy.

\begin{figure*}
 \centerline{
 \includegraphics[width=\textwidth]{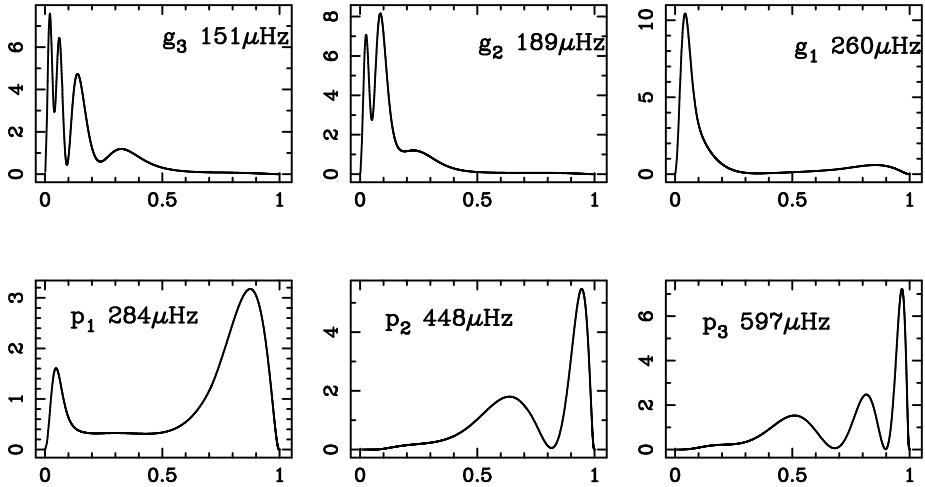}}

 \caption[]{Kinetic energy density (i.e $\rho r^2 |{\vec\xi}|^2$
normalized by the inertia ${\cal I}_{n, l}$), as function of the
radius for modes g$_3$, g$_2$, g$_1$, p$_1$, p$_2$ and p$_3$ of degree
$l$=1, as computed for a reference solar model. From
\citet{JPGB2000}.}

\label{fig:mixed}
\end{figure*}

In the high-frequency range, which corresponds to observed solar
oscillations, the modes are acoustic modes, and their
properties depend mainly on the variation of the sound speed $c(r)$.
Their radial wave number $k_{\rm r}$ is 
\begin{equation}
k_{\rm r}\sim \sqrt{ {\omega^2 \over c^2} \left(1
-{{S_l}^2\over\omega^2}\right)}.
\end{equation}
Thus these modes obey the classical
dispersion relation $k_{\rm r}^2+k_{\rm h}^2\sim {\omega^2\over c^2}.$ and the 
frequencies increase with the radial order $n$. Owing to the increase of
the sound speed towards the solar interior, the p modes are refracted
at the level where $\omega = S_l$. Thus, they are trapped in a cavity
between that level and the surface. The lower the degree of the mode,
the more deeply it penetrates the solar interior. An example of a
trapping region is indicated in Figure~\ref{fig:Vais} for the p mode
with $n= 3$ and $l= 1$. Figure~\ref{fig:mixed} shows how the
amplitudes of various modes vary within the solar interior.

In the low frequency range, we are dealing with gravity modes (g
modes), for which buoyancy in radiative zones ($N^2$ positive) provides the restoring force.
Their radial wave number is given by: 
\begin{equation}
k_{\rm r} \sim {k_{\rm h}}{ N\over
\omega}.
\end{equation}
The properties of these modes therefore depend
principally on $N$, and the frequencies decrease with increasing
radial order $n$.  Because g modes oscillate in regions where their
frequencies are smaller than the Brunt-V\"ais\"al\"a frequency,
g modes are confined in the solar radiative zone and core
(see Figure~\ref{fig:Vais} and Figure~\ref{fig:mixed}).  Low-degree
gravity modes have much larger displacements in the horizontal than the vertical, the opposite of what is shown by p
modes.

The g modes are attenuated in the convection zone with a factor
proportional to ${r_{{\rm cz}}^{L}}$, $ r_{{\rm cz}}$ being the radius
at the base of the convection zone \citep{dalsg80,GB90}, so that
low-degree g modes may be easier to detect than their higher-degree
cousins.  We shall see in the next subsection that the frequency
spectrum shown by the g modes is manifested as a pattern of roughly
equidistant periods for modes of given degree $l$.

\begin{figure*}
 
\centerline{
\includegraphics[width=0.7\textwidth]{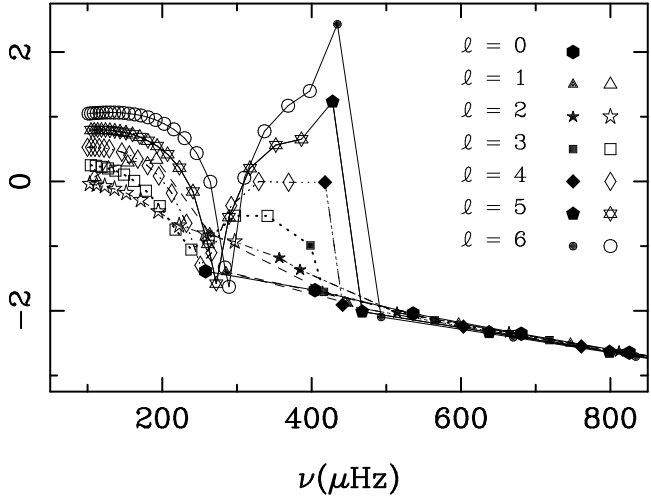}
}

 \caption[]{Logarithm of the normalized energy $\omega^2{\cal I}_{n,l}$ of low-frequency g and mixed modes (open symbols) and
low-frequency p modes (filled symbols) for a reference solar model M1
of \citet{JPGB2000}, plotted as a function of frequency, for modes of
degree $l$=0, 1, 2, 3, 4, 5 and 6.  (The normalization is taken assuming that the modes have equal amplitudes  at the
photospheric level, where the temperature equals the effective
temperature level).  Note the transition from p to g modes around
450\,$\mu$Hz, and the existence of a set of modes of mixed character
around 280\,$\mu$Hz, having lower energies than modes in the
neighbouring frequency regions \citet{JPGB2000}.}

\label{fig:energie1}\end{figure*}

\begin{figure*}
\centerline{
\includegraphics[width=0.7\textwidth]{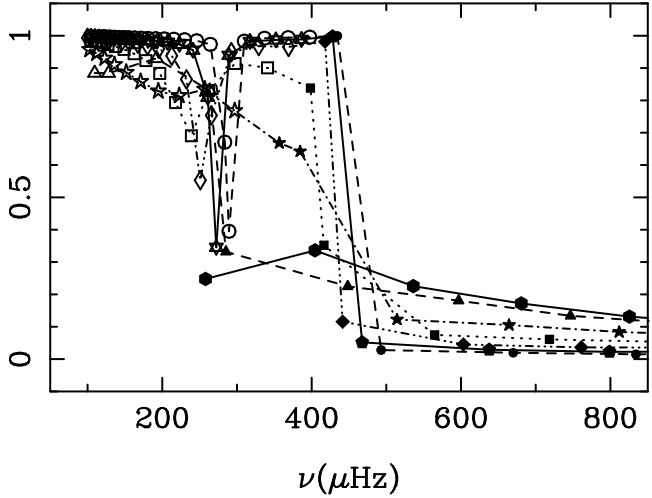}
} 

\caption[]{Relative contribution of the buoyancy restoring force to
the frequency, ${\cal K}_2/ ({\cal K}_1+ {\cal K}_2)$.  ${\cal K}_1$
and ${\cal K}_2$ are defined in Eq.~(\ref{eq3J}), for the same
modes, model, and symbols in Figure~\ref{fig:energie1}.  From
\citet{JPGB2000}.}

\label{fig:energie2}\end{figure*}

In an intermediate frequency range between 200 and 400\,$\mu$Hz, there
are low radial-order g modes, f modes (surface gravity oscillations that have
no radial nodes) and p$_{1}$ modes (a mode with one radial
node). Some of the modes, called mixed modes, have a pronounced mixed
character, i.e. they have amplitude both in the central layers, like g
modes, and in the external layers, like p modes (e.g. mode p$_{1}$ ($l=1$)
in Figure~\ref{fig:mixed}).  The properties of mixed modes have been
studied in detail by \citet{JPGB2000}.  The mixed character of these
modes appears very clearly when we look at their inertiae, ${\cal
I}_{n,l}$. The inertia may be written
 \begin{equation}
 {\cal I}_{n, l}=  
  \int_V\rho\,{\boldsymbol\xi}\cdot{\boldsymbol\xi}\,{\rm d}V\, ,
 \label{eq2J}
 \end{equation}
where the eigenfunctions $\boldsymbol\xi$ are normalized at the surface of the Sun and where the integral is taken over the whole volume of the Sun.
Figure~\ref{fig:energie1} plots the normalised kinetic energy of the
modes, which is proportional to the mode inertia, as a function of
frequency.  Under the assumption of equipartition of the energy in the modes,
the surface amplitudes would be approximately inversely proportional to the
square root of the mode energy \citep[e.g.][]{GB90}.

For g modes with frequencies less than 200 $\mu$Hz, the lower the
degree, the smaller is the energy, and the higher is the surface
amplitude.  Around 280~$\mu$Hz, modes of mixed character have smaller
energies and therefore higher surface amplitudes than modes adjacent
in frequency. We discuss the excitation and damping mechanisms of g
modes in detail in Section~\ref{sec:excite}.

The frequencies of the oscillation modes may be expressed using a
variational principle \citep[e.g.][]{Unno89}.  Neglecting surface
terms and the perturbation of the gravitational potential, we have:
 \begin{equation}
 \left({2\pi\tilde{\nu}}\right)^2 \sim
 \frac{1}{{\cal I}_{n, l}}
 \int \left(\frac{\tilde{p}^2}{\gamma_1 p}\right) {\rm d}r+
 \frac{1}{{\cal I}_{n, l}} \int {N^2   \xi_{\rm r}^2 \rho r^2 {\rm d}}r 
     \equiv \frac{( {\cal K}_1+{\cal K}_2)}{{\cal I}_{n, l}},
 \label{eq3J}
 \end{equation}
where the Eulerian pressure perturbation, $\tilde{p}$, is given by
$\tilde{p}=\rho r \omega^2 L^{-1} \xi_{\rm h}$.  The quantities ${\cal K}_1$
and ${\cal K}_2$ correspond, respectively, to the contributions to the
frequency of pressure and buoyancy forces.

Figure~\ref{fig:energie2} shows the variation of ${\cal K}_2/ ({\cal
K}_1+ {\cal K}_2)$ as a function of the frequency, for the modes of a
reference solar model.  As expected, for p modes there is a dominant
contribution to the frequency from the ``acoustic'' term ${\cal K}_1$,
while for g modes it is the the ``buoyancy'' term, ${\cal K}_2$, which
dominates. Mixed modes, between 200 and 450\,$\mu$Hz, have frequencies
defined by significant contributions from both terms.

\subsubsection{Asymptotic approximation}

Low-degree g modes with cyclic frequencies less than $\nu_{n,l}\le 200 \mu$Hz
may be described by an asymptotic relation
 \citep{Vandakurov1968,TASSOUL80,Olver1956}. For small frequency, the second-order asymptotic expression for the period $P_{n,l}$
 ($P_{n,l}=1/\nu_{n,l}$) can be represented by:
 \begin{equation}
 {P}_{n,l}\sim \overline{P}_{n,l} = {P_0\over
 L}\left(n+l /2- {1\over 4}
 +\vartheta\right)
 +{P_0^2\over \overline{P}_{n,l}}{L^2V_1+V_2\over L^2}, 
 \label{eq4J}
 \end{equation}

 \begin{equation}
 \hbox{with}\ \ \ \ \ P_0={{2\pi^2}\over 
 \int_0^{r_{\rm cz}}{({N}/{r})dr}
 },
 \end{equation}

 \begin{equation}
 V_1={1\over P_0}\lim\limits_{r_{\varepsilon} \to 0}
 \left(\int_{r_\varepsilon}^{r_{\rm cz}}{dr\over N(r)} -{1\over N(\varepsilon)}\right)
 \ \ \ \ \hbox{and}\ \ \ \ L^2=l (l +1).
 \end{equation}

 \noindent
Here, $\vartheta$ is a phase factor sensitive to the properties of the
layers lying below the convection zone
\citep{1986A&A...165..218P,Ellis86}.  If $N^2$ is assumed to vanish
proportionally to $(r_{cz}-r)^p$ for some power $p$ at the base of the convection zone,
$\vartheta$ tends at low-frequency to a constant $- 0.5/(p+2)$
\citep[see also for details][]{1991SoPh..133..127B}. For the standard solar model, a linear behaviour of $N^2$ may be assumed and $\vartheta$
tends to $-1/6$.  The second-order coefficient $V_1$ depends on the
Brunt-V\"ais\"al\"a frequency.  $V_2$ depends in a more complicated way on
the stratification \citep{1991SoPh..133..127B}.

In sum, the above tells us that the frequencies of g modes are related
closely to the Brunt-V\"ais\"al\"a frequency, particularly through
$P_0$ and $V_1$.  Typical values of $P_0$, $V_1$, and $V_2$ for a
reference solar model are, $P_0 \sim 35$ to $36\,\rm min$, $V_1\sim
0.4$ and $V_2\sim 5.7$.

The periods of g modes of a given degree $l$ are proportional, in the
first order, to $P_0/L$. The separation in period between modes
adjacent in $n$ is therefore almost equidistant. In the low-frequency
asymptotic limit, the relative frequency difference of two modes of
the same degree, $l$, and same radial order, $n$, from two different
models is equal to:
 \begin{equation}
  {\delta\nu_{n,l} \over \nu_{n,l}} =  -{\delta P_{n,l}  \over P_{n,l}}
 \sim -{\delta P_0  \over P_0}.
 \label{eq4Jbis}
 \end{equation}
Detailed comparisons of the numerical periods and the asymptotic
periods \citep{1986A&A...165..218P,jcdgb91} have allowed checks to be
made of the validity of the asymptotic formula. A least-squares fit
minimising $\Sigma_{n,l}(P_{n,l}-\overline{P}_{n,l})^2$, for g modes
of degrees $l= 1, 2, 3, 4$ and radial order $n_{\rm min}\le n\le
n_{\rm max}$, allows the values of $P_0$, $V_1$, $V_2$, and
$\vartheta$, to be estimated It has been shown that by varying the
values of $n_{\rm min}$ and $n_{\rm max}$ -- for example $n_{\rm min}$
from 10 to 15 and $n_{\rm max}$ from 27 to 35 -- and by taking the
mean values of the results for the global quantities, it is possible
to determine from a set of numerical periods the quantities $P_0$ and
$V_1$, related to the structure of the inner radiative zone.  The
values of $P_0$ and $V_1$ are then shown to be very close to their
asymptotic values.

Similar good agreement is shown between the numerical periods
${P}_{n,l}$ and the asymptotic periods $\overline{P}_{n,l}$. The
fractional agreement is of order 10$^{-3}$ or less for radial orders
larger than 10, that is for frequencies less than 60~$\mu$Hz at $l=1$,
and for frequencies less than 100~$\mu$Hz at $l=2$.  At lower radial
orders, the effects of the third-order terms and of the non-constant
$\vartheta$, which are neglected here, may account for fractional
differences of up to 5~10$^{-3}$.  

These asymptotic properties of g modes may be exploited in attempts to
detect the modes, i.e., by searching for signatures of near-regular
patterns in period, as we shall discuss in detail in Section~4.


\subsubsection{Effect of rotation : splittings, rotation kernels }

\begin{figure*}
\begin{center}
 \includegraphics[width=0.75\textwidth]{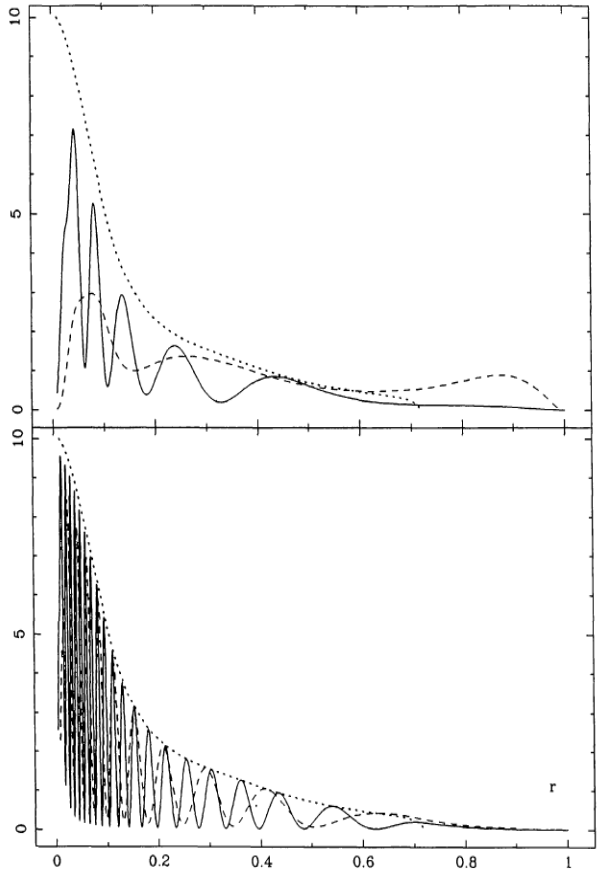}
 \end{center}

\caption[]{Rotational kernels ${\cal K}_0^{n, l}$, caused by a rotation constant on spheres (cf. Eq. \ref{eq6J}), plotted as a function of radius $r/R_\odot$.
The quantity $R_\odot \times {\cal K}_0^{n, l}$ is plotted for modes $l=2$ with radial order $n= 5$ (full line) and $n= 1$
(dashed line) in the upper panel and $n= 20$ (full line) and $n= 10$
(dashed line) in the lower panel.  The dotted lines represent the
asymptotic envelope given by the function $(P_0/\pi^2)(1-1/L^2)N/r$
\citep[from][]{1991SoPh..133..127B}.}
\label{fig:noyrot}
\end{figure*}

The rotation of the Sun induces a splitting of each multiplet frequency $\nu_{n,
l}$ into 2$l$+1 separate singlet frequencies $\nu_{n, l, m}$, where $m$ is the
azimuthal order of the oscillation ($-l \le m \le l$). We define the
splitting as:
\begin{equation}
\sigma_{n, l, m} = (\nu_{n, l, m} - \nu_{n, l, 0}).
\end{equation}
The effects of the Coriolis force and distortions of the model must
also be considered. The angular velocity $\Omega$ is small compared with
the characteristic frequency $\Omega_{\rm g}=\sqrt{G M/R^3}$ (${\Omega /
\Omega_{\rm g}} \sim 5 \times 10^{-3} $), so that the distortion of the model, which
depends on ${(\Omega/\Omega_{\rm g})}^2$, may usually be neglected.  Generally a
first-order analysis in the ratio $\Omega / \omega$ is adequate to
compute the rotational splittings \citep{Cowling1949,Ledoux51}. In this
first-order approximation, the splittings in an inertial frame are
related to the rotation by
 \begin{equation}
 \sigma_{n, l, m} = {
m \int{\Omega \vec{\xi}\cdot\vec{\xi}^* \rho {\rm d}V}
  -{\rm i}\int{ (\vec{\xi} \times \vec{\xi}^*)\cdot\vec{\Omega}  \rho {\rm d}V}
 \over
 \int{ 2 \pi \vec{\xi}\cdot\vec{\xi}^*  \rho {\rm d}V}},
 \label{eq5J}
 \end{equation}
where the integrals are taken over the whole solar volume, d$V$
being an element volume.  Moreover, it is sufficient to adopt the
corresponding (unperturbed) eigenfunctions $\xi$ of the nonrotating
stellar model to evaluate the integrals.  If the magnitude $\Omega$ of the
angular velocity $\Omega$ is constant on spheres the splitting can be
represented by the simple weighted averages:
 \begin{equation}
 \sigma_{n, l, m}= m\int{
 \Omega_0(r){\cal K}_0^{n, l}(r) {\rm d}r },
 \label{eq6J}
 \end{equation}
 \noindent
with 
\begin{equation}
 {\cal K}_0^{n, l}= \rho r^2 
 {
 \xi_{\rm r}^2+\xi_{\rm h}^2- 2 L^{-1}\xi_{\rm r}\xi_{\rm h} -L^{-2}\xi_{\rm h}^2
 \over
 {\int_0^{R_\odot}}{(\xi_{\rm r}^2+\xi_{\rm h}^2) \rho r^2 {\rm d}r}},
\end{equation}
Some typical examples of ${\cal K}_0^{n, l}$ are plotted in
Figure~\ref{fig:noyrot}; they have large amplitudes in the solar
core, implying a high sensitivity to core rotation.

As noted previously, low-frequency g modes of low degree have much 
larger displacements in the horizontal direction than
in the radial direction, so that the simplified expression for
the rotational kernels depends only on $\xi_{\rm h}^2$. Using the asymptotic
expressions for the eigenfunctions
\citep[e.g.][]{TASSOUL80,1986A&A...165..218P},
\citet{1991SoPh..133..127B} have shown that the asymptotic kernels
have a sensitivity envelope in the radiative zone that is proportional
to $(1-1/L^2)N/r$. This demonstrates the sensitivity of the frequency
splittings of the gravity modes to the solar core rotation
(Fig~\ref{fig:noyrot}).

We know from the inversion of observed p-mode frequency splittings
that the solar rotation is not dependent only on radius, but that the rotation in the convective
zone varies with latitude as much as the rotation observed at the solar surface \citep[]{Mike1996}.

If we take a simple law that reflects this differential rotation namely 
$\Omega=\Omega_0(r)+ \Omega_1(r)\cos^2(\vartheta)$, we may then
express the frequency splittings $\sigma_{n, l, m}$ of the g modes by 
 \begin{equation}
 \sigma_{n, l, m}= m\left[
 \overline{\Omega_0}\left(1-{1\over L^2}\right)+\overline{\Omega_1}\left(C_{l, m}(1-{6\over
   L^2})+{1\over L^2}\right)
 \right],
 \label{eq7J}
 \end{equation}
with 
\begin{equation}
C_{l, m}={1\over 2l+1} \left[ 
 {(l +1)^2-m^2\over 2l +3} +{l^2-m^2\over 2l-1}
 \right],
\end{equation}
where  $\overline{\Omega_{0}}$ and $\overline{\Omega_{1}}$ are mean values of the
rotation coefficients seen by g modes, and
 \begin{equation}
 \overline{\Omega_{0,1}}={\int{\Omega_{0,1}(r)(N/r){\rm d}r}
 \over
 \int{(N/r){\rm d}r}}. \label{eq8J}
 \end{equation}

An important point to be made about the equations above is that
$\sigma_{n, l, m}\sim m \overline{\Omega_0}(1-{1\over L^2})$, so that,
at low frequencies, the splittings of $l= 1$ g modes are about half the
size of the splittings of $l= 2$ g modes. This must be taken into
account when attempts are made to identify g modes in the
low-frequency range. When g modes with very low frequencies are
considered, the ratio $\Omega / \omega$ is no longer small and another
asymptotic analysis has to be made to take into account the fact
that Coriolis effects cause the oscillation eigenfunctions to be no longer described by a single spherical
harmonic \citep[e.g.][]{GB78,WDAK1987,Dintrans2000,Mathis2008}.  These modes can be expanded in terms of a
hierarchy of unperturbed eigenfunctions $\xi$, a procedure which is valid
because $\Omega/\Omega_g$ is small even though $\Omega/\omega$ is not.  However the usual first-order
expression for the frequency splittings is relevant for modes of
frequency larger than about 20$\mu$Hz \citep{1991SoPh..133..127B}.

\subsection{g-mode  frequency predictions and their reliability}

In order to aid observational attempts to detect and identify
low-frequency p and g modes in the frequency spectrum, several groups
have computed theoretical g-mode frequencies
\citep[e.g.][]{JPGB2000,couvidat2003,Mathur2007,cox04}. The reliability
of these predicted frequencies depends first on the uncertainties of
the solar modelling and on the sensitivity of the low-frequency
oscillations to these uncertainties, and second on the numerical
precision of the frequency calculations. Another important uncertainty
lies in the estimation of the frequency splittings of the g modes,
because these splittings depend of course on the rotation close to the
centre of the Sun, which is not well constrained observationally.

\subsubsection{Modern solar models and their uncertainties}

The large number of p-mode frequencies already detected lead to a
{\it seismic} model of the Sun \citep[e.g.][]{Gough2004} that is found to have a structure that
is very close to the structures shown by modern solar models
\citep[e.g.][]{JCD96,DOGKetal96}. The only regions where the structure
inferred by p-mode observations of the p modes remains uncertain is in the
central few percent of the solar core. (Uncertainty in the angular velocity, as inferred from
the frequency splittings of p modes extends through a large part of the solar core.)


Standard solar models are assumed to be spherically symmetric, in hydrostatic
equilibrium with no macroscopic motion except convection, no mass loss
nor accretion, no rotation and no magnetic field.  The structures are
determined by the assumed microscopic properties of the solar material
(equation of state, opacity, nuclear reaction rates etc.), and by
the description of energy transport by convection usually 
\citep[mixing-length prescription, e.g.][]{BV58,CM91,DOGKetal96,STC93}.  The numerical codes leading to these solar models have been carefully compared \citep{Monteiro2009}.  Solar
models are calibrated by adjusting the initial abundances $Y_{\rm i}$ and $Z_{\rm i}$ 
of helium and heavy elements, and (usually only a single) mixing-length
parameter which determines the efficacy of the convective transport
in a mixing-length description, to obtain, for a one-solar-mass model,
the solar radius and the solar luminosity at the solar age, and the
observed surface metallicity $(Z/X)_\odot$, where $X=1-Y-Z$ is the abundance of hydrogen.   \citet{2006ApJ...644.1292A} attempted to determine the solar heavy-element abundance without reference to spectroscopy by
using helioseismic data.

The introduction into the models of microscopic diffusion and settling of chemical
elements decreases the surface helium content,
reducing the discrepancy between the observed and modelled sound-speed
profiles, and also bringing the surface helium content, and the
convection zone depth, of the models into better agreement with the
observations.  \citep[e.g.][]{Burgers, CoxGuzKid89, ProfMich91,
MichProf93, JCD93, Thoul94, Morel97, ASB98}. But what of the heavy-element abundances?

The quantity of heavy elements assumed in the models is extremely
important, because these elements dominate the radiative
opacities. Usually one considers only the total heavy element
abundance $Z_\odot$ or the so-called metallicity $(Z/X)_\odot$.  Until
2004, the solar abundances were taken from results given by two rather
close spectroscopic analyses, namely \citet{grev93} (GN) and
\citet{grsauv98} (GS).  However, since then new heavy element
abundances have been inferred using a 3D, NLTE analysis of the
spectroscopic data, which takes hydrodynamics into account in the
transfer calculations as in \citet[][hereafter AGS]{AGS05,AG2009} and \citet[][]{Caffau2009}.

Amongst other differences, the AGS abundances of C, N, O, Ne are substantially lower than the older GN and GS abundances, by $0.16$~dex, $0.19$~dex, $0.21$~dex and $0.24$~dex. The
reduction lowers the total surface metallicity of
$(Z/X)_\odot$ (0.0166, compared to 0.0245 for GN and 0.023
for GS abundances). These changes have led to an extensive discussion
in the literature on the robustness of the new results, which is
beyond the scope of this paper, \citep[e.g.][]{basuantia07,caffau08}.  In particular, it led to an attempt by
\citet{2006ApJ...644.1292A} to determine the heavy-element abundance
seismologically without reference to spectroscopy,  by calibrating the
average effect of ionization on $\gamma_1$.   However,
\citet{2006ApJ...644.1292A} simply adopted a particular equation of state which, like
all others, was known not to be exact \citep[cf.][]{Baturin2000}, so their
result cannot be relied upon.

Solar models constructed with the GN or GS abundances in the generally accepted opacity formulae
are in good agreement with helioseismic constraints. For example, the agreement
for the sound-speed profile is within 0.2 percent, as shown in
Figure~6 \citep[e.g.][]{JCD96,DOGKetal96}.  In attempts to account for the discrepancies that remain, several non-standard
physical processes have been included in the solar modelling, in
particular processes to improve the description of the layers located just
beneath the convection zone where the most marked differences between the
observations and the models occur.  We know from helioseismology that
the transition from differential rotation in the convection zone to
almost uniform rotation in the radiative zone below occurs in a
shallow layer called the tachocline
\citep[e.g.][]{Spiegel92,AK1996,Mike1996,Corbard99}.  The
shear of the rotational velocity  necessarily induces mixing \citep{HRW2005}.
Various descriptions of this mixing have been included in some models,
leading to reductions in the differences between the observations and
models.  \citep[e.g.][]{Richard96,MG97,Morel98,ASB99,Elliott1999}.


Other non-standard processes, such as overshoot of convective elements
at the boundary of the convective core which appears at the end of
the pre-main-sequence evolution, or mass-loss occurring during the
initial stages of solar evolution, have been studied, with special
emphasis placed on the changes induced in the most central parts of
the models which are not well constrained by observations of p modes
(see below).

The AGS abundances have caused considerable concern.  This issue is simple to state:
Except in the near-surface layers, which are of no import to this discussion, we already know from seismology the sound speed and the density stratification throughout the Sun to a precision easily good enough to be confident in the inference.  If one then accepts the basic tenets of solar evolution theory - namely that a helium-abundance augmentation has occurred in the core due to nuclear transmutations on the main sequence and that the Sun is in thermal balance, energy being transported throughout the entire region beneath the convection zone by radiative transfer - and if one accepts also the equation of state (which we surely know to well within the limits set by the abundance change) and the nuclear reaction, then one can infer the absolute value of the hydrogen-helium abundance ratio and hence the temperature stratification (by demanding that the energy-generation rate in the core is equal to the luminosity observed at the surface) from which can be inferred the {\it value} of the opacity beneath the convection zone \citep{Gough2004}.  The issue, therefore, is simply how to reconcile that value with the photospheric abundances.  At first sight, the apparently most straightforward resolution is that the abundances in the convection zone differ from these in the radiative interior, presumably as a results of accretion of metal-deficient material after the Sun has (almost) reached the main sequence and had a relatively shallow convection zone, which is contrary to standard assumptions.  We should point out, however, that \citet{Guzik06} and \citet{Castro2007} have failed to construct such a model that is in adequate agreement with helioseismology.

\begin{figure}
\begin{center}
\includegraphics[width=0.45\textwidth,angle=0]{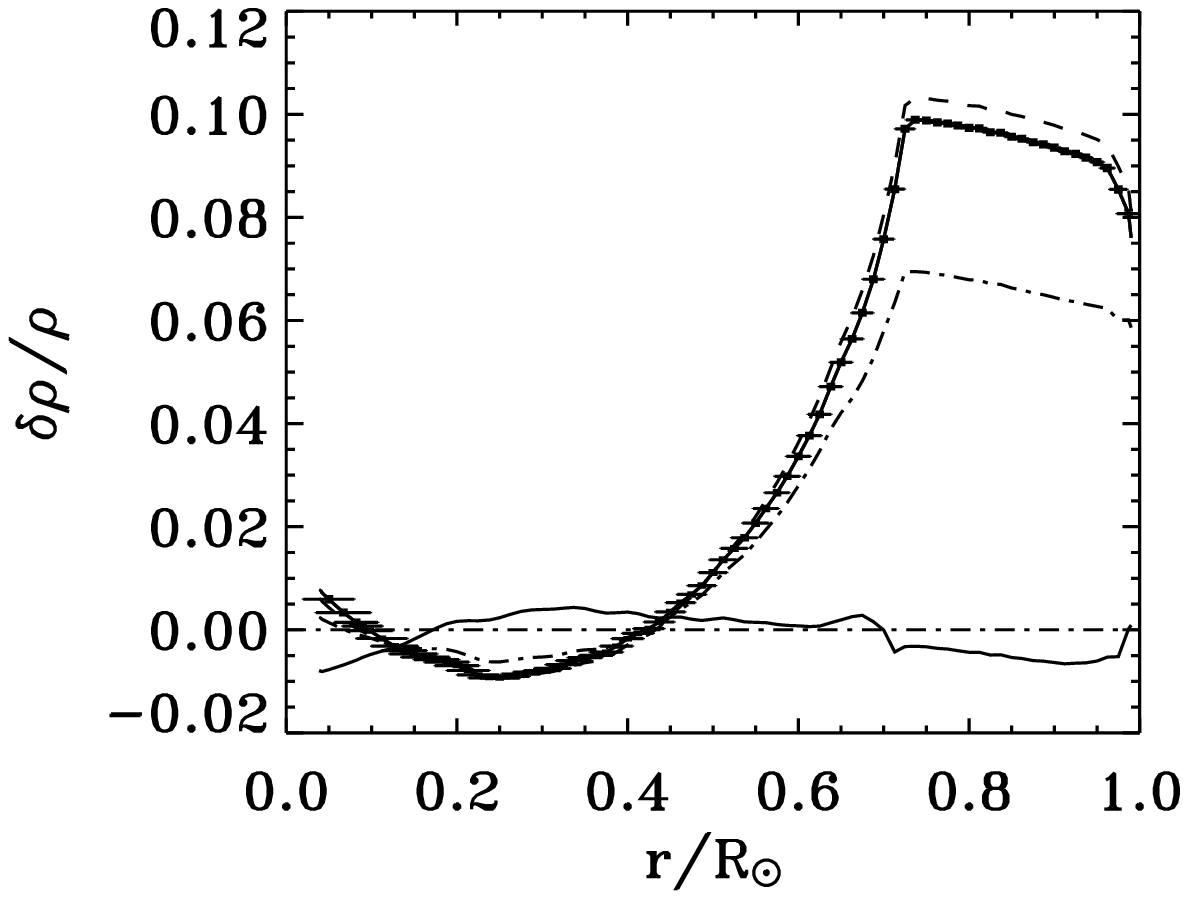}
\includegraphics[width=0.47\textwidth,angle=0]{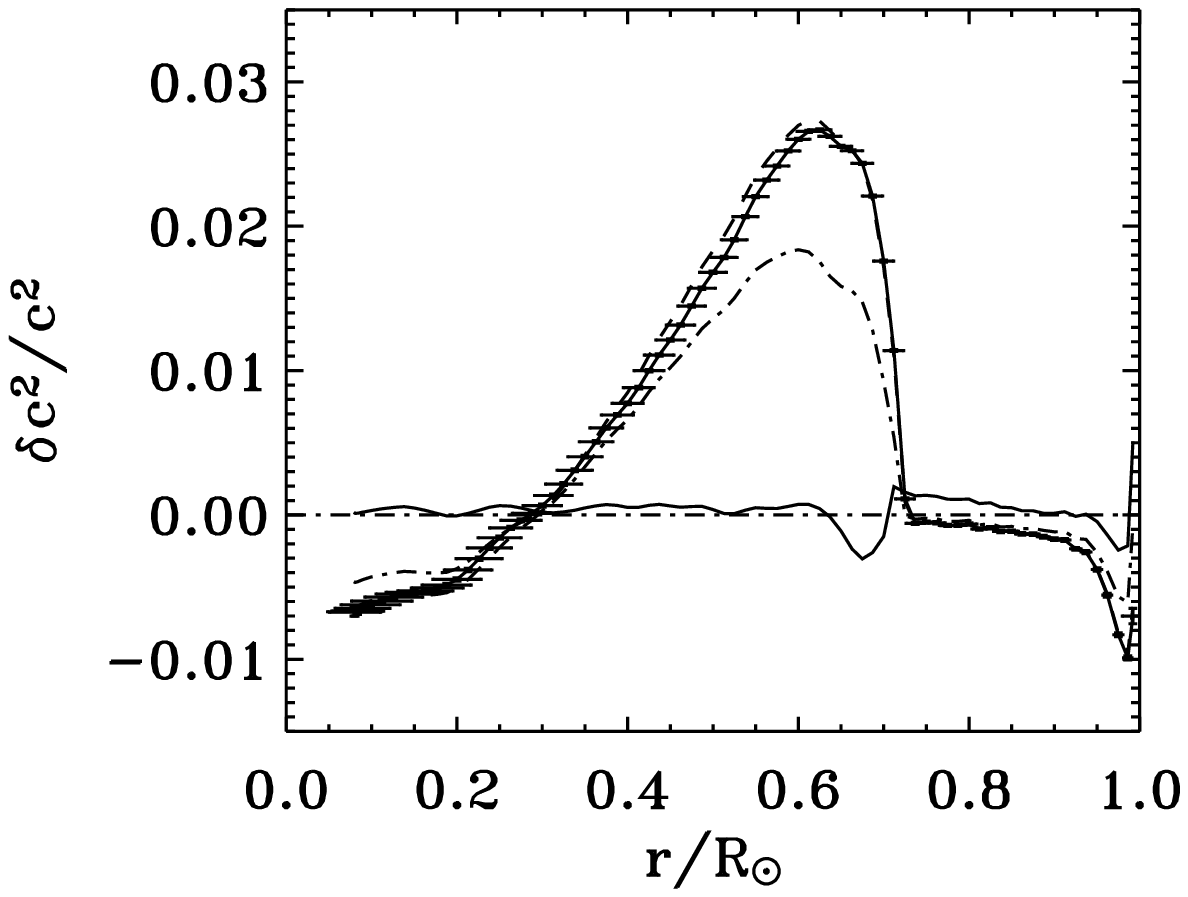}
\end{center}

\caption[]{
Relative density profile and relative squared sound-speed differences between standard models using a new estimate of the CNO composition proposed by \citet{AGS05} (full line with seismic error bar) and the helioseismic observations. The dot-dash line corresponds to a model computed with  abundances from \citet{Holweger2001}, intermediate between AGS and GN.  The seismic model (thin line) is an model which reproduces properly the observed sound speed and is used to predict gravity modes and neutrino fluxes \citep[After][]{tc-couvidat04}.}


\label{fig:bac.bas}\end{figure}

It follows that using the AGS abundances in the accepted opacity formulae leads to solar models which disagree
with the helioseismically determined constraints on the solar interior
properties \citep[e.g.][]{tc-couvidat04, mmng04, Bahcaletal05, guz05,
zpbmc06, zpbmc07}.  There is a large discrepancy in the sound speed
(e.g. Figure~\ref{fig:bac.bas}). The surface helium content of these
new models is too low, and  the convection zone is too
thin, compared to the observed (helioseismic) values.

Many other attempts, and suggestions, have been made to try to reconcile the
AGS abundances with the helioseismic results \citep[e.g.][]{Guzik06}. Evidently, increasing the radiative opacities below the convection zone
\citep[e.g.][]{sbbp04,mmng04,bsp04} decreases the discrepancies if one insists on lowering the heavy-element abundances to the AGS values.  Using artificially enhanced diffusion rates
for He and heavy elements \citep{basu04,mmng04,guz05,YangBi07}, does
improve the agreement with seismic sound speed, though such changes by the amount required have not been justified. Rotational mixing,
acting in the tachocline just beneath the convection zone can lead to
models with acceptable values of the surface helium abundance, but
their convection zones are too shallow
\citep[e.g.][]{tc-couvidat04}. Taking account of tachocline mixing (perhaps calling it convective overshoot) helps to bring the convection zone depth closer into
line, but the discrepancies shown in the sound-speed profile are not
necessarily reduced in size \citep[e.g.][]{mmtng06}. 

In sum, at the time of writing the solar abundances remain a major
source of uncertainty in solar modelling \citep[e.g.][and references therein]{basuantia07}.


Finally in this section we consider mechanisms to describe transport
of angular momentum and chemical elements which are not fully
accounted for in solar and stellar evolution codes.  The respective
r\^oles in angular momentum transport of dynamical processes involving
rotation, magnetic fields
\citep[e.g.][]{1998Natur.394..755G,Garaud2002} and gravito-inertial
waves have been much discussed in the solar context
\citep[][]{Schatzman1993, Schatzman1996, TalonZahn98,
2002ApJ...574L.175T, Mathis2008,Gough2009}.  When only the meridional
circulation and the ``classical'' hydrodynamic instabilities are
invoked, models predict a Sun with a radiative interior and core
rotating much faster than the surface
\citep[e.g.][]{Pinsonneault-et-al1989}, which is incompatible with p-mode helioseismology \citep[e.g.][]{Mike1996}.
\citet{MathisZahn2005} have recently derived, in a self-consistent
manner, a formalism to describe the effects of an axisymmetric
magnetic field on the meridional circulation and the turbulence
generated by the shear of differential rotation. This formalism is now
implemented in some evolution codes.

\subsubsection{Numerical precision}

The numerical precision of the frequencies depends both on the
physical and numerical accuracy of the evolutionary models and on the algorithms
used to compute the oscillation frequencies themselves.  Many
evolution codes have been compared, in the 1990s for the Sun (GONG
solar model comparison) and more recently in a stellar context as part
of the CoRoT Evolution and Seismic Tools Activity (ESTA)
\citep{Lebreton07,Monteiro2009}.  The numerical precision of the frequencies may be
estimated for a given stellar model by considering the range of
values obtained from different oscillations codes, which compute the
mode frequencies \citep{Moya07}.

For a given oscillations code, an internal accuracy may also be
estimated by comparing a numerical eigenfrequency $\nu_{n, l}$ with
its integral value, as obtained from Eq.~(\ref{eq3J})
\citep[e.g.][]{JP07}.  Any inconsistency, either in the computation
of the oscillation frequencies or in the equilibrium models, gives 
rise to a non-zero value of $\tilde{\nu}_{n, l} -\nu_{n, l}$ as, of course, does the neglect of the surface term in Eq.~(\ref{eq3J}) which are actually negligible only at low frequency.

\begin{figure}  
\begin{center}
\includegraphics[width=0.65\textwidth]{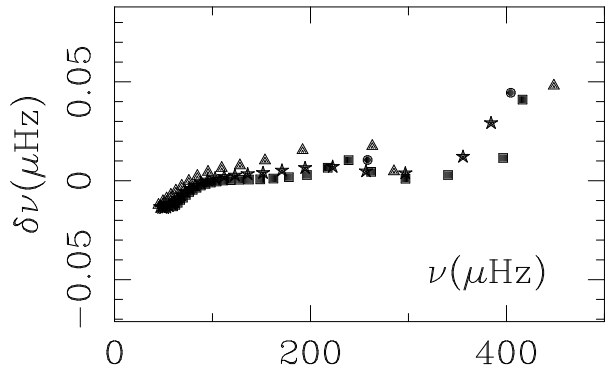}
\end{center}

\caption[]{Differences $\delta\nu= \tilde{\nu}_{n, l} - \nu_{n, l}$
between eigenfrequencies $\nu_{n, l}$ computed by an oscillations
code, and predictions of the frequencies given by the integral
expressions for a reference solar model, $\tilde{\nu}_{n, l}$ [cf.
Eq. (\ref{eq3J})] for modes of degree $l$=0 (dots), 1 (triangles), 2 (full
stars) and 3 (squares).}

\label{fig:consistency}\end{figure}

For g-mode calculations one also needs a solar model with a
sufficiently large number of mesh points (where the model quantities
are given explicitly), well distributed in solar radius, in particular
close to the centre of the model where g modes have large amplitudes
\citep[e.g.][]{jcdgb91}.  Figure~\ref{fig:consistency} shows that, for
a reference solar model, a numerical precision of g-mode frequencies
-- as measured by $\tilde{\nu}_{n, l} -\nu_{n, l}$ -- better than
0.01 $\mu$Hz for frequencies lower than 200~$\mu$Hz has been achieved. This difference
is much larger than uncertainties we expect to obtain from
observations (should the g modes be detected). Therefore the numerical procedures should be improved.  However, as we shall
show below, this error is much smaller than errors introduced by
uncertainties in the theoretical structure and the dynamics of the
models.  Therefore g modes have diagnostic potential.


\subsubsection{Sensitivity of low-frequency g- and mixed mode frequencies to solar  models}

\begin{table}[t]

\caption{Asymptotic coefficients of g modes, i.e., $P_0$, $V_1$,
$V_2$, and $\vartheta$. Plotted are mean values of the coefficients
for $n_{\rm min}$ from 10 to 15 and $n_{\rm max}$ from 27 to 35, as
computed for models M1 to M8 from \citet{JPGB2000} (1), M-GN \& M-AGS
from \citet{zpbmc07} (2).  The estimation of the uncertainty, in
percent, was obtained from the scatter of the least-squares fit values
around their mean values.  In this table the following differences
show the following effects (for details see \cite{JPGB2000}): (M6-M1)
shows the effect of increasing age from 4.65 to 4.75\,Gyr; (M2-M1)
shows the effect of increasing the surface metallicity from 0.0245 to
0.0260; (M3-M1) and (M8-M1) show the effects of mass loss, set to a
rate of $\dot{M}= -5\times 10^{-10}M_\odot y^{-1}$, occurring for 200\,Myr
at the beginning of the evolution, with additional turbulent mixing
below the convection zone in model M8, according to
\citet{2000A&A...354..943M}; (M4-M1) and (M5-M1) show the effects of
overshoot of, respectively, 0.2 and 0.25 pressure scale heights at the
edge of the convective core which appears at the end of the pre-main
sequence evolution; (M1-M7) shows the effects of improved nuclear
reaction rates; (M-AGS - M-GN) shows the effects of changing the
chemical abundances from GN to AGS (see text); and finally (M-AGS -
M-GN) shows the effects of changing the chemical abundances from GN to
AGS (see text).}

\centering
\begin{tabular}{lllllllllll}
\hline\noalign{\smallskip}
   & $P_0$  & \%  &  $V_1$   &  \%    &  $V_2$  &  \%    & $\vartheta$  &  \% \\
&(mn)&&&&&&&&&\\
\tableheadseprule\noalign{\smallskip}
 M1      &35.08  &0.04  &0.437  &1.6  &5.69  &4   &-0.160  &11&(1)\\
 M2      &34.94  &0.04  &0.439  &1.6  &5.67  &4   &-0.161  &11&--\\
 M3      &34.63  &0.04  &0.444  &1.6  &5.72  &4   &-0.161  &11&--\\
 M4      &35.39  &0.04  &0.430  &1.6  &5.64  &4   &-0.161  &11&--\\
 M5      &36.75  &0.03  &0.399  &1.4  &5.46  &3   &-0.164   &8&--\\
 M6      &34.77  &0.04  &0.440  &1.6  &5.71  &4   &-0.161  &11&--\\
 M7      &35.42  &0.04  &0.433  &1.6  &5.61  &4   &-0.162  &11&--\\
 M8      &34.78  &0.07  &0.437  &1.3  &6.12  &6   &-0.217  &14&--\\
&&&&&&&&\\

M-GN    &35.09  &0.03  &0.442  &1.5  &5.77  &3   &-0.166  &9&(2)\\
M-AGS   &35.69  &0.04  &0.433  &1.4  &5.96  &3   &-0.157  &11&--\\

\noalign{\smallskip}\hline
\end{tabular}
\label{tab:asymp}       
\end{table}

As a first step, the sensitivity of g-mode frequencies to
uncertainties in the solar models may be studied by taking advantage
of the asymptotic properties of g modes.  As noted previously, the
three global coefficients $P_0$,
$V_1$, $V_2$, and the phase factor $\vartheta$, which characterize the frequency may be obtained by a least-squares fit to a set of numerical frequencies. The values of $P_0$ and $V_1$ are
typically very close to those derived directly from the integrals (cf.
Eq.~(\ref{eq4J})), and results for some models are given in
Table~\ref{tab:asymp}. In this table one can see the effects of
changing some fundamental model parameters, such as the age to which
the model is evolved, the metallicity, and also some important physics such as the nuclear reaction rates.

We find that the values of the asymptotic coefficients are very
similar for models with slightly different fundamental parameters and
different physics, including changes between the old and new
metallicities.  Ignoring gravitational settling increases $P_0$ by about 0.7\,min \citep{Morel97a}.  \citet{Mathur2007} have found
that the theoretical value of $P_0$ changes by no more than 1\,min when models with
plausible differences in treatments of microscopic diffusion,
turbulence in the tachocline, and the chemical composition, are
considered.  The results give an idea of the possible
range of values for these asymptotic coefficients.

\begin{figure}
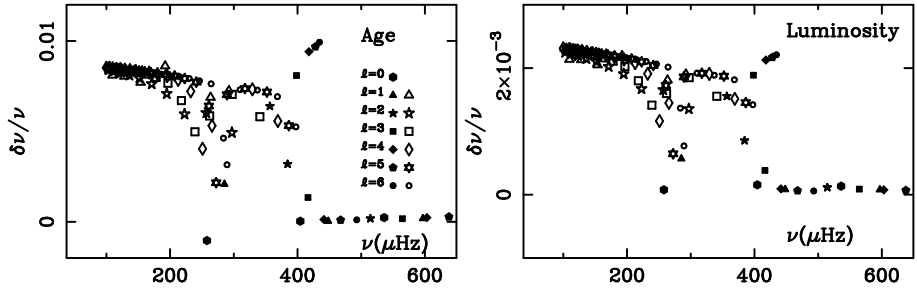

\vbox{
\includegraphics[width=0.49\textwidth]{fig8_AA_age.ps.rev}
\vspace{0.5truecm}
\includegraphics[width=0.49\textwidth]{fig8_AA_lplus.ps.rev}
}

\caption[]{Relative frequency differences, $\delta\nu/\nu$, between a
solar model and the reference model M1. Results are shown for g modes
(open symbols) and low-frequency f and p modes (full symbols) with
degrees $l$ = 1 to 6, to test the sensitivity to the solar parameters:
age (model M6 - $\delta t = 0.1Gy$), and luminosity ($\delta$ L/L=3.6
10$^{-3}$).  Adapted from \citet{JPGB2000}.}

\label{fig:param}\end{figure}

\begin{figure*}
\centering
\small
 \includegraphics[width=1.1\textwidth]{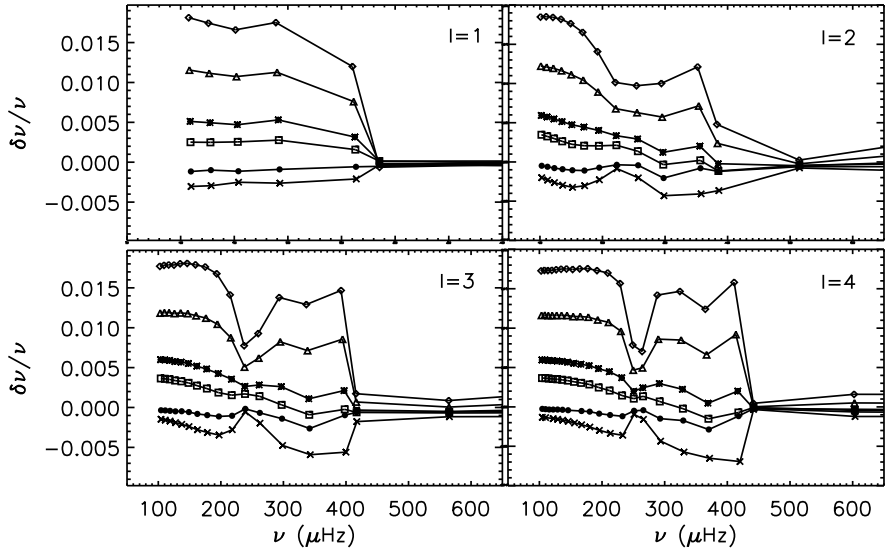}
\caption{Relative frequency differences between the g modes and the
gravest low-frequency p modes of a calibrated solar model computed with
the old GN abundances (M-GN) and a model computed with the AGS
abundances($\diamond$). The other symbols, ($\triangle$, $\ast$,
$\square$, $\times$, $\bullet$) correspond to models computed with
increasing changes to the the neon abundance of the models with a dex of 8.10, 8.29, 8.35, 8.47 and 8.35, respectively.  From \citet{zpbmc07}.}
\label{fig:AGS}
\end{figure*}

\begin{figure}
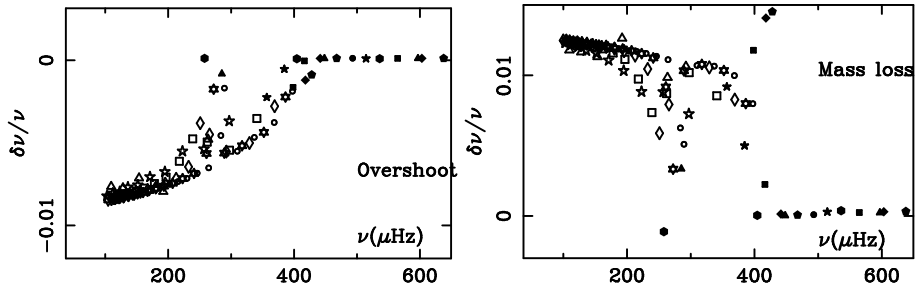

\begin{center}
\hbox{\includegraphics[width=0.49\textwidth]{fig8_AA_u.ps.rev}
\includegraphics[width=0.49\textwidth]{fig8_AA_rap.ps.rev}}
\end{center}

\caption[]{Relative frequency differences, $\delta\nu/\nu$, between a
solar model and the reference model M1, shown to test the sensitivity
to the following physics (see caption of Table 1): overshoot of 0.2
pressure scale heights (model M4); and strong mass loss rates (model
M3) (same symbols as in Figure~\ref{fig:param}).  Adapted from
\citet{JPGB2000}}.


\label{fig:physical-proc}
\end{figure}

The sensitivity of the frequencies to uncertainties in the models can
also be studied in more detail by looking at the relative frequency
differences computed for two different evolutionary models
\citep[e.g.][]{JPGB2000,Mathur2007}. Some general properties can be seen
in Figures~\ref{fig:param}, \ref{fig:AGS}, and
\ref{fig:physical-proc}: (i) at low frequencies, the frequency
differences between two models obey approximately the asymptotic
relation of Eq.~(\ref{eq4Jbis}); (ii) the largest frequency
differences appear in the range 200 to 300 $\mu$Hz, where there are
mixed modes, and the behaviour is more irregular as a function of
frequency than the behaviour shown at very low frequencies; and (iii)
in contrast, the sensitivity of the low-frequency p-mode frequencies
to changes in the solar models is much smaller than for the g modes.

The sensitivity of low-degree g-mode frequencies to changes in the age
and luminosity of solar models is shown in Figure~\ref{fig:param}.
The frequencies are seen to increase with increasing age, increasing
metallicity, $Z/X$, and increasing luminosity.  For instance, an
increase of either 1\,\% in age, 6\,\% in metallicity increases the frequencies by about 0.4\%.  An increase in luminosity
of 3.5 \%, which corresponds to the upper value reported in
\citet{gdkp92}, increases the g-mode frequencies by 0.2\%. A
frequency augmentation of about 1\,\% arises from augmenting the nuclear
reaction rates above the values in the tables of \citet{Caughlan88} to the most
recent tables of \citet{adelb98}.

A reduction in the solar abundances from the old (GN/GS) to the new (AGS) values
reduces the frequencies of the order of 1.5\,\% or more,
as shown in see Figure~\ref{fig:AGS} \citep{Mathur2007,zpbmc07}.  This effect
is in agreement with the global asymptotic parameter results shown in
Table~\ref{tab:asymp} for models M-AGS and M-GN. We also note that
\citet{zpbmc07} have reported that the characteristic low-frequency
g-mode period $P_0$ decreases by increasing the neon abundance in the
new solar mixture (Figure~\ref{fig:AGS}) which is consistent with inferences from the work of \citet{Mathur2007}.

The effects of different descriptions of mixing in the tachocline
have been considered by \citet{Mathur2007}, who showed that the
introduction of horizontal diffusion, as suggested by
\citet{Spiegel92}, would decrease the g-mode frequencies by about
0.25\,\%.  Core overshooting and mass loss during initial stages of
evolution decreases and increases the g-mode frequencies,
respectively, by about 1\,\% (Figure~\ref{fig:physical-proc}).

In conclusion, the sensitivity of the g-mode frequencies to
uncertainties in the solar models appears to be of order at least 1\%
for the cases considered here, which corresponds to frequency
differences of the order 1~$\mu$Hz around 100~$\mu$Hz and 2~$\mu$Hz
around 200~$\mu$Hz, i.e. much larger than the numerical precision of
the frequency computations (down to 100~$\mu$Hz).

\subsubsection{Sensitivity to the rotation and dynamics in the core}

\begin{figure}
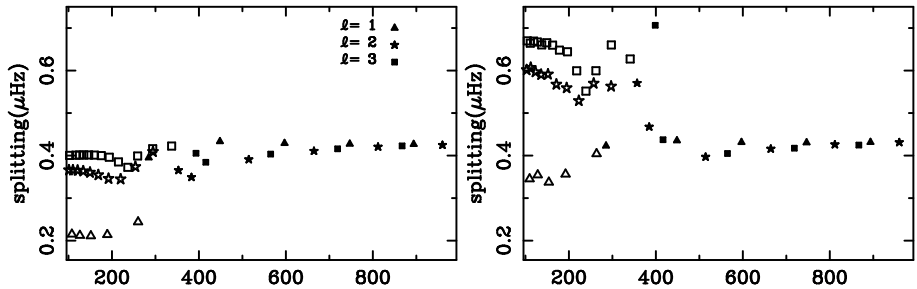

 \vbox{
\includegraphics[width=0.49\textwidth]{figAA_split.ps}
\vspace{0.5truecm}
\includegraphics[width=0.49\textwidth]{figAA_split1.ps}
}

\caption[]{Splittings of low-degree, low-frequency sectoral $m=l$ modes as a
function of the frequency for a reference model and for a simplified
rotation law, which mimics for $r>0.2~R_\odot$ the solar rotation
inferred by inversion of the observed p-mode frequency
splittings. Shown are predictions for different values of the rotation
$\Omega_{\rm c}$ below 0.2~R$_\odot$.  Left panel: $\Omega_{\rm c}$=
0.433\,$\mu$Hz.  Right panel: $\Omega= 2 \Omega_{\rm c}$ for $r < 0.2
R_\odot$ (open symbols: g modes; full symbols: f and p modes).  From
\citet{JPGB2000}.}


\label{fig:rot}\end{figure}


The sensitivity of g-mode frequency splitting to the
dynamics of the solar core may be investigated by the forward method,
using a solar rotation law, which, for $r\ge 0.2 \,R_\odot$, mimics
the rotation rate obtained by inversion of the p-mode frequency
splittings, i.e. rigid rotation below the convection zone and latitudinal differential rotation inside the zone \citep[see
e.g.][]{corbard97,Dikpati02}.  Since the core rotation is very
uncertain, analyses of this type assume a constant rotation in the
core $r < 0.2 \,R_\odot$ equal to $\Omega_{\rm c}$ \citep[e.g.][]{JPGB2000,savita07b}.
Figure~\ref{fig:rot} shows the variation with the frequency of the
model-predicted splittings, $\sigma_{n,l,l}$, of sectoral g modes, for
two different values of $\Omega_{\rm c}$: either the core rotation is
assumed the same as in the radiative zone ($\Omega_{\rm c}= 0.433 \mu$Hz;
left panel), or it is assumed to be twice this value (right panel).
At low frequencies, the splittings depend on the degree $l$, with a
behaviour close to the asymptotic behaviour of g modes, i.e. a
splitting proportional to $1-1/L^2$. At high frequencies the
splittings instead tend to the asymptotic p-mode values, with a very
weak dependence on the degree $l$ arising from differential rotation
in the convection zone.  As expected, if the core rotation is
increased by a factor of two, the g-mode splittings are increased
significantly by a factor of about 1.75, while the p-mode splittings
are almost unchanged.  The splittings of the mixed modes have a
complicated dependence on degree, frequency and the shape of the rotation
law below 0.2 $R_\odot$.  If the rotation of the solar core differs
notably from that of the radiative zone, such behaviour will make
their detection and identification in the observed spectrum harder.

\begin{figure}
\begin{center}
\includegraphics[width=0.6\textwidth,angle=90]{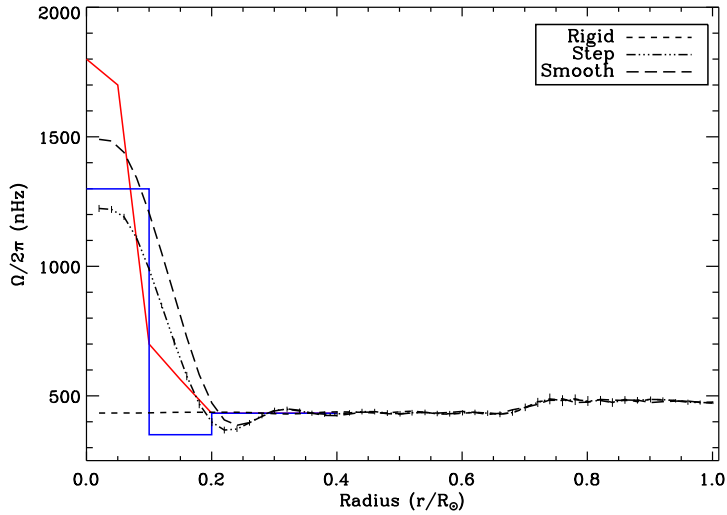}
\end{center}

\caption[]{Equatorial rotation profiles below 0.4~R$_\odot$
reconstructed by inversion of artificial data, using eight g modes,
each with an assumed frequency splitting uncertainty of 7.5\,nHz. Results are shown for three different rotation laws in the solar
core (dashed line rigid, full line step profile, long dashed line
smooth profile).  The continuous blue, red and dotted lines represent,
respectively, the input step, smooth and rigid artificial rotation
profiles.  From \citet{savita07b}.}

\label{fig:inversion}\end{figure}

The internal magnetic field, which is often neglected, can also split
the frequencies into several components and thus it can affect the
reliability of predicted g-mode frequencies.  \citet{rashba07} attempt
to estimate analytically the effect of a magnetic field in the
radiative zone on the g-mode frequencies, using a MHD perturbative
correction \citep[e.g.][]{Unno89}. They show that a 1-per-cent g-mode
frequency shift can be obtained for magnetic field as low as 300kG in the radiative zone,
for modes of radial order $n$=20 and degree $l$=1 ($\nu\sim 33 \mu$Hz),
and they argue that a similar shift for the case of low-order g-mode
frequency cannot result from a central magnetic field unless its
strength exceeds 8 MG which would be a very large magnetic field.

The rotational kernels would help to infer the core rotation by
inversion, once the splitting of g modes could be measured.
Preliminary work using inverse methods has been carried out to try to know how measured
splittings of some g modes would influence our knowledge of the
rotation of the solar core \citep{Mathur2007,2008SoPh..251..119G}. Numerical inversions using artificial data, corresponding to
different laws of rotation in the solar core (see
Figure~\ref{fig:inversion}) were performed.  It was shown that the introduction of only one
g mode gives a much better idea of the rotation profile. The
introduction of eight g-mode frequencies with an error bar of 7.5\,nHz
in the data allows one to recover the rotation profile in the core quite well, with
the low-degree p modes contributing to constrain the solar
rotation profile down to 0.2 $R_{\odot}$. As expected, without g
modes, it is not possible to distinguish between the three assumed
core rotation laws

\subsection{Conclusion}

It appears that g-mode frequency predictions are affected by two
sorts of uncertainty.  First, the faults in the solar modelling
induce uncertainties in the solar stratification, hence in the
Brunt-V\"ais\"al\"a frequency which determines mainly the g-mode
frequencies.  Some contributions have been estimated to 1\% (see
subsection 2.2.3), but the consequences of the lack of a reliable description of the dynamical and magnetic effects remains to be estimated.  Second, the
prediction of the splittings depends on the core rotation, which
remains poorly known.  A rotation twice larger below 0.2$R_{\odot}$
would increase the low-frequency g-mode splittings by a factor about
1.75.


\section{Excitation and damping of g modes}
\label{sec:excite}

In the past, several attempts have been made to estimate solar g-mode amplitudes 
\citep{Gough85, GB90,BA96,Kumar96}. 
The common result of these predictions is that g-mode amplitudes are not 
larger than a few mm$\,$s$^{-1}$ \citep{Elsworth06}.

The investigations of g-mode amplitudes first focused on the
linear stability of the modes so as to determine whether or not they could 
be excited by overstability ({i.e.} thermal instabilities). 
Results published to date suggest that g modes are most likely linearly stable 
(Section~\ref{sec:linear_stab}). Consequently most of the theoretical efforts 
have therefore assumed that g modes are intrinsically damped and 
excited stochastically by turbulent convection (Section~\ref{sec:stochastic}), 
as it is also believed to be the case for p modes, whether the excitation is
predominantly in the body of the convection zone 
\citep{Gough85,GB90,Kumar96,KB2009},
or in the lower overshooting layers
\citep[e.g.,][]{BA96,Dintrans05}. 
Note also that other mechanisms such as mode coupling \citep[e.g.,][]{WD83,Guenther84,Ando86,Wentzel87,Wolff07}, 
or excitation by magnetic torques \citep{Dziembowski85} have been investigated.

The resulting amplitudes are determined by a balance between the intrinsic
linear stability of the modes, which are characterised by the damping
rates $\eta$, and the energy input via nonlinear interactions with the
turbulence. These will be discussed separately in Section~\ref{sec:linear_stab} 
and \ref{sec:stochastic}.

\subsection{The separation of the oscillation modes from the convection}
\label{sec:separation}
Before describing some of the details of what has appeared in the
astrophysical literature, it is instructive first to ask just what the 
question is that we wish to ask. The glib response is simply: to what
amplitudes does the turbulence drive the modes? But before we can address
that matter, we should try to be clear in our minds how to separate the
motion into convection and modes of oscillation. There have been several
attempts to accomplish that task, mainly unpublished because unfortunately
unambiguous success has never been achieved. Therefore we shall be brief.

In the case of radial (p) modes, superficially the issue is reasonably clear:
the modes are horizontally uniform, and convection, which is motion driven by
buoyancy, is intrinsically horizontally nonuniform.  Therefore one simply
takes a horizontal average, and regards the average motion as the oscillation
mode and the remainder as convection. This is a fairly safe procedure, at least
for isolating radial modes (and was justified to some degree by 
\citealt{DOG69}), provided that one takes the 
horizontal average of $\rho w$, where $\rho$ is density and $w$ is the vertical
(radial) Eulerian component of the convective velocity, rather than $w$, to 
define the oscillation mode.
Incidentally, convection, even when approximated by a mixing-length-type 
theory of eddies \citep[e.g.][]{DOG65, DOG77a, DOG77b, Unno67}, is most 
naturally described in (local) Eulerian coordinates; since the eddies 
maintain a degree of integrity, they are advected by the large-scale 
pulsational flow, which is therefore most appropriately described in 
Lagrangian coordinates.

Separating nonradial modes from convection is a more difficult task. It can
be accomplished only approximately, and that only when there is a good 
separation
of scales. Modes of low degree have horizontal scales comparable with the
radial coordinate $r$, and because most of the energy in the convective
motion is in scales rather smaller than that (at least on timescales
comparable with the periods of the p modes and grave ($n\leq3$) g modes) one can at least 
separate the motion, this time by taking appropriate large-scale spatial 
and temporal averages. 
In this way, a mixed Eulerian-Lagrangian coordinate system can be set up
\citep{Gabriel75, Unno89} similar to that defined originally for radial 
pulsations \citep{DOG69}.
However, such a separation cannot be achieved for the very-high-order
g modes that resonate with the turbulence at the base of the convection zone.
In that case it is necessary to resort to more involved procedures. For example, one may try,
following \citet{Poyet83}, to decompose the motion
into a superposition of linear p modes and g modes, regarding the function 
space spanned by the direct, so-called g$^+$, modes
as convection; that procedure suffers from ambiguity when trying to estimate the 
overlap integrals that couple the two kinds of motion, because there can 
remain some freedom in how the decomposition is carried out. Most of the 
analytically based studies simply separate the motion by fiat.  And now, 
having issued our warning, we shall do likewise.

Before proceeding, a word concerning the analysis of numerical
simulations is in order. Simulations have been carried out to shed light 
directly on the mode-excitation issue, although in some cases also with 
the intention of calibrating analytical approximations to the energy-input rate 
\citep{Rosenthal98, SteinNordlund01,Rogers05, Rogers08}. Because convective motion cannot be
unambiguously separated from what we imagine to be the oscillation modes, the
details of what one envisages to be energy conversion cannot be isolated.
But what one can do is to look in the far field where convection is ignorable, 
provided the domain of the simulation is extensive enough, and there determine 
the spectrum of the g-wave and p-wave radiation by projection onto linearized
wave functions \citep[e.g.,][]{Dintrans05}.

\subsection{Linear g-mode stability}
\label{sec:linear_stab}

The discussion of the stability of g modes in a star has a long history. 
\cite{LS-G50} were probably the first to address 
the general problem, although they limited their discussion to white dwarfs.
As is the case for acoustic modes, the stability of gravity 
modes depends on the integrated effect of various physical mechanisms which
can either drive or damp the oscillations. The excitation and damping
processes usually take place in layers with rather small radial extent: these 
are the ionisation regions of hydrogen and helium where the
Eddington valve can operate, the energy-generating core where 
temperature-sensitive nuclear reactions can feed energy into the oscillatory 
motion, and also the highly superadiabatic upper convective boundary layer 
and, for g modes, the vicinity of the boundaries of convection zones where 
dynamical interaction between the convection and the oscillations is 
relatively strong.   In the following section we review these mechanisms. 

\subsubsection{Linearized equations governing g-mode oscillations}

One first takes a statistical (temporal) average of the governing
equations (which determines the basic, background state of the star)
and subtracts it from the full equations to yield the fluctuation
equations. Having separated the fluctuating variables into an oscillatory
mode contribution and convection, one linearizes the fluctuation equations
in the mode variables, and then typically arranges the outcome with the mode 
variables on the left and the fluctuating terms associated with the convection 
on the right. In all studies to date the mode amplitudes are regarded as 
being so small that they do not influence the inhomogeneous convective driving 
term on the right. Therefore each mode can be treated separately. The equation 
of motion can then be written formally as
\begin{equation}
{\cal L}\bm{\xi}=\myf(\bm{v},T^\prime)\,,
\label{eq:inhomo-wave}
\end{equation}
where $\bm{\xi}$ is the displacement eigenfunction of the mode, with $(\bm{v},T^\prime)$ is the 
convective velocity and temperature fluctuation, and complex
frequency $2\pi\nu_{n,l,m}\equiv\omega-{\rm i}\eta$ (See Eq.~\ref{Eq1}), in which $\omega$ is real and positive, $\eta$ is real. 
 The term $\myf$ represents inhomogeneous stochastic driving and damping  terms that depend only on convective quantities (\emph{i.e.} $\bm{v}$ and $T^\prime$) unpertubed by the oscillations.
The spatial differential wave operator $\cal L$ \citep[e.g.,][for details]{Unno89}
 depends upon $\omega-{\rm i}\eta$ and the background state 
of the star, which itself depends on the convective fluxes of heat and 
momentum; it depends also on the perturbation to those fluxes that is 
produced by the oscillations. Ignoring the right-hand side of 
\eq{eq:inhomo-wave} yields the equation of free oscillation (\emph{i.e.} free from the inhomogeneous damping and driving, but including in principle the linearized momentum and convective momentum flux perturbations that are induced by the oscillations), which in the
adiabatic Cowling approximation can be approximated by Eq.~(\ref{wave}).
 

\subsubsection{Excitation by the $\epsilon$ mechanism}
The first dynamical investigation of g-mode instability of the Sun in which the
nuclear reactions were perturbed consistently was carried out
in the quasiadiabatic approximation
by \cite{DG72} who approximated the eigenfunctions in the radiative interior
by those of a polytrope of index 3. Soon afterwards \cite{RB73} discussed the
thermal instability of hydrostatic disturbances. The aim in both papers
was to question the assumptions upon which standard solar models are built 
with a view to addressing the still-existing solar neutrino problem. 
Dilke \& Gough found that low-order g modes could have been dynamically 
unstable in the early evolutionary stages of the Sun owing principally to 
the strong temperature dependence of particularly the 
$^3$He($^3$He,2p)$^4$He reaction which manifests itself in the energy budget 
when the reactions of the p-p chain are thrown out of balance by the 
dynamical oscillations, and also to the fact that the Sun has a convective
envelope deep enough to provide an effective evanescent shield; earlier
investigations confined to stars with radiative envelopes
\citep[e.g.][]{Aure71} had indicated that g-mode amplitudes are so high in
the surface layers that radiative damping would overwhelm any driving in the
core.  Rosenbluth \& Bahcall found that their solar models 
are stable to non-radial thermal instability. 

\subsubsection{Radiative damping}
\label{Rad_damp}

\begin{figure}
\begin{center}
\includegraphics[width=0.75\textwidth,angle=0.3]{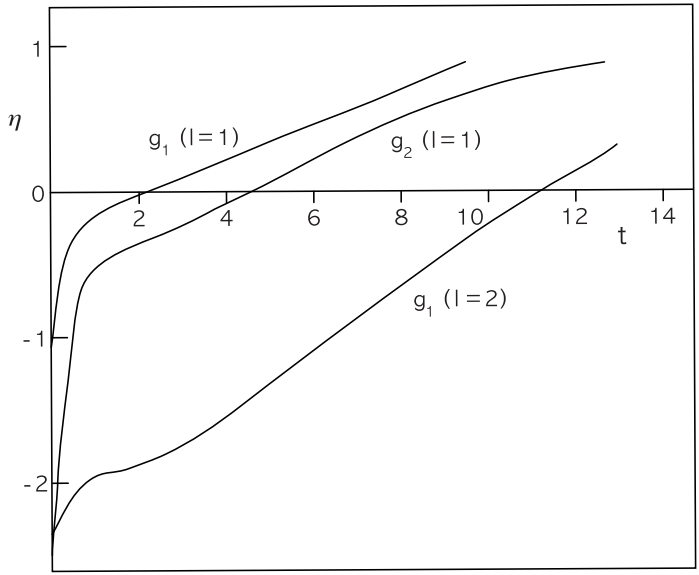}
\end{center}
\caption[]{Theoretical growth rates $\eta$ of three low-order g modes as
a function of age $t$ in the early stages of the main-sequence evolution of the Sun. The units are $10^{-7}\,$y$^{-1}$ for $\eta$ and the age 
is measured in units of $10^8\,$y of the equilibrium models \citep{CDG74}.  The current age of the Sun is about 4 10$^{9}$ years.  Modes
are found linearly unstable for $\eta < 0$ and stable for $\eta > 0$.
}
\label{fig:c-ddg74}
\end{figure}

Following \citet{DG72}, \cite{CDG74}, 
\cite{BouryEtal75}, and \cite{SOU75} addressed the g-mode instability using
eigenfunctions of solar models, although still in the quasi-adiabatic 
approximation. 
%
%
All authors found unstable low-order g modes in models younger than
the present Sun. The instability arose because the temperature sensitivity of nuclear reactions perturbed on a g-mode timescale is substantially greater than it is on a stellar-evolution timescales -- the latter is too weak to overcome radiative damping \citep[e.g.][]{DS73}.  For the present Sun, however, all authors found 
all g modes to be stable, at least if the evolutionary consequence of
earlier instability could be ignored. 
It is particularly so in the early stages of evolution when the star
was less centrally condensed that the g-mode amplitudes are
relatively large in the core \citep[e.g.][]{NoelsEtal74}, and are
therefore more receptive to nuclear driving. 
Figure~\ref{fig:c-ddg74} shows low-order g-mode growth rates 
for an evolving 1\,$\,{\rm M}_\odot$ star 
computed in the quasi-adiabatic approximation. 

The first fully nonadiabatic stability computations of solar p and g modes 
were undertaken by \cite{CG75}. The authors 
concluded that nonadiabatic processes in the very outer layers of the present 
Sun dominate over the destabilising influence of the temperature-dependent 
$^3$He($^3$He,2p)$^4$He reactions in the core, suggesting that low-order g modes
are very likely stable. An important deficiency in these computations, however, 
is that the modulation of the convective fluxes of heat and momentum 
were omitted.  \citet{AG85} investigated the g-mode instability for solar models
with different metallicities and concluded that the instability is stronger in low metallicity models.

\subsubsection{The $\kappa$ mechanism}
In stellar layers where partial ionisation of chemical elements takes place,
opacity can increase with temperature 
during the compression phase of a stellar pulsation, tending to produce
an Eddington valve that can lead to instability.
The radiative (energy) flux through such a layer is then absorbed by the 
stellar matter more effectively than it would have been otherwise. 
After the subsequent expansion phase the previously
absorbed excess energy is released, the net effect over the whole cycle being  
to feed energy into the pulsation. 
This mechanism is responsible for pulsations of some variable stars such as 
$\delta$ Scuti, $\beta$ Cephei and Cepheids \citep[e.g.][]{Dziembowski93,Gautschy95a,Gautschy95b}.
This process is effective only if much of the heat flux in the relevant layer
is carried by radiation, which in the Sun is not the case. However, the valve
can operate also as a result of a perturbation to the convective heat flux.
Irrespective of whether the flux is predominantly radiative or convective, 
in any star the valve is most effective if the thermal relaxation time of the 
material in and above the relevant
layer is similar to that of the period of the pulsation. Because the thermal
relaxation time increases with depth, low-period oscillation modes are driven 
in deeper layers than are corresponding short-period oscillations 
\citep[e.g.][]{Pamyatnykh99,HG03}. 

\subsubsection{The effect of convection}
\cite{Saio80} reported the first
non-adiabatic solar g-mode stability computations that incorporated 
the modulation of the convection, adopting the 
treatment of convection by \cite{Gabriel75}, which is
based on Unno's (1967) time-dependent mixing-length formalism.
In these computations, only the heat flux was taken into account: the momentum 
fluxes (Reynolds stresses) were omitted from both the equilibrium structure 
and the stability analysis. 

\cite{Saio80} confirmed the earlier findings by \cite{CDG74}
of g-mode instability during the
early evolutionary phases of the Sun. Interestingly, for solar models with
an age of $4.5\times10^9\,$y he found that some of the g modes are 
additionally driven by the hydrogen ionization zone in the very outer
layers of the star. However, as discussed by \cite{DG72}, 
the main agent driving grave low-degree g modes appears to be the 
$^3$He$\;$--$\!\;^3$He reaction in the energy-generating core. 

It was reported by \cite{DOG65, DOG80} that the dynamical
effects arising from the turbulent momentum flux perturbations
contribute significantly to the growth rates of solar oscillations. 
Stability computations of a star in which both the convective heat 
flux and the Reynolds stresses were included, in both the 
equilibrium structure and the pulsation calculations, using the time-dependent 
mixing-length formalism by \cite{DOG65, DOG77a}, were reported by 
\cite{BG79} for acoustic oscillations in RR Lyrae stars.  A subsequent 
detailed analysis of the Sun by \cite{B92a} revealed the manner in which the 
phases of the fluctuating Reynolds stresses and the density perturbations 
differ, causing radial p modes to be damped. Similar findings were reported 
also by \cite{HoudekEtal99} in other main-sequence stars.
Different results were obtained from the computation of \cite{MAD06c}, which suggest that the perturbation 
of the convective heat flux is dominant. The formalism that was adopted is based on the approach of \cite{Unno67} 
generalized to non-radial modes by \cite{MAD05}. In addition, \cite{MAD06c} found that 
the effect of the turbulent pressure modulation is partly compensated by the perturbation of the 
dissipation rate of turbulent kinetic energy into heat \citep[see also][]{Ledoux58}.

For high-order g modes ($\nu < 100 \mu$Hz), \cite{KB2009} have shown, using 
a partially non-adiabatic non-radial computation \citep{MAD05}, that 
the damping rates are dominated by radiative losses and are insensitive to the convection-pulsation interaction. 
This result is in agreement with previous findings of \cite{Kumar96}.

It is perhaps useful to point out the fundamental approximations 
that have been used to model the temporal modulation of the
convection by the oscillations. Formulae for these modulations
are needed when estimating the influence of convection on
the frequencies and, more importantly, the growth rates of
the oscillations. We limit ourselves here to those procedures
that have been used for estimating solar g-mode amplitudes.

The most naive procedure is to presume that the convective
fluxes simply relax exponentially on a timescale $\tau_{\rm c}$
towards the time-independent formula evaluated for the current
(time-varying) environment. Thus, if $F_{\rm c}$ is a
component of a flux, be it of heat or momentum, it is taken
to obey an equation of the type
\begin{equation}
\frac{{\rm d}F_{\rm c}}{{\rm d}t}=\frac{F_{\rm c0}-F_{\rm c}}{\tau_{\rm c}}\,,
\label{eq:flux_decay}
\end{equation}
where $\tau_{\rm c}$ is a multiple of $\ell/w$, $\ell$ being the
mixing length and $w$ a characteristic convective velocity, both
of which are evaluated typically for the non-oscillating background
state. The function $F_{\rm c0}$ is some chosen formula for the
corresponding flux evaluated for the instantaneous state of the
star as though that state were not pulsating. Evidently, there is
some formal freedom in the choice of that formula, because it
was derived under the assumption that the background state is in
(statistical) hydrostatic equilibrium, whereas when the star is
oscillating that is not the case. Therefore different
representations that are equivalent in the non-oscillating state
can differ for the oscillating star. It is also evident that the
phase lag of the oscillatory perturbation $F^\prime_{\rm c}$ to
$F_{\rm c}$ behind that of the perturbation to $F_{\rm c0}$, which 
controls the damping rate of the oscillations, is determined 
directly by the (arbitrary) value adopted for the constant of
proportionality between $\tau_{\rm c}$ and $\ell/w$.

Other approaches address the dynamics of the turbulent perturbations.
In the mixing-length approach, the fluid is considered to be composed
of an ensemble of eddies, which somehow attain an existence, grow in
the unstable environment and then break up, presumably through a
cascade of scales. Two extreme approximations have been adopted.
In one \citep{DOG65, DOG77a}, the growth is estimated from linear
theory, thereby ignoring nonlinear interactions, and eventually the
nonlinearities are posited to act instantaneously to break up the
eddy and destroy its correlations between velocity and temperature
fluctuations. The instants of creation and breakup are determined
statistically, with probability distributions that are determined
solely by the background state. In the other extreme 
\citep{Unno67}, the nonlinear interactions are presumed to 
dominate that aspect of the dynamics that determines the turbulent
fluxes.  In a statistically steady state, the eddies are presumed to
be steady too, the buoyancy driving being balanced by an algebraic 
representation of nonlinear transfer to other scales.
This obviates the need to discuss creation and destruction. Both
approaches can easily be worked out for a statistically steady
convection zone, and yield essentially the same formulae for the 
fluxes. And both can be perturbed for studying stellar
oscillations, although now the resulting equations for the
perturbed fluxes lead to very different results. A careful comparison of these
two extreme approaches has never been undertaken.

One major deficiency resides in the reduction of the whole 
turbulent cascade into a single length-scale, as it is assumed in 
mixing-length approaches. As the physical link with the cascade is lost, no 
realistic and reliable predictions for the perturbation of the mixing-length can be obtained. 
Unfortunately, it appears to play a major role in mode driving and damping. 
In all the formulations it is necessary to make assumptions,
either explicitly or implicitly, about the anisotropy of the
convective motion, and how it is modulated by the oscillations.
In so naive an approach as mixing-length theory, this
inevitably entails the introduction of new parameters, which,
together with the parameters in the theory for steady convection,
should somehow be calibrated against astrophysical observations,
laboratory convection or numerical simulation, where possible.
This is a difficult task, partly because we do not have
enough experience with modelling and analysing convection 
and partly
because we probably do not have enough data in a form in which
we can comprehend their implications. The manner in which the
anisotropy is taken into account affects the formulae for the
turbulent heat flux and Reynolds stress, and thereby influences
the computed damping rates of the oscillations.

It is important to recognise that the form adopted for the anisotropy
of the convective motion reappears in the evaluation of the excitation
integral of Eq.~(\ref{eq:expect-ampl}) discussed in the next
section, although it is rare for the formulation
of the forcing term $\myf$ in the integrand to be consistent with the
calculation of the damping rate. The rationale, usually unstated,
appears to be that because the calculations of both the damping and
excitation are uncertain, consistency between the two is hardly called
for.

All currently reported stability computations suggest 
that low-degree solar g modes of order $n>3$ are stable. 
For these modes, the damping is dominated by radiative losses \citep{Kumar96} 
and is found to be insensitive to the pulsation-convection interaction \citep{KB2009}.
For the stability of g modes with $n\leq 3$, however, detailed agreement has 
not been achieved, largely because the balance between nuclear
driving, radiative damping and energy exchange with the convection
is quite delicate \citep{CDG74, BouryEtal75, SOU75}.
If grave g modes were indeed unstable some nonlinear 
process must have limited their exponential growth.

\subsection{Amplitude limitation}
\label{sec:amp_limit}
\citet{WD82, WD83} studied the effect of  three-mode interactions on the
g-mode amplitudes. In particular, \cite{WD83} considered the special case of 
parametric resonance, in which a parent linearly unstable mode is coupled with 
two linearly damped daughter modes.  Once a mode is sufficiently mature, it spawns two
resonating g-mode daughters who sap energy from their parent at a rate
proportional to the product of the amplitudes of the three modes. The
parent is thereby unable to continue to evolve unencumbered, and eventually
her growth is stunted. There are various ways in which the interactions can 
proceed, but the most effective for grave low-degree parents appears to be
a coupling with a pair of similar high-degree twins leading to a state in which the amplitudes of all three modes are steady. The most
intricate part of the calculation is the evaluation of the interactions.
However, the basic overall limiting process is generic. Therefore we illustrate 
it with a much simpler, one-dimensional acoustic process described in Appendix~\ref{app:nonlinear}.
 In that discussion the structure of the background state was assumed not to vary with time.  Yet
to achieve amplitude limitation appears to require frequency
mismatches $\Delta\omega$ of order $10^{-6}\omega_0$, which
would need to be maintained for times of order $\eta^{-1}$. That
could be several decades or centuries. Can such precise resonance
be maintained in the face of solar-cycle variation, or any other change?
And if not, what are the consequences regarding g-mode amplitudes?
These questions have been addressed by \citet{JG00}, who merely
reported that the reduction in the amplitude limitation caused by
drifting out of resonance is offset by the increased probability
of actually encountering a resonance, leaving Dziembowksi's
estimate unchanged.

Dziembowski estimated a likely actual mode amplitude for g$_1$($l$=1)
in the photosphere to be about 20$\,$cm$\,$s$^{-1}$. That value is
considerably greater than the upper bounds set by observation. Of
course that discrepancy presents no actual serious conflict, because
the estimate is probabilistic: Eq.~(\ref{eq:ac2}) does not
necessarily preclude a precise resonance with daughters of moderate
degree whose damping rates are somewhat smaller than those of the
`most likely' resonance. Moreover, provided solar-cycle activity is
confined essentially to the convection zone, associated structure
variations are unlikely to have a large impact on mode resonance,
which is determined by conditions in the region in the radiative
interior in which the daughters are trapped.

But perhaps instead it is more likely that there is no intrinsically
unstable g mode. All calculations to date have found models of the
present Sun to be stable to all g modes, although a consistent
modern calculation, carried out since the solar structure was determined
seismologically, appears not to have been reported. So the
possibility of there being grave self-excited g modes should not
yet be wholly abandoned.

What is it, one naturally might ask, that is so different in more massive
stars, such as SPB stars or $\gamma$ Doradus stars, that enables them to
undergo g-mode oscillations at much higher amplitudes? The principal
difference is that they have only shallow convection zones, and therefore
daughter g modes extend much higher in the envelope and dissipate much
more strongly, thus being limited themselves to much lower amplitudes
and thereby being less able to extract energy from their parent.

If all g modes were indeed stable today,
their amplitudes would be determined from a balance between stochastic 
driving by turbulent convection and damping. We therefore discuss 
in the coming sections the physical mechanisms responsible for 
solar g-mode damping and excitation, and the resulting theoretical 
amplitude estimates that have been published to date.

\subsection{Stochastic excitation by turbulent convection}
\label{sec:stochastic}
\subsubsection{Amplitudes of oscillations}
If the general solution to the homogeneous equation of free oscillation 
were known, then the solution of \eq{eq:inhomo-wave} for forced 
oscillations could be written down in terms of $\myf$. The general 
solution for free oscillations of realistic stellar models 
is not known, but because the adiabatic oscillation eigenfunctions computed 
in the absence of the interactions with convection form a complete set 
\citep{Eisenfeld69, DysonSchutz79}, they can be used as a basis for 
what is probably a reasonable approximation to the spatial structure 
of the oscillatory motion.  What then 
remains is to characterise the fluctuating driving term $\myf$ on the 
right-hand side of \eq{eq:inhomo-wave}.

The first serious attempt to address this general problem for the Sun 
was by \cite{Stein66,Stein67}. This pioneering work addressed the 
generation of acoustic waves based on the method developed by 
\cite{Lighthill52} to derive the acoustic radiation from an isolated 
unstratified free-turbulent region, a region in which no external body force is 
imposed on the fluid. Because acoustic waves can exist in and propagate from the turbulent 
region, the local wave-turbulent interaction can be estimated from the 
turbulent field in terms of simple waves. Lighthill showed that, in the 
relatively straightforward situation in which he applied his analysis, 
conservation conditions required that the basic wave emission is quadrupolar. 
But in a stellar envelope stratified under gravity, the situation is more 
complicated.  First, one cannot even define universal multipole emission 
components in general circumstances, although in simple cases, such as an 
isothermal atmosphere under constant gravity, a multipolar decomposition is 
possible \citep{Unno66}: the stratification produces monopolar (mass fluctuations) and dipolar 
emission (force fluctuations), in addition to quadrupolar (shear fluctuations), although, as \cite{Unno66}, 
\cite{Stein67} and subsequently \cite{Osaki90} pointed out, there is a 
tendency for the monopolar and dipolar components to cancel. 

A direct generalisation of Lighthill's method is not possible, because gravity
waves cannot propagate through a convectively unstable medium. Indeed, it is 
probably not most prudent even to work in terms of simple emitted waves, 
because one would still have to impose resonance conditions on them in order 
to represent the normal modes, which, as Stein appreciated, involves some 
subtle argument. Instead, it is more straightforward to solve 
\eq{eq:inhomo-wave} as a nonsingular perturbation, regarding $\myf$ as a 
forcing term on a spatially unperturbed oscillation eigenfunction $\bm{\xi}$ of oscillation having angular frequency $\omega$
and (positive) damping rate $\eta$. 
Basically, one represents the forced motion as 
$A(t)\bm{\xi}(\bm{r})\exp(-{\rm i}\omega t-\eta t)$, where $A$ varies slowly
compared with $\exp(-{\rm i}\omega t)$. Substituting this into the 
spatio-temporal wave operator from which $\cal{L}$ was derived, and 
retaining only terms proportional to $A$ and its first derivative, yields 
a differential equation for $A$ whose solution can be written down explicitly 
as an integral depending on $\bm{\xi}$ and $\myf$ 
\citep{GK77, B92c, MRSU94, Samadi00I, Chaplin05}. 
For statistically stationary turbulence exciting modes having characteristic wavelengths much greater than the
typical scale $k^{-1}$ of all convective eddies and with $\eta<<\omega$, 
the integral expression for the mean-square amplitude $A$ reduces to
\begin{equation} 
\overline{A^2}\simeq\frac1{4\eta\omega^2{\cal I}^2}
\int_0^\infty G(\tau)\,{\rm e}^{-\eta\tau}\,\cos\left(\omega\tau\right)\,{\rm d}\tau
\label{eq:expect-ampl}
\end{equation} 
\citep{Chaplin05}, where $G(\tau)=\overline{Q(t^\prime)Q^*(t^{\prime\prime})}$
is a global correlation function with argument $\tau=t^\prime-t^{\prime\prime}$, the overbar denoting ensemble (or temporal) average, and 
\begin{equation}
Q(t)=\int_{\cal V}\bm{\xi}^*(\vec{r})\cdot\myf(\vec{r},t)\,{\rm d}V\, ,
\label{eq:overlap-function_Q}
\end{equation}
the integral being over the volume $V$ of the star.  The asterisk denotes complex conjugate, and ${\cal I}=\int \rho \xi.\xi^* {\rm d} V$ is the inertia of the mode.  The ensemble average $\overline{Q(t')Q^*(t'')}$ can itself be written as a double integral over the volume of the star.
$Q(t)$ depends on a two-point correlation function between $(\vec{r}',t')$ and $(\vec{r}'',t'')$ which can be represented by its (local) Fourier transform $\chi(\nu,\vec{k};\vec{r})$ with respect to $\tau$ and $\vec{r'-r''}$ referred to the mean position $\vec{r}=(\vec{r'+r''})/2$.


The principal uncertainties in the estimation of 
$\overline{A^2}$ lie in the evaluation of the damping rate 
$\eta$ and the correlation function $G$.  The mean-squared surface velocity is computed, for each mode, as
\begin{equation}
\label{Pobs}
\overline{v_{\rm s}^2(h)}=
     \overline{A^2}\left[v^2_{\rm r}(h)+v^2_{\rm h}(h)\right] 
\end{equation}
where $h$ is the height in the atmosphere at which the complex oscillation 
velocity $(v_{\rm r}, v_{\rm h})={\rm i} \omega_0(\xi_{\rm r},\xi_{\rm h})$ is observed 
\citep[for details see][]{HG95,Baudin05}. 

\subsubsection{Application to g-mode amplitudes}
There have been several attempts to estimate the stochastic excitation of
g modes, all of which stem from the early formulation by
Goldreich \& Keeley (1977a,b) of the mechanism of the excitation of global p modes.
This work was initially extended by \citet{GMK94} and then by others.
The analyses differ in detail, as do the results. The differences
in the assumptions adopted in the analyses
occur in several places, and it is not possible to determine from the
published work just how they influence the conclusions.
Therefore we shall only summarise very briefly the principles behind the
work.


The intention of Goldreich and his colleagues was to improve
the earlier work on the excitation of solar p modes.
Subsequently \citet{Kumar96} addressed the issue of g-mode observability.
The first step in the endeavour was to derive an explicit form of the 
linearized momentum equation governing the oscillations in order to
obtain an explicit expression for the forcing term in 
Eq.~(\ref{eq:inhomo-wave}).
\citet{GMK94} approached that task by implicitly assuming 
that the convection also satisfies the linearized momentum 
equation.
The outcome is
\begin{equation}
\myf=-\nabla\left[\left(\frac{\partial p}{\partial s}\right)_{\!\!\rho} s^\prime
             -\ob{\left(\frac{\partial p}{\partial s}\right)_{\!\!\rho} s^\prime}
           \right]
     -\nabla\cdot\left(\rho\vec{v}\vec{v}-\ob{\rho\vec{v}\vec{v}}\right)\,,
\label{eq:gmk94}
\end{equation}
where $p$ is pressure, although \citet{GMK94} did not explicitly separate the fluctuations
from the means. This result generalises a similar expression presented
by \citet{GK90} for an isentropically stratified plane-parallel
atmosphere of perfect gas. The function $\myf$ was entered into the
integrals in Eq.~(\ref{eq:overlap-function_Q}) and the terms
estimated from mixing-length theory.


In deriving the inhomogeneous wave equation \citet{GMK94} omitted
terms containing the Eulerian density perturbation (even when coupled
with the gravitational acceleration, a term that must not be neglected
even in the Boussinesq approximation) and the inertia terms of the
convection.  Furthermore, the authors assumed that the effect of the
oscillations is to destroy the balance of the convective momentum
transfer in such a way that the linearized momentum equation is
satisfied in the absence of Eulerian entropy fluctuations, leaving the
unbalanced convective entropy fluctuations and nonlinear inertia terms
to drive the oscillations. This assumption caused the entropy
fluctuations to be an order of magnitude more effective than the
fluctuating Reynolds stress in driving the oscillations.
\citet{GMK94} ignored products of oscillation variables and convective
fluctuations, and the effect of the oscillations on the convection.
Consequently an adiabatic linear wave equation (with homogeneous terms
untrammelled by convection) is supplied just with an inhomogeneous
forcing term to account for all the effect of convection.  The
anisotropy of the convective motion was acknowledged merely by
multiplying the entire inhomogeneous term $\myf$ by a single constant
scaling factor, which was determined by calibrating the energy supply
rate to the oscillations against observation.

The theory was reformulated by \citet{Samadi00I}, who combined aspects
of the analyses by \citet{Stein67} and \citet{GK77}. Perturbations due
to convection and global oscillations were separated in the equation
of motion, and approximated according to their dominating dynamics,
adopting a Boussinesq-like approximation for the convection.
\citet{Samadi00I} have shown that the linear term due to entropy
fluctuation [Eq.~(\ref{eq:gmk94})] introduced by \citet{GMK94}
gives no significant contribution to the driving but the advection of
Eulerian entropy fluctuations by turbulent velocity.  The analysis
represents what is probably the most careful recent attempt to
generate a consistent forced wave equation. Unfortunately, however,
there is subsequently an error in the treatment of the frequency
correlations in the expression for the driving by the fluctuating
Reynolds stress, corrected by \citet{Samadi05c}.

The predictions for the maximum amplitude for $\xi$ Hydrae made by \citet{HoudekGough02} and \citet{SamadiEtal07}
are in good agreement with the observations by \citet{FrandsenEtal02}, while their predicted 
maximum amplitude for Procyon are overestimated by a factor 2 to 4. 
This discrepancy for Procyon is serious and must be
understood; the structure of Procyon is rather different from other 
stars that have been modelled, indicating that the theory, anchored by 
calibration against solar p modes, cannot not be extrapolated reliably to
very different stars. By the same token the reliability of extrapolating
from p modes to g modes even in the Sun must be exercised with due caution, 
and may not be as reliable as \citet{Kumar96} suspect.

One of the controversial conclusions of the discussion by
\citet{GMK94} is that driving from the first term in
Eq.~(\ref{eq:gmk94}) for $\myf$ is an order of magnitude greater
than driving from the fluctuations in the Reynolds stress, at least
for acoustic emission. This contradicts the earlier findings of
\citet{GK90} who concluded that, in a convection zone that appears to
the oscillations (but not to the convection) to be adiabatically
stratified, the contribution from the two terms are comparable. The latter
conclusion had been drawn from the realisation that the emission from
what \citet{GK90} termed the monopole and dipole sources
(contributions which in a stratified envelope are not unambiguously
defined, cf. \citet{Unno66} from the buoyancy (entropy fluctuation)
terms largely cancel as a result of the manner in which the eddy
motion is correlated with them. Such cancellation results in
predominantly quadrupolar emission, as \citet[][see also
\citealt{Houdek06}]{Osaki90} also pointed out, and is consistent with
Stein's (1967) earlier findings.  It is interesting to note that simulations by \citet{SteinNordlund01, Stein2004}, and the model calculations by \cite{Belkacem06b} and
\citet{SamadiEtal07} indicate that emission is dominated by the
fluctuating Reynolds stress and entropy fluctuations as had been implied by Balmforth's (1992b) discussion.


The treatment of the turbulent velocity spectrum in evaluating the
correlation integral of Eq.~(\ref{eq:expect-ampl})
deserves some comment.  In this context, \citet{Stein67} first adopted
the quasi-normal approximation to relate the fourth-order correlation
to a sum of three independent products of second-order correlations,
as have others who have explicitly considered the velocity correlation
after him, notwithstanding \citet{Kraichnan57} warning of the danger
of so doing.  This decomposition would be correct if the probability
distribution function of turbulent quantities were Gaussian, which is
not what was subsequently assumed \citep[see][]{Belkacem06a,Kupka07}.  Note that a
decomposition of the fourth-order moment that takes the effect of
plumes into account has been proposed by \citet{Belkacem06a} and
permits to better reproduce the observations \citep{Belkacem06b}.  The
dominant term (a product of single-point correlations) is balanced by
the Reynolds stress that appears as the last term in
Eq.~(\ref{eq:gmk94}), leaving the forcing term $\myf$ to depend
only on two-point correlations. The next approximation adopted by
\citet{Stein67} is to write the energy spectrum as a product of a
function of local wavenumber and a function of frequency, relating the
three-dimensional velocity correlation to the energy spectrum by the
formula valid for incompressible, homogeneous isotropic turbulence
\citep{Batchelor53}. This result is approximately valid for isotropic
turbulence in the Boussinesq approximation - almost universally
adopted, either explicitly or implicitly, in mixing-length theory -
and which can be generalised to the axisymmetric case expected of
convection in a nonrotating star \citep{Chaplin05}, although the
generalisation to anisotropic turbulence has been adopted
\citep[e.g.][]{KB2009}.  Similar approximations have been used in
p-mode studies by \citet{B92c}, \citet{Samadi00I} and
\citet{Samadi02I}.



\subsubsection{The energy-equipartition principle}

A simple way of crudely estimating the mode amplitudes, without recourse 
to an explicit model for the excitation and damping of oscillation modes, is 
to adopt the equipartition ansatz 
discussed first by \citet{GK77}. 
It is obtained from a rough estimate of the integral in \eq{eq:expect-ampl}, 
which
is proportional to the ratio of the rate of forcing to the energy lost by 
dissipation \citep[cf][]{Batchelor53}. The energy supply rate was estimated 
from a mixing-length-like description of turbulence assuming that the 
motions of all the convective eddies are independent of one another, and that, 
as usual, the spatial scale of the eddies is the smallest of all scales, 
being less than both the scale height of the background state and the inverse 
wavenumber of the oscillation eigenfunction. The temporal spectrum of the 
turbulence was taken to be Gaussian, with variance proportional to the square 
of the characteristic timescale $\tau_{\rm c}$ of the energy-bearing eddies
(i.e. the largest eddies, with spatial scale equal to the mixing length $\ell$),
and with an autocorrelation that is significant only over the spatial scale of
a single eddy. The intrinsic damping rate $\eta$ was estimated by \citet{GK77a}
from a radial-pulsation calculation in which they presumed, as had
\citet{CoxEtal66} previously, that the phase of the perturbed heat 
flux $F^\prime_{\rm c}$ is determined by the equation 
${\rm d}F^\prime_{\rm c}/{\rm d}t=-F^\prime_{\rm c}/\tau_{\rm c}$,
and that the Reynolds stress, whose influence dominates the energy loss, can 
be represented by a time-independent scalar turbulent viscosity
$\nu_{\rm t}$ whose value is either $v^2_{\rm c}\tau_{\rm c}$, where
$v_{\rm c}$ is a characteristic velocity of an energy-bearing eddy, provided
$\omega\tau_{\rm c}<1$, or $v^2_\lambda\tau_\lambda$, where $v_\lambda$ is
the velocity of an eddy in the turbulent cascade (assumed to satisfy Kolmogorov
scaling) that resonates with the pulsation: $\omega\tau_\lambda=1$. This
discontinuous change in behaviour had been adopted previously by
\citet{GoldreichNicholson77}, and takes some account of the fact that eddies
with timescales much greater than $\omega^{-1}$ contribute very little to 
the dissipation of the pulsational motion; more sophisticated mixing-length
descriptions of convection \citep[e.g.][]{DOG65, DOG77a, Unno67} predict 
a continuous transition.

From simple scaling arguments \citet{GK77} concluded that the
turbulent fluctuating momentum fluxes with timescales comparable with
the pulsation period dominate the driving term $Q$
[Eq.~(\ref{eq:overlap-function_Q})], as had \citet{Stein67}
before them, and, because the dissipation is also dominated by the
same fluxes, via $\nu_{\rm t}$, there is a cancellation of a squared
turbulent velocity in the power-to-dissipation ratio, leading to an
equipartition between the energy $E_{\rm osc}$ in a mode of
oscillation and the energy $\rho\ell^3v^2_{\rm c}/2$ in a single
resonating eddy satisfying $\omega\tau_{\rm c}=1$, if such an eddy
exists, or the most energetic resonating eddy in the turbulent cascade
if no such energy-bearing eddy exists.  In the case where one does
exist, this balance may be written
\begin{equation}
E_{\rm osc}=\Lambda p_{\rm t}\ell^3\equiv E_{\rm c}\,,
\label{equi_principle} 
\end{equation}
where $p_{\rm t}$ is the $(r,r)$ component of the Reynolds stress
(also known as the turbulent pressure), and $\Lambda$ is a factor
of order unity whose value depends on uncertain properties of the 
turbulent flow, and is therefore itself uncertain.
 
This equipartition principle was adopted by \cite{CD83} for 
estimating p-mode amplitudes in main-sequence stars, and was 
applied for estimating g mode amplitudes in the Sun by 
Gough (1985; see also \citealt{GB90}).
In these calculations it was necessary to estimate the convective
timescale $\tau_{\rm c}$, which in the Sun has a minimum in the 
superadiabatic boundary layer \citep[see e.g.][]{Chaplin05}. 
Hence, for a mode with a period $\Pi$ 
that is longer than min$(\tau_{\rm c})$ there are at least two radii
in the convection zone where $\tau_{\rm c}=\Pi$, because formally
$\tau_{\rm c}$ tends to infinity towards the boundaries of the
convection zone. The kinetic
energy in a mode is then balanced against the sum of the values 
of $E_{\rm c}$ at these points.

Because the value of $\Lambda$ is uncertain, 
\cite{Gough85} calibrated the calculation by equating the maximum value of the 
amplitudes of the 5-minute p-mode oscillations with the value observed. 
He then obtained a maximum velocity amplitude of about 
$0.5 \, \textrm{mm\,s}^{-1}$ 
for the gravest quadrupole g mode and $1 \, \textrm{mm\,s}^{-1}$ for the
gravest dipole g mode, and about $3 \, \textrm{mm\,s}^{-1}$ for 
gravest of low-degree p modes. The result is very nearly a function
of frequency alone; it is depicted in Fig.~\ref{fig:amp_dog_kumar_kevin}.

\subsubsection{The results of Kumar, Quataert and Bahcall}
\label{Kumar}

\cite{Kumar96}, motivated by a claim 
of g-mode detection in the solar wind \citep{DT95}, 
carried out computations using the formalism developed by 
\cite{GMK94} originally to estimate the amplitudes of p modes.
In this formulation, \cite{Kumar96} assumed a simplified description of 
turbulence in which the velocities, length scales and timescales cascade
consistently according to the Kolmogorov spectrum. 
They assumed that the modes are driven solely by 
the fluctuating Reynolds stress.
Their calculation was identical to that of \citet{GMK94}.

Of particular interest is the way in which the eddies and the standing waves 
are temporally correlated. 
In the \cite{GK77} approach, from which was derived the procedures of 
\cite{B92c, GMK94}, it was assumed that the temporal correlation between eddies is Gaussian. 
As we shall see in the following sections, variations in the manner in which 
that function is chosen can lead to very different estimations of g-mode 
amplitudes. 

Concerning the damping rates, both turbulent and radiative contributions 
to the damping rates were included as derived by \cite{Goldreich91}. 
Turbulent damping was approximated purely as a momentum-diffusion process, 
with a time-dependent scalar diffusion coefficient derived from mixing-length 
estimates of the motion of eddies with lenghtscales smaller than the local 
wavelength 
of the oscillation. Temporal modulation of the Reynolds stresses, considered 
to be as important as momentum diffusion, if not more so, at least for grave 
p modes \citep[e.g.,][]{BG79, B92a}, was ignored, as was the modulation of the 
energy generation in the core. For higher-order p modes of low degree, the 
diffusive contribution was estimated by \cite{Goldreich91} to be comparable 
to the radiative losses, but for low-order p and g modes it is smaller 
\citep{Goldreich91, B92a}.

\cite{Goldreich91} found mode lifetimes of about $10^6\,$y, comparable with 
previous estimates. The resulting surface mode amplitude, based on \eq{Pobs}, 
was found to be greatest, at about $1\,$mm$\,$s$^{-1}$, for $l=1$ modes near
$\nu=\omega/2\pi=200\,\mu$Hz; modes with $\nu<100\,\mu$Hz were all found to 
have amplitudes less than $10^{-2}\,$mm$\,$s$^{-1}$.

\begin{figure}
\begin{center}
\includegraphics[width=0.75\textwidth,angle=0.3]{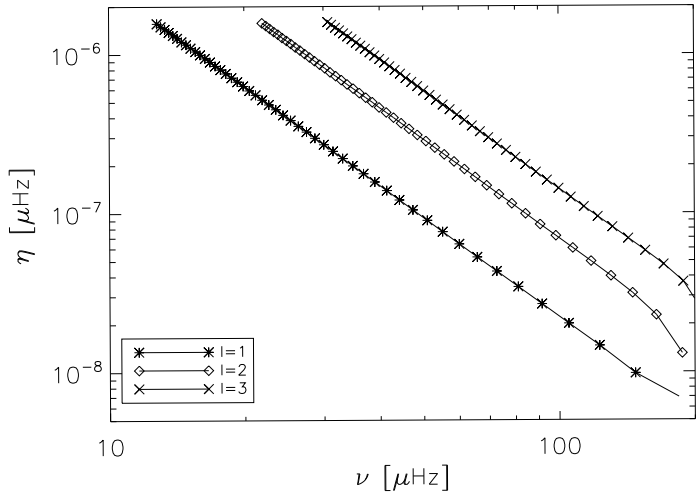}
\end{center}
\caption[]{Theoretical damping rate $\eta$ of g modes of low degree plotted
versus cyclic frequency $\nu$ according to the predictions 
by \cite{KB2009}.}
\label{fig:kevin_damp}
\end{figure}

\begin{figure}
\begin{center}
\includegraphics[width=0.6\textwidth,angle=90]{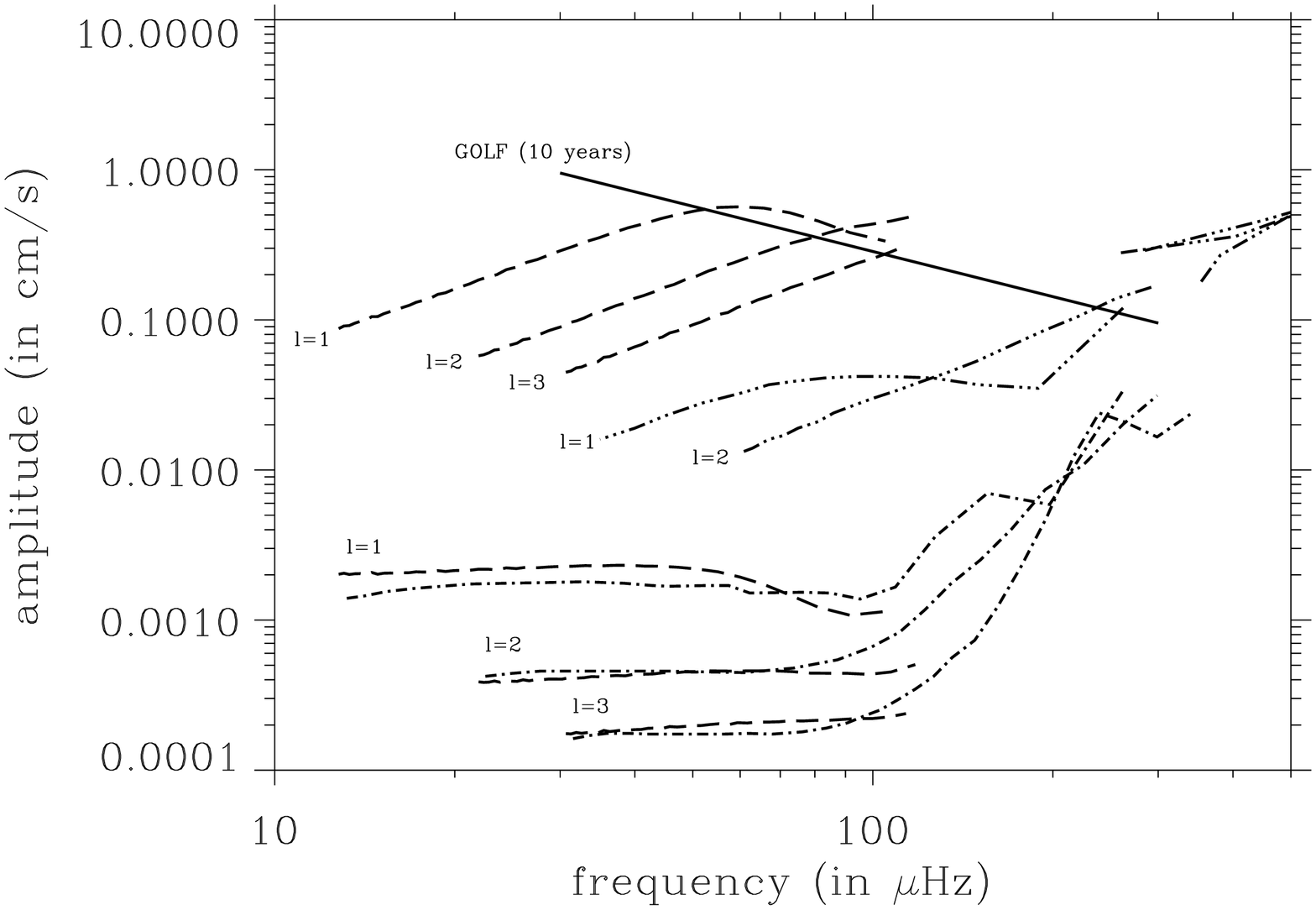}
\end{center}
\caption[]{Estimated amplitudes of stochastically excited 
g modes of low degree plotted against cyclic frequency $\nu$. The estimates
are rms surface values for singlet modes (single values of $n$, $l$ and $m$),
joined by lines: triple-dot-dashed for \citet{Gough85}, dot-dashed for \citet{Kumar96} 
and dashed for \citet{KB2009}. The continuous line is an 
estimated from 10 years GOLF data \citep{RAG2007}.  For proper comparison, the effect of the spatial instrumental filter must be included, i.e. visibilities \citep[See][]{KB2009} (See Section on Observables).
}
\label{fig:amp_dog_kumar_kevin}
\end{figure}

\subsubsection{The results of Belkacem et al.}

\cite{KB2009} investigated the particular case of the amplitude of asymptotic $g$ modes.
 The formalism used by \cite{KB2009} to compute excitation rates of non-radial modes 
was developed by \cite{Belkacem08} who extends to non-radial modes the work of \cite{Samadi00I}. 
The turbulent-mean stresses were neglected since the excitation takes place in the innermost convective 
layers where the Mach number is very small. The driving is shown to be dominated by the Reynolds-stress 
fluctuations.  In this formalism, the turbulent kinetic energy spectrum as well 
as the eddy-time correlation function are derived from 3D numerical simulation 
\citep[see][ for details]{KB2009}.  
The Lorentzian function ($\chi_k$) is found to better reproduce the eddy-time correlation function from 
the 3-D numerical simulation than a Gaussian function in the frequency range 
$\nu \in [20 \, \mu\textrm{Hz}; 110 \, \mu\textrm{Hz}]$. In addition, the eddy-time correlation function is poorly represented by a Gaussian function, which underestimates $\chi_k$ by many order of magnitudes.

Damping rates are computed with a fully non-radial non-adiabatic 
pulsation code MAD \citep{MAD06a,MAD06b,MAD06c}, including a non-local treatment of convection based 
on \cite{Unno67} formalism extended to non-radial modes by \cite{Gabriel1996,MAD05}. 
It takes into account the role played 
by the variations of the convective flux, the turbulent pressure and the dissipation rate of 
turbulent kinetic energy. 
\cite{KB2009} have found that for high frequency $g$ modes ($\nu > 110 \, \mu\textrm{Hz}$), 
the work integrals and thus the damping rates are sensitive to the 
convection/pulsation interactions because the role of the surface layers in the work 
integrals becomes important. 
In contrast, for low-frequency $g$~modes ($\nu < 110 \, \mu\textrm{Hz}$),  
the work integral and then the damping rates are found to be {\it insensitive} to the convection/pulsation interactions 
as well as the non-local parameters.  
Eventually, the damping rates dominated by radiative losses behaves as $\nu^{-3}$.

Taking  visibility factors as well as the limb-darkening into account, 
\cite{KB2009} finally found that the maximum of apparent surface velocities of asymptotic
 $g$-modes is $\approx 3$ mm\,s$^{-1}$ for $\ell=1$ at $\nu\approx 60\,\mu$Hz and $\ell=2$ at 
$\nu\approx 100\,\mu$Hz. Those results then put the theoretical $g$-mode amplitudes near the GOLF 
observational threshold.

The principal differences between the calculations \citet{KB2009} and \citet{Kumar96} 
stems from the turbulence spectra that were adopted. 
It would create enormous differences in g-mode amplitudes. 
They are produced by convection-oscillation
interactions deep in the convection zone, which are off resonance, and whose
intensity depends crucially on the assumed frequency dependence of the
turbulent spectrum in the high-frequency tail.
\citet{Stein67} and \citet{GK77} had favoured a Gaussian distribution. 
A Gaussian was preferred also by \citet{Chaplin05}, for although the 
energy-bearing eddies of the convection theory they used (that of
Gough 1965, 1977a) has a Lorentzian time-dependence.  We discuss the disparity in Section \ref{KGD-disc}.


\subsubsection{High-order waves generated at the base of the convection zone}
A series of papers using a wave-mechanical consistent numerical approach with a simplified two-dimensional solar model attempted to quantify the excitation of gravity waves in the solar interior as well the transmission of gravity waves through the solar convection zone \citep{Press81,Zahn91,BA92,BA93,BA94,BA96,Talon2003}. The results indicate that the efficiency of generating gravity waves in the interior by convective overshoot is slightly less than 0.1\% for waves with a horizontal extent equivalent to $l$=6-18. The overshoot layer where the excitation occurs is very shallow, 0.02-0.06 pressure scale heights. The energy transmission through the convection zone of a gravity wave equivalent to an $l$=6 mode is found to be 3-8 10$^{-4}$ in the 50-250\,$\mu$Hz. Combining these results indicate that the convective overshoot into the interior provides enough energy to sustain an amplitude of 1-5\,mm s$^{-1}$ at the surface of a single gravity mode. If the energy is distributed into more modes the amplitudes are reduced by the square root of the number of modes.

These results should not be taken as serious physical predictions of amplitudes of solar g-modes, but rather as a demonstration that there is enough energy available in the convective overshoot to maintain a single g-mode at an observable level at the solar surface. Assuming that a thousand modes are excited and scaling the amplitudes to an $l$=1 mode implies an amplitude of about 0.02-0.08\,cm s$^{-1}$ in the 50-250\,$\mu$Hz. As seen from Fig.~\ref{fig:amp_dog_kumar_kevin}, this is approximately the same values provided by \citet{Gough85}. 

\subsubsection{Prediction of g-mode amplitudes}
\label{sec:gmodeamp}
Amplitudes of singlet g modes (single values of $n$, $l$ and $m$) predicted 
by \citet{Gough85}, \citet{Kumar96} and \citet{Belkacem08,KB2009} are 
depicted in Fig.~\ref{fig:amp_dog_kumar_kevin}. Shown are the root-mean-square 
total velocity amplitudes in the photosphere; instrumental filtering
\citep[e.g.][]{Dziembowski77, JCDDOG82} must be taken into account when 
converting these values into observables.

\subsection{Discussion}
\label{KGD-disc}
The principal difference between the calculations of \cite{Kumar96} and \cite{KB2009} 
stems from the turbulence spectra that were adopted and, in particular, the way the  two
lead to enormous differences in g-mode amplitudes. 
Those differences are produced by convection-oscillation
interactions deep in the convection zone, which are off resonance, and whose
strength depends critically on the assumed frequency dependence of the 
turbulent spectrum in the high-frequency tail\footnote{ In this section, we denote by time-correlation function the Fourier transform $\chi_k$ of the local correlation function entering the evaluation of the global correlation function $G(\tau)$ appearing in Eq. (\ref{eq:expect-ampl}) \citep[see][for details]{Samadi00I,Chaplin05}.  We discuss Gaussian and Lorentzian forms, both of which have characteristic widths that depend on the lengthscales and
timescales of the convective eddies, and therefore vary with position $\vec{r}$ in the star.}.

For solar p modes, \citet{Stein67} and \citet{GK77} had used a Gaussian time-correlation function. 
A Gaussian was preferred also by \citet{Chaplin05}, for although the 
energy-bearing eddies of the convection theory they used (that of
Gough 1965, 1977) has a Lorentzian time-dependence, \citet{Chaplin05} reported that there is experimental evidence for more rapid decline in the tail. 
However, \citealt{Samadi02II} had pointed out that the three-dimensional simulations by \citealt{SteinNordlund01} were more nearly Lorentzian. Indeed, \citet{Samadi02II} and \citet{Belkacem06b} reported that the use of a Lorentzian function reproduces the observations for solar p modes better.  A similar conclusion was reached by \cite{Samadi08a} in the case of the star $\alpha$ Cen A. 
In contrast, \cite{Chaplin05} mention that a pure Lorentzian results in an over-estimation 
for the low-frequency modes. 
They explained that the over-estimation of the mode excitation rates at low frequency results from the Lorentzian function 
decreasing too slowly with depth compared to a Gaussian. Consequently, a substantial fraction of the excitation rate of the low-frequency modes arises from large eddies situated deep in the Sun. A Gaussian time-correlation function gives much less weight to the large off-resonance eddies.  Which is more nearly correct remains an open issue. 

\citet{KB2009} favoured a Lorentzian modelling for solar g modes, on the ground that 
it reproduces the 3D numerical convection simulations by \citet{MieschEtal08} more closely. 
More specifically, the best fit was found using a sum of Lorentzian and Gaussian functions. The Gaussian shape is found by the authors to reproduce the very low-frequencies while the Lorentzian the highest frequencies, which are of particular interest. 
There is yet no definite issue, but what is sure is that it has a crucial impact on predicted g-mode amplitudes. As demonstrated by \cite{KB2009}, Gaussian and Lorentzian time-correlation functions lead to differences in mode surface velocities of several order of magnitude.

It is worthwhile mentioning that experimental studies of turbulent convection can provide some clues towards the resolution of this issue.  In particular, the recent work of \citet{MordantEtal04} demonstrates that the two-point second-order time-correlation relevant to our discussion is reasonably well reproduced by an exponential decrease with separation time in the inertial range (see their Fig.~10) which is in agreement with \citet{Kolmogorov1941} phenomenology \citep[see also][]{Sawford91}.  It corresponds to a Lorentzian power spectrum with respect to frequency, which lends support to the use of a Lorentzian in the Sun, even though the Prandtl number in the experiment (6.8) is very far from solar.  At frequencies beyond the inertial range dissipation causes the spectrum to decline more rapidly \citep[see Fig~13 of][]{MordantEtal04}.  It is interesting to record that simulations by \citet{GeorgobianiEtal06} indicate that the spectrum is actually not separable in frequency and wavenumber, as \citet{Kraichnan57} had already pointed out, and suggested an empirical improvement.

Another important matter for solar g modes is the way in which the damping rates are computed. 
\cite{KB2009} did not calibrate their predictions against observations of p modes,
because they considered that because the g modes of higher-frequency
($\nu\!>\!110\,\mu\,$Hz), whose damping rates depend on the uncertain modelling of the interaction with convection, could not be predicted
reliably, a meaningful connection with the p-mode regime of
current observation could not be made. 
Instead they restricted attention to low-frequency modes whose damping rates $\eta$ are dominated 
by radiative losses, and scale with $\nu^{-3}$,
a behaviour which follows from straightforward asymptotic
argument \citep[e.g.][]{KB2009}, and which is evident
in Fig.~\ref{fig:kevin_damp}. 
But for grave g modes, the situation is less clear. It should perhaps be pointed
out in passing that the $\epsilon$~mechanism is significant for the grave g modes, and the influence of the dynamically induced thermodynamic perturbations must consistently be taken into account. 
A predictive description of the interaction between convection and oscillations when 
both are strongly coupled is mandatory to obtain reliable mode amplitude estimates 
in this frequency domain.

\subsection{Some concluding remarks}

Concerning the damping rates, the failure to detect g modes favours
that they are linearly stable. Theoretical amplitude estimates investigated by
\citet{WD83,WD82} and \citet{JG00} are well above the current detection threshold, although it must be appreciated that the results are somewhat uncertain.

In the high-order asymptotic regime, radiative diffusion is thought to be the dominant contribution to g-mode damping.  
For low-order g modes the situation is less clear since
those modes are sensitive to the interaction with convection, as are the p modes; there is no consensus
about the dominant contribution to the mode line-width
\citep[e.g.,][]{Houdek06}.  This issue is very important and prevents
an unambiguous theoretical determination of low-order g-mode
amplitudes.

For the driving mechanism, turbulent convection is thought to be
responsible for g-mode excitation \citep{DOG80,Kumar96,KB2009}, and
is probably dominated by the contribution of the Reynolds stresses. Indeed, the
so-called entropy contribution to the driving is found in the current theoretical computation to be negligible for both
low- and high-order g modes since they are not sensitive to the
very upper layers where entropy fluctuations are important.

Quantitative estimates of mode amplitudes differ from each
other by orders of magnitude, depending principally on the way the
turbulent eddies are time-correlated. While \citet{Kumar96,Chaplin05}
favours a Gaussian profile to describe this function,
\cite{Samadi00I,Samadi02II,SamadiEtal07,KB2009} used a Lorentzian
profile.  It should be clear that the eddy correlation function
cannot be analytically derived, and therefore is adhoc taken either from 3D simulations or turbulence experiment; as such it is a scale parameter.  In fact the truth is probably
somewhere in between, the challenge is now to determine where exactly.
The detection of g modes would provide a definite answer to this
burning question.


\section{Detection techniques}
\label{sec:det}

In this section, we present different detection techniques that
have been applied in searches for g modes.  We discuss the
conceptual design of each technique using the metaphor of
bricks.  We consider each {\it detection technique} to be built or
comprised of several bricks or {\it categories}. An appropriate
combination of various bricks constitutes a given {\it detection
technique}.  The categories we consider are as follows:
  \begin{itemize}
  \item Observables
  \item Spatial and temporal filters
  \item Spectrum estimators
  \item Statistical testing
  \item Patterns
  \item Data combinations
  \end{itemize}

The choice of the solar observable (or perturbation) to be observed
plays a key r\^ole in determining the likely success, or otherwise, of
a given technique (i.e., it can affect sources of noise, observed mode
amplitudes etc.).  Since we analyse time series, the temporal filter of
these series (e.g., low-pass, high-pass, band-pass) also merits
careful attention . Observations of the Sun may be made by imaging the
surface onto several or many detector pixels, or by observing the
``Sun as a star''. When resolved observations are made, it is advisable to apply
appropriate spatial masks to extract the signatures of the modes of
interest.

The detection of potential frequencies of interest requires that we
obtain estimates of the frequency spectrum, using Fourier transform or
other techniques.  Then, a thorough statistical assessment of what is
observed plays a key part in deciding whether or not something of interest
has been detected.  If the identification of individual
frequencies fails, one can then bring more {\it a priori} information
to bear in the detection procedures, for example by searching for
signatures of patterns expected from theoretical modelling of the g
modes (e.g., near constant spacings in period of modes in the
asymptotic regime, rotational frequency splittings etc.). Use of a
priori knowledge in this way suggests a Bayesian approach to the
analysis.  Finally, if the detection is not successful with a single
instrument, one can bring data from different instruments to bear, for
example by searching for coincidences in the different datasets.

\subsection{Observables}

Perturbed parameters such as temperature and displacement can be
deduced from observables.  The displacement can be either physically
observed or inferred from the solar radial velocities.  

The solar
oscillations were first detected in radial velocity observations made by
\citet{Leighton1962}.  This method is now widely used for observing
solar p modes.  Solar radial velocities can either be
measured from resolved images of the Sun \citep[][and references
therein]{Unno89}, or from full-disc integrated Sun-as-a-star
observations.  The instrument aboard SOHO all rely on measuring the intensity
distribution in an absorption line for deriving the solar radial velocity either locally such as with the SOI/MDI\footnote{Solar Oscillations Investigation / Michelson Doppler Imager} instrument \citep{PS95} or globally with the GOLF\footnote{Global Oscillations at Low Frequency} instrument \citep{Gabriel95}.  This set of instrument is also complemented by ground-based networks such as the high-resolution GONG\footnote{Global Oscillations Network Group} instrument \citep{Harvey1996}, and the full-disk integrated BiSON\footnote{Birmingham Solar Oscillations Network} instrument \citep{WJC1996}.  The lengths of the usable time series are about 14 years for MDI and GOLF, 15 years for GONG and nearly 30 years for BiSON.


Temperature perturbations induce intensity fluctuations that can be
detected.  Intensity fluctuations may also be induced by changes in
the size and shape of the solar surface produced by the oscillation
modes \citep{Toutain99}.  Intensity fluctuations were first observed
from space by \citet{Woodard1983} in solar irradiance data collected
by the ACRIM\footnote{Active Cavity Radiometer Irradiance Monitor} instrument \citep{Willson1979}.  Similar
observations have been made with sunphotometers from
stratospheric balloons \citep[e.g.][]{Fro84a} which confirmed
the ACRIM results.  Ground-based observations with the
SLOT\footnote{Solar Luminosity Oscillation Telescope} instrument produced rather marginal p-mode detections \citep{Jimenez1987}, while
rather higher signal-to-noise data were collected by the
LOI\footnote{Luminosity Oscillations Imager}
instrument, observing from El Teide, Tenerife \citet{TABA97,TA95a}.
Following the success of ACRIM, the IPHIR\footnote{InterPlanetary
Helioseismology by IRradiance measurements} instrument onboard the
PHOBOS mission lead to the first long-duration intensity measurements
of the solar p modes \citep{TTCF92}.  The ACRIM, SLOT and IPHIR
instruments lead to the design of the VIRGO\footnote{Variability of
IRradiance and Gravity Oscillations} instrument for the SOHO
mission, which was composed of two radiometers (DIARAD and PMO6),
three Sun PhotoMeters (SPM) and the LOI \citep{CFJR95}.  The lengths of the usable time series are about 14 years for VIRGO.

Intensity fluctuations produce a much lower signal-to-noise ratio in
the p modes than do solar radial velocity observations.  The
signal-to-noise ratio for the former is at most 30, while the latter
is at most 300; examples of the various signal-to-noise ratios for the
different observables can be found in \citet{Toutain1997}.

The detection of the physical displacement is more difficult.  For
instance a putative g-mode amplitude of 1 mm s$^{-1}$ at a frequency
of 100 $\mu$Hz would induce a peak-to-peak variation of 4.3
$\mu$arcsec as seen from the Earth, whereas for a p-mode amplitude of 30\,cm\,s$^{-1}$ (at a frequency of $3000\,\rm \mu Hz$) would induce a variation of 43 $\mu$arcsec.  In the 1970's, the SCLERA\footnote{Santa Catalina Laboratory for Experimental
Relativity by Astrometry} instrument was used in an attempt to
detect p modes and g modes alike \citep{SCLERA1974,Hill1985,Hill1992}.  The
detection of displacement was performed by observing the change in
diameter of the Sun \citep{Brown1978}, and signals with typical
amplitude of the order of 5 marcsec were observed. These amplitudes
were about a factor of 100 larger than the expected amplitudes for p
modes, based on high signal-to-noise detections of the p modes made in
radial velocity observations \citep{Claverie81,Fossat81}.  This
discrepancy cast some doubt on the identification of the detected
SCLERA peaks as p modes.

The PICARD mission, due for launch in late
2009 \citep{LD99}, will be sensitive not only to the displacement, but also to
intensity fluctuations induced by temperature perturbations, like
those detected by \citet{TA98d} with the VIRGO/LOI or by \citet{CT99}
with SOI/MDI. PICARD will have the capability of making observations of
displacement and intensity fluctuations at the limb of the Sun, where
the perturbations appear to have higher amplitude than those observed
in the disc.  This amplification, theorised by \citet{Toutain99}, is
due to the fact that at the solar limb the atmosphere becomes more
transparent, providing a larger contribution from the other
hemisphere.

 Another potential observable is the perturbation of the gravitational field caused by the g modes producing
tidal perturbations of the Newtonian fields, and generating detectable 
gravitational waves \citep{JCDDOG80,GG99}.  The perturbation is non-zero only 
for modes with $l \ge 2$; for $l=1$ it is exactly 
zero.  The advantage is that the solar noise caused by convection would have a negligible impact on the perturbation of the gravitational field.  Unfortunately, as \citet{JCD2002a} explicitly mentions, signal from galactic 
binary stars could be the limiting factor for g-mode detection and not the Sun itself.

\subsection{Spatial and Temporal filters}
\subsubsection{Temporal filters}

The analysis of discretized signals dates back to the start of the
telephone and telegraphic system, when Harry Nyquist devised the
so-called Nyquist frequency \citep{Nyquist24}, and when Claude Shannon
enunciated the Shannon theorem concerning the sampling of band-limited
signals \citep{Shannon49}. Both worked at Bell Telephone Laboratories.

Although, we do not always follow the Shannon-Nyquist prescription,
one should not forget the limitation inherent in the sampling of a
band limited signal.  The main difficulties encountered in
astrophysical applications are as follows:
 \begin{itemize}
 \item the signal is {\it not} band limited
 \item the signal is undersampled
 \item the signal is aliased
 \end{itemize}
These three factors are not unrelated to one another.

In our case, the observations of intensity fluctuations and of solar
radial velocities do not present a clear high-frequency cut-off
\citep{AK97,CF97}.  The frequency drop-off of the solar background
noise at high frequencies is proportional to $\nu^2$ \citep[][ and references
therein]{Harvey85,Aigrain2004}, or possibly even a higher power law
\citep{TA2002}.  Since the astrophysical signal is {\it not} band
limited, the highest frequencies will not be properly sampled,
resulting in an {\it undersampled} signal. This leads to
high-frequency signal leaking back into the spectrum, below the
Nyquist frequency, i.e., aliasing of the signal.  The imaging effect
of aliasing is to produce the Moir\'e effect when the image is not
properly sampled.


It is also worth pointing out that the Shannon theorem states that the
signal should not only be properly sampled, but also properly
integrated over the sampling interval.  In other words, one should
have a 100\% fill cycle (the fill cycle being the ratio of the
integration time to the sampling time).  If this is not the case,
frequencies from the non-band limited signal above the Nyquist
frequency will not be suppressed.  It is then very critical to make
sure that the fill cycle is as high as possible, thereby helping to
reduce the aliasing effects of the high-frequency power contained in
the non-band limited signal.  The aliasing effects will severely
affect the derivation of how the power drops at high frequency,
implying that extreme care should be taken when recovering the power
law of the background at high frequencies.  From this point of view,
the VIRGO instrument aboard SOHO has, by design, a fill cycle very
close to 100\% \citep{CF97}, while that of GOLF is 80\% \citep{Gabriel95}.

Last, but not least, since the signal is digitised, it is possible to
use digital filters that are routinely used in the electronic world
for many digital applications \citep[telephone, images and so
forth;][]{Antoniou}.

\subsubsection{Spatial filters: Sun-as-a-star observations}

When the Sun is observed as a star, there is a cancellation effect due to the integration over the disk.  The net effect is that not all modes can be detected.  The calculation of the attenuation as a function of the degree of the modes is given by \citet{JCDDOG82} for solar radial velocities and by
\citet{Toutain99} for intensity; mainly modes with $l \le 4$ can be
detected.  To the first order, when the solar inclination angle is 90 degrees with respect to the line of sight, only the odd $l+m$ modes are detectable.  Nevertheless, there is a small effect with the solar inclination angle discussed recently by \citet{Gizon2003}, rendering all $2l+1$ singlet modes of a multiplet detectable \citep[See also][]{Ballot2006,Ballot2008}.  Indeed, several attempts had already been made
to determine from the relative amplitudes of the observed p-mode singlets
in the Sun the orientation of the principal axis of rotation \citep{DOG1995,Gizon98}  For g modes, the sensitivity to the degree of the modes was
derived by \citet{GB90} for both intensity and solar radial
velocities.  They showed that for g modes, sensitivities of different
degrees depends strongly on frequency.  The large difference in
sensitivities between solar radial velocities and intensities can in
principle be exploited to help identify detected g modes.




\subsubsection{Spatial filters: spatially resolved solar images}

The observation of resolved images of the Sun makes possible the
decomposition of the spatial structure of the modes onto functions
that are related to the modes.  The spatial filters may be
classified in terms of different types of mask:
 \begin{itemize}
 \item spherical harmonics masks
 \item g-mode specific masks
 \item optimal masks
 \item time-distance masks
 \end{itemize}
For p modes, the obvious choice for the masks is to use the spherical harmonics that
 match the spatial dependence of the modes when the star is
spherically symmetric \citep{Unno89}.  When the star rotates steadily or has a magnetic field, the spatial
dependence becomes more complicated \citep{DOGAK93,Gizon98} partly because the star is no longer spherically symmetrical, and partly because the modes must be referred to a frame of reference, if one exists, in which the state of the star is steady.  If no such state exists, a well defined frequency of oscillations does not exist either.  One has
to bear in mind that such effects could be important for the visibilities
of g modes if there is a large magnetic field in the central core
\citep{goode1992,rashba2006}.

Spherical-harmonic masks correspond to displacement or
temperature perturbations (velocity or intensity observations).  These
masks were applied first in the 1980s by \citet{Brown1985} (using an
algorithm based on the Fast Fourier transform).  The degree
sensitivities of these masks have been derived by \citet{JS92},
\citet{TA2000} and \citet{Corbard2008} for imaging instruments such as
SOI/MDI, GONG, LOI, or SODISM\footnote{SOlar Diameter Imager and
Surface Mapper} on the PICARD mission.

Unfortunately, although the spherical-harmonic masks are rather well
adapted for most p modes, they are not well adapted for the g modes.
For velocity observations, the significant horizontal g-mode displacement 
must be taken into account (the perturbations
induced by p modes are predominantly radial), the spatial dependence of which differs from that of the radial displacement.  For intensity observations, the oscillations perturb the figure of the Sun modulating the
light emitted from the surface \citep{GB90,Toutain99}.  For either type of observation, the
additional contribution is a non-spherical harmonic function
\citep{Unno89,TABA90}. Specific g-mode masks can be
devised \citep[for intensity observations see ][]{TABA90}.  In some cases inclusion of the most appropriate spatial dependence of the g-mode perturbation
still remains to be done.  A large uncertainty lies in the way
nonadiabatic effects are accounted for.

Masks that are optimal for some specific signals have been developed.  For
instance, \citet{TABA90} developed p-mode intensity masks that minimise both the
leakage from other degrees and also the leakage from other $m$
\citep[see also][]{JCD1984,Sasha1986a,Sasha1986b}.  Similar masks have been
derived by \citet{Latour1984}, \citet{TALG98} and by \citet{TT2000}.  Recently,
\citet{RW2002} produced g-mode masks optimising (i.e. minimising) the
noise contribution from the supergranulation noise across the solar
disc (which has a strong horizontal component).

Other kinds of masks derived from the properties of the observations
themselves have been used by \citet{Vecchio2005}.  These masks are
based upon a Proper Orthogonal Decomposition (POD) of the velocity
field, which is nothing less than the computation of eigenvalues and
eigenvectors of that vector field.  This POD analysis showed
different power spectra at disc centre compared to the solar
limb. These differences could be related to the dependence of the
perturbation with distance to the centre of the solar disc.

The aforementioned masks are defined for observing perturbations at
the surface.  With the advent of time-distance helioseismology
\citep{Duvall1993}, there is the possibility of constructing special
masks that will be sensitive to perturbations located deeper in the
Sun, where the motions induced by gravity modes are much larger than at
the surface.  The idea is to detect motions induced by the g modes on
the solar p modes.
  
\subsection{Spectrum estimators}
\subsubsection{For data with no gaps}

The estimation of the spectrum of the time series is of prime
importance when one wants to detect eigenmodes.  The lifetimes of the
modes should be taken into account when using these estimators.
The following estimators are at our disposal:
 \begin{itemize}
 \item Fourier spectrum (power spectra)
 \item Lomb-Scargle periodogram and sine wave fitting
 \item Average, smoothed and multitapered spectra
 \item Cross spectrum
 \item Random Lag Singular (Cross) Spectrum analysis
 \item Frequency matching (oversampling and bin shifting)
 \item Time-frequency spectrum
 \item Varying time base
 \end{itemize}

Fourier spectrum estimation is widely used in helioseismology.  Its
properties are well known and quite often well understood
\citep{Bracewell}, as are its statistics \citep{Davenport}.  In short, the Discrete Fourier spectrum estimation
is widely used for time series being sampled at a regular sampling cadence ($\Delta t$) during an observing time ($T$) 
that is an integer number $N$ of the sampling cadence.  The Nyquist frequency of the Discrete Fourier spectrum is then $\nu_{\rm Nyquist}=1/\Delta t$, and the so-called frequency bin is the $N$th part of that or $\Delta \nu=1/N/\Delta t$.  The increase in frequency resolution makes the detection of sine waves extremely effective because the power spectrum level of any stationnary noise decreases like $1/T$.  If a sine wave would change many times its phase during the observation time $T$, it would not be advised to compute the Fourier transform of the whole observation.  In that latter case, the optimal computation would be to perform the Fourier transform for the period of constant phase, if these are known, and then to add the power spectra.  An other incentive for not computing the whole Fourier transform would be be that the frequency of the sine wave changes gradually with time (See later in this Section).

Other tools have been developed for time series that are not evenly
sampled.  Then, the Fast Fourier Transform should be replaced
by an adaptation of the Discrete Fourier Transform, the so-called
Lomb-Scargle (LS) periodogram \citep{Scargle82}.  The LS periodogram
is widely used in astrophysics for stars having modes with very long
lifetimes.  The LS periodogram is strictly equivalent to sine-wave
fitting, as shown in Appendix C of \citet{Scargle82}.  The main
difference between the two methods is more in the approach than in the
result: with the LS periodogram the frequency of the sine wave is
found from the spectrum itself, while for the sine wave fit the frequency
is obtained from the fit provided that a proper initial guess is
given, coming from the LS periodogram for instance.  For speed, the LS
periodogram is usually computed using the implementation prescribed by
\citet{Press1989} which is an approximation of the LS periodogram
based upon {\it extirpolation}\footnote{Reverse interpolation or
extirpolation replaces a function value at any arbitrary point by
several function values on a regular mesh} on a regular mesh and the
use of the FFT.  The prescription is then very close to interpolating
onto a regular mesh.  It is worth noting that most scientists using
the LS periodogram for unevenly sampled data are in fact computing the
FFT of the original data resampled onto a regular mesh, but with only
a proper normalization as given by \citet{Scargle82}.

Fourier spectrum estimation is well adapted for periodic signals (pure sine waves or 
stochastic waves) but not necessarily well suited for estimating the spectral
density of pink or red noise.  For that purpose, one can: 
 \begin{itemize}
 \item average the power spectrum over an ensemble of $q'$ sub-series,
 \item smooth the power spectra over $q'$  frequency bins or,
 \item use multitapered spectra using the full time series for deriving a similar average.  
 \end{itemize}

Fourier spectrum estimation is now being replaced by multitapered
spectra that are widely used in geophysics \citep[for a review
see][]{thomson82}.  Multitapered spectra are generated by applying a
set of {\it slepian} tapers to a single time series, and an estimate
of the mean power spectrum is derived from an average of these
spectra.  The multitapered spectra are statistically independent from
one another, and the statistics of the mean spectrum follows a
$\chi^{2}$ distribution with 2$q$ degrees of freedom \citep[where $q$
is the number of tapers,][]{thomson82}.  While the statistics of the
average power spectrum (or smoothed power spectrum) also follow a
$\chi^{2}$ distribution with 2$q'$ degrees of freedom, the resolution
of the average spectrum is $q'$ times lower than that of the
multitapered spectrum.  In helioseismology the use of these slepian
tapers has been replaced by more practical (but less accurate) sine
tapers \citep{Komm99}.  Unfortunately, for long-lived modes, tapers
tend to broaden the peaks, as shown by \citet{thomson82}.  Tapers as
such provide more benefit for p modes having a lifetime shorter than
the observation time.

The use of cross spectrum estimation can also be useful for detecting signals
being present in two different time series  \citep{Sturrock2005}.  This can be done either from the same observable or from different
observables (e.g. intensity and radial velocity).  This technique has been used by \citet{RG1999} for improving 
the signal-to-noise ratio in the p-mode region, but it is not useful for g-mode detection as shown by \citet{TA2007}.

Random Lag Singular (Cross) Spectrum analysis is an elaborate
technique based on Singular Value Decomposition \citep[RLSSA,
RLSCSA][]{Varadi99,Varadi2000}.  Although the technique claimed successful
detection of low-frequency p modes \citep{LB2000,RAG2001}, although a proper assessment
of its statistical properties is still lacking: i.e. it cannot be
excluded that the technique produces large peaks solely due to noise
\citep{Couvidat02}.  The spectrum produced by the RLSSA technique can
be compared to a power spectrum raised to an unknown power (greater
than one). As such, small peaks with a signal-to-noise ratio slightly
larger than unity are amplified, while those with a lower
signal-to-noise ratio are damped.  The fact that this exponentiation
is unknown (dependent upon the data) makes the RLSSA and RLSCSA
techniques in our view unsuited for robust signal analysis.

Frequency matching has been developed by \citet{AG2002} and by
\citet{WJC2002} using zero padding and  frequency-bin shifting, respectively.
When one wants to detect signals from oscillators having a lifetime
longer than the observing time, there is a significant chance that the
frequency bin will not match the frequency of the oscillator.  As a
result, power from the signal may be split between  frequency bins and the main
peak could be reduced in power by up to 60\,\%, as shown by
\citet{AG2002}.  In order to alleviate this problem one can either
oversample the data by using zero padding \citep{AG2002} or try to
tune for the frequency by using  frequency-bin shifting \citep{WJC2002}.  This
latter technique involves creating many different time series of
similar but slightly different lengths, but without zero padding
\citep{WJC2002}.  In either case, the statistics of the observation are indeed affected.  For oversampling, the analytical calculation has been replaced by Monte-Carlo simulation
showing that there are typically three independent frequency bins when
the Fourier spectrum is oversampled by a factor five or more
\citep{AG2002}.  The  frequency-bin shifting method also produces three
independent power spectra from the many spectra generated
\citep{WJC2002}.  Both approaches have the added benefit that they
give access to different ``realizations'' of the noise background, as
if we had three actual, independent realizations of the noise
available.

Time-frequency spectrum or wavelet analysis has been used by
\citet{Gabriel99}, by \citet{Finsterle2001} and by \citet{AJM2008} to look for signatures
that are affected by the lifetimes of candidate g modes.  The
statistical properties of these wavelets can be derived from those of
the Fourier spectrum. The potential influence of solar activity on the
size of the resonant cavity, which would affect the frequency of long-lived modes, has also been studied by \citet{Gabriel2006}.  If the
frequency of the mode changes slowly with time, the frequency of the
mode will be spread over several  frequency bins.  The idea followed by
\citet{Gabriel2006} is to change, or distort, the sampling time to
follow the variations in the mode frequency induced by solar activity.

\subsubsection{For data  with gaps}
All the estimators described above have different properties when gaps appear in the data.  Hereafter, we shall discuss mainly the following:
 \begin{itemize}
 \item Fourier spectrum (power spectra)
 \item Lomb-Scargle periodogram and sine wave fitting
 \item Average, smoothed and multitapered spectra
 \item Frequency matching (oversampling and bin shifting)
 \end{itemize}
 The impact of the gaps on the Fourier spectrum has been described by \citet{MG94} providing as a result that the correlation matrix of the frequency bins is not longer diagonal.  The gaps introduce correlation between the frequency bins that require to be taken into account when one wants, for example, to fit the power spectrum \citep[See for applications][]{Stahn2008}.
 
The LS periodogram does not provide a better solution to coping with the presence of gaps.  The reason is that although the Fourier transform {\it explicitly} includes gaps as zeros, adding zeros is also {\it implicitly} performed with the LS periodogram.  As a consequence, correlations between frequency bins also exist with the LS periodogram, but these are frequently ignored. 

The impact of the gaps for the average will have the same origin as for the Fourier spectra.  As for the smoothed spectra, there is an intrinsic correlation between the $q'$ bins; the number of truly independent frequency bins is then divided by $q'$.  In that latter case, the gaps do not have a large influence on the smoothed spectrum unless the fraction of gaps is about $1/q'$.  For multitapered spectra, the influence of gaps can be taken into account for obtaining optimised tapers that match the structure of the gaps \citep{Fodor1998}.  In that case, the correlation between the frequency bins, although reduced, will be not negligible, especially if several tapers are used for the estimation of the mean power spectrum.

\subsection{Statistical testing}

Statistical testing is essential when one wants to decide: {\it have
we found g modes} or {not}?  This is related to {\it decision theory},
which can be summarised as {\it how do we choose between one
hypothesis versus another in the presence of uncertainties?}  In this
area, there are two schools of thought: the frequentist school and the
Bayesian school.

The difference between a Bayesian and a frequentist relates to his or
her views of {\it subjective} versus {\it objective} probabilities.  A
frequentist thinks that the laws of physics are {\it deterministic},
while a Bayesian ascribes a belief that the laws of physics are true
or {\it operational}.  The {\it subjective} approach to probability
was first coined by \citet{Finetti}.  For the rest of us, the
difference in views can be summarised by this quote from the Wikipedia
encyclopaedia: {\it Whereas a frequentist and a Bayesian might both
assign a probability $\frac{1}{2}$ to the event of getting a head when
a coin is tossed, only a Bayesian might assign a probability
$\frac{1}{1000}$ to personal belief in the proposition that there was
life on Mars a billion years ago, without intending to assert anything
about any relative frequency.}  In short, frequentists assign
probability to measurable events that can be measured an infinite
number of times, while Bayesians assign probability to events that
cannot be measured, like the outcome of sport-related bets, for
instance. or the survival time of the human race \citep{Gott94}.

With this word of caution, one should never forget that we wish to use
this armada of statistics because we know that we are at the limit of
detection for g modes.  In what follows, we will try to give an
overview of what {\it we believe} we know on a subject that is fast
evolving; and what we write is certainly {\it not} gospel.

\subsubsection{Frequentist approach}

For a frequentist, statistical testing is related to hypothesis
testing.  In short, we have two types of hypotheses:
 \begin{itemize}
 \item ${\rm H}_0$ hypothesis or null hypothesis: what has been observed is pure noise?
 \item ${\rm H}_1$ hypothesis or alternative hypothesis: what has been observed is a signal?
 \end{itemize}
For the ${\rm H}_0$ hypothesis, we assume a known statistics for the
random variable $X$ observed as $x$ and assumed to be pure noise; and then set a {\it false alarm probability}
that defines the acceptance or rejection of the hypothesis
\citep{Scargle82}.  The so-called {\it detection significance} (or
{\it p-value}, terms not widely used in helioseismology) is the
probability of having a value as extreme as {\it the one actually
observed}.  There is an on-going confusion because statisticians call
{\it the significance level} what astronomers call the {\it false
alarm probability}; and they call the {\it p-value} what is set in
astronomy as the {\it detection significance} (which is {\it not} the
significance level).  Here we shall use the current vocabulary
understood in astronomy. For example, the {\it false alarm
probability} $p$ for the ${\rm H}_0$ hypothesis is defined as:
 \begin{equation}
 	p = P_0(T(X) \geq T(x_{\rm c})),
 \end{equation}
where $T$ is the statistical test, and $P_0$ is the probability of
having $T(X) \geq T(x_{\rm T})$ when ${\rm H}_0$ is true; and $x_{\rm c}$ is the cut-off threshold derived from the test $T$ and the value $p$.  For example, take
the case of a random variable $X$ distributed with $\chi^2$, 2 degrees
of freedom (d.o.f) statistics, having a mean of $\sigma$. If we
further assume that $T(X)=X$, we then have that:
 \begin{equation}
	p = P_0(X \geq x_{\rm c}) = e^{-\frac{x_{\rm c}}{\sigma}}.
 \end{equation}
If one observes a value $\tilde{x}$ of the random variable $X$ that
is larger than $x_{\rm c}$, the ${\rm H}_0$ hypothesis is rejected.  The
value that is quoted in this case is the {\it detection significance}
${\cal D}$, i.e.,
 \begin{equation}
	{\cal D} = e^{-\frac{\tilde{x}}{\sigma}}.
	\label{p-value}
 \end{equation}
The ${\rm H}_0$ hypothesis was used by \citet{TA2000} to impose an
upper limit on the g-mode amplitudes.  The method was based on the
knowledge of the statistical distribution of the power spectrum of
full-disc integrated instruments, namely the $\chi^2$
distribution with 2 d.o.f.

For the ${\rm H}_1$ hypothesis, we assume given statistics both for
the noise and for the signal that we wish to detect, and set a level
that defines the acceptance or rejection of that hypothesis.  The
${\rm H}_1$ hypothesis was used in \citet{AG2002} to determine the
probability of detecting a sine wave given noise in the GOLF data
\citep{AG2002}.  In this latter case, the statistics follow a
non-centred $\chi^{2}$ distribution with 2 d.o.f.  Both
hypotheses have also been used to calculate significance levels for
detecting stellar p modes \citep{Appourchaux2004}.

Taking a decision based on the result given by a single test, for
either hypothesis, could lead to errors in the decision process.  For
instance, the null hypothesis could be wrongly rejected when it is
true ({\it false positive} or wrong detection), but could also be
wrongly accepted while it is false ({\it false negative} or no
detection in presence of a signal).  The {\it false positive} results
in a {\it Type I} error, while {\it the false negative} results in a
{\it Type II} error (see Table 2).  The ideal case would be to set a
test that would minimise the occurrence of both types of errors.

In helioseismology, it has been customary when applying the ${\rm
H}_0$ hypothesis to set the decision level arbitrarily at 10\%
\citep{TA2000}.  From the frequentist view point there is nothing
wrong in setting {\it a priori} the decision level before the test is
applied.  There are three types of result we might obtain from the
test:
 \begin{enumerate}
 \item ${\rm H}_0$ always rejected
 \item ${\rm H}_0$ rejected or accepted at a level very {\it close} to 10\%
 \item ${\rm H}_0$ always accepted
 \end{enumerate}
Decision 1 will lead to the mention of a {\it detection being
statistically significant} at a level provided by the {\it detection
significance} [from Eq.~(\ref{p-value})].  The next step would be
the application of a test for the ${\rm H}_1$ hypothesis, very likely
resulting in the detection of signal.  Decision 2 is the more
difficult borderline case, forcing us to either accept or reject ${\rm
H}_0$.  Here, we might ask: are things really that clear-cut?  What
are the chances that if we accept ${\rm H}_0$ is it actually wrong
(Type II error), or truly right if rejected (Type I error)? Decision 3
seems straightforward, i.e., noise dominates, but might one then be
tempted to lower, {\it a posteriori}, the decision level?

These potential actions result from the application of a frequentist
test trying to answer the following question: what is the likelihood
of the observed data set $\tilde{x}$, given that ${\rm H}_0$ is true
or $p(\tilde{x} | {\rm H}_0)$?  The {\it detection significance}
mentioned when the test rejects the ${\rm H}_0$ hypothesis is nothing
but $p(\tilde{x} | {\rm H}_0)$, when actually what we want to know is
the likelihood that ${\rm H}_0$ is true given the data, i.e., $p({\rm
H}_0 | \tilde{x})$ ($\neq {\cal D}$).  The frequentist view does
provide a useful answer when one can repeat the observations ad
infinitum.  But when we have only one observations, another approach
we may apply is the Bayesian approach, which in principle gives access
directly to $p({\rm H}_0 | \tilde{x})$.

\begin{table}[t]

\caption{Types of error obtained for different decisions, based upon
the statistical test performed, and how the error relates to the
status of the ${\rm H}_0$ hypothesis.}

\centering
\begin{tabular}{cccc}
\hline\noalign{\smallskip}

&&\multicolumn{2}{c}{{\bf Status of H$_0$}}\\[1ex]
&&True&False\\[1ex]
\hline\noalign{\smallskip}
&Reject& Type I& Correct\\
\raisebox{1.5ex}{{\bf Decision}}&Accept&Correct&Type II\\
\noalign{\smallskip}\hline
\end{tabular}    
\end{table}

\subsubsection{Bayesian approach}

{\bf On the posterior probability} We should never forget the {\it two
sides of the coin}: if probability (likelihood) can justify {\it
alone} the rejection or acceptance of an hypothesis, this probability
{\it is not} the significance that the hypothesis is rejected or accepted.
The decision levels discussed above are related directly to a
well-known controversy in the medical field, concerning improper use
of Fisher's p-values as measures of the probability of effectiveness
of a medicine or drug \citep{Sellke2001}.  The {\it detection
significance} (or p-value) is improperly used as the significance of
the evidence against the null hypothesis.  It is far from trivial at
first sight to understand what is wrong with the {\it detection
significance}. Let us recall the example of a noise with 2 d.o.f statistics.  In that
case the {\it detection significance} is given as:
 \begin{equation}
	{\cal D} = e^{-\frac{\tilde{x}}{\sigma}} \not\equiv P_0(X \geq \tilde{x}).
	\label{p-value1}
 \end{equation}
 The latter statement ($\not\equiv$) is fundamental.  The observation is performed only once providing a value of $\tilde{x}$.
It is not correct to assume that this observation if it were repeated would provide the same level $\tilde{x}$.
 The mistake is to ascribe a significance to a
measurement performed only once, i.e., not repeated, and spanning just
a very small volume of the space of the parameters (e.g. $X \in
[\tilde{x},\tilde{x}+\delta x]$).  If one makes a measurement $\tilde{x}$ of the random
variable $X$ which is above $x_{\rm c}$, the significance of that measurement
is {\it not} $e^{-\tilde{x}/\sigma}$.  In the framework of Bayesian
statistics, we are not interested in the {\it detection significance}
but in the posterior probability of the hypothesis, in other words as
already stated above $p({\rm H}_0 | \tilde{x}) \neq {\cal D}$.  A similar description of this misunderstanding has been presented by \citet{Sturrock2009}.

In order to derive the posterior probability $p({\rm H}_0 | x)$, let
us first recall the Bayes theorem.  The posterior probability of a
hypothesis H, given the data D and all other prior information I
is stated as:
 \begin{equation}\label{bayes}
    P({\rm H} | {\rm D,I}) = \frac{P({\rm H} | {\rm I})P({\rm D} | {\rm H,I})}{P({\rm D} | {\rm I})}.
 \end{equation}
where $P({\rm H} | {\rm I})$ is the prior probability of H given I, or
otherwise known as the prior; $P({\rm D} | {\rm I})$ is the
probability of the data given I, which is usually taken as a
normalising constant; $P({\rm H} | {\rm I})$ is the direct probability
of obtaining the data given H and I. \citet{Berger1987} obtained,
using the Bayes theorem, $p({\rm H}_{0} | \tilde{x})$ with respect to
$p(\tilde{x} | {\rm H}_0)$ and $p(\tilde{x} | {\rm H}_1)$, where ${\rm
H}_{1}$ is the alternative hypothesis.
 \begin{equation}
 p({\rm H}_{0} | \tilde{x}) =\frac{ p({\rm H}_{0}) p( \tilde{x} | {\rm
 H}_{0})}{p({\rm H}_{0})
 p( \tilde{x} | {\rm H}_{0})+p({\rm H}_{1}) p( \tilde{x} | {\rm H}_{1})}.
 \end{equation}
We set $p_0=p({\rm H}_{0})$, and since we have $p({\rm H}_{1})=1-p_0$,
they finally obtained:
 \begin{equation}
 p({\rm H}_{0} | \tilde{x})= \left(1+\frac{(1-p_0)}{p_0}\cal{L}\right)^{-1},
 \label{bayes}
\end{equation}
with $\cal{L}$ being the likelihood ratio defined as:
 \begin{equation}
 {\cal L}=\frac{p(\tilde{x} | {\rm H}_1)}{p(\tilde{x} | {\rm H}_0)}.
 \end{equation}
Here, $p({\rm H}_{0} | \tilde{x})$ is the so-called posterior
probability of ${\rm H}_{0}$ given the observed data $\tilde{x}$.
Naturally there is no way to privilege ${\rm H}_0$ over ${\rm H}_1$,
or vice versa, otherwise our own prejudice would most likely be
confirmed by the test, i.e. $p_0=0.5$.  Subsequently,
\citet{Berger1997} recommended to report the following when performing
hypothesis testing:
 \begin{equation}
 {\rm if\,\,} {\cal L} \leq 1, {\rm \,\,reject\,\,} {\rm H}_0 {\rm \,\,and\,\,report\,\,} p({\rm H}_0 | \tilde{x})=\frac{1}{1+ {\cal L}},
 \label{prescrit1}
 \end{equation}
 \begin{equation}
 {\rm if\,\,} {\cal L} > 1, {\rm \,\,accept\,\,} {\rm H}_0 {\rm \,\,and\,\,report\,\,} p({\rm H}_1 | \tilde{x})=\frac{1}{1+ {\cal L}^{-1}}.
 \label{prescrit2}
 \end{equation}
The advantage of such a presentation is that even for a borderline
case, say when the ratios above are close to unity, it is clear that
there is only a 50\,\% chance that the H$_0$ hypothesis is wrongly
accepted, or wrongly rejected.  This presentation is more honest and
better encapsulates human judgement and prejudice.

\begin{figure*}
\center{
 \hbox{ \includegraphics[width=0.35\textwidth,angle=90]{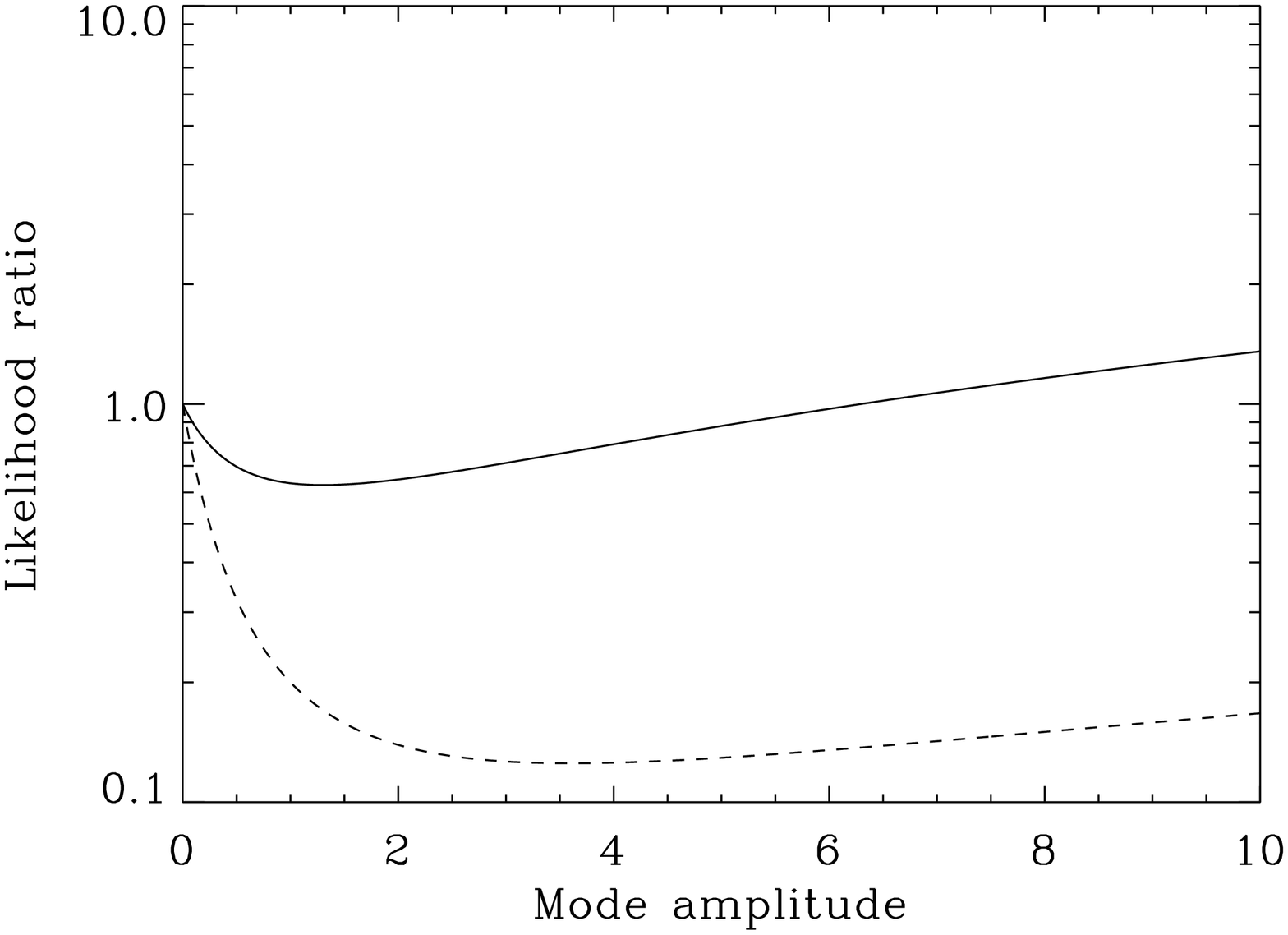}
\vspace{0.0truecm}
\includegraphics[width=0.35\textwidth,angle=90]{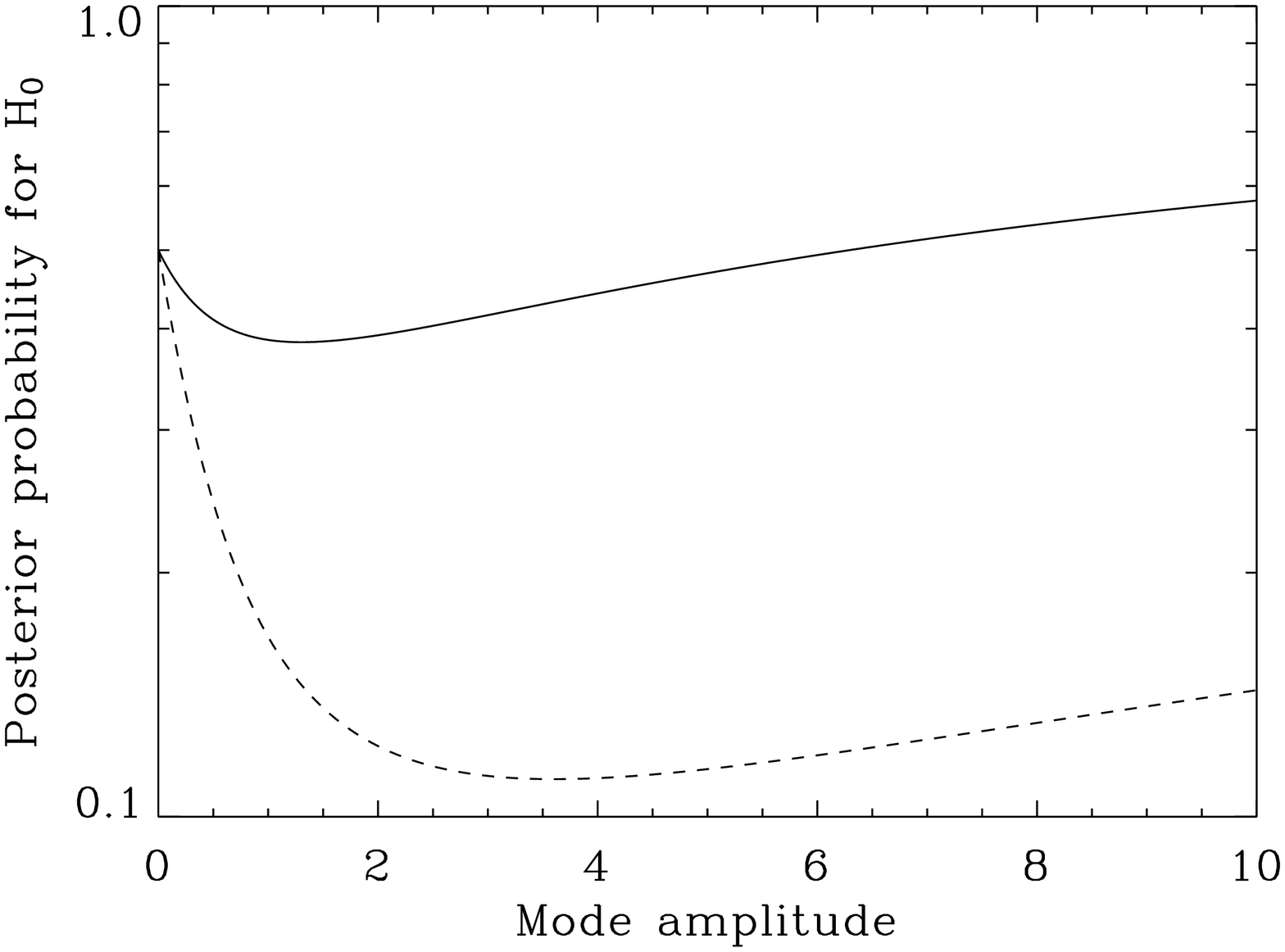}}
}

 \caption[]{On the left-hand side, likelihood ratio ${\cal L}$ as a
 function of the mode amplitude for {\it detection significances} of
 10\,\% (solid line), and of 1\,\% (dashed line); the noise is set to
 unity.  On the right-hand side, the posterior probability of ${\rm
 H}_0$ as a function of mode amplitude for {\it detection
 significances} of 10\,\% (solid line), and of 1\,\% (dashed line)
 (from Eq.~\ref{posterior})].}

\label{bayes-stat}
\end{figure*}

The work of \citet{Berger1997} can be applied to the problem of
detecting modes with very long lifetimes, i.e., those modes whose
peaks are predominantly restricted to one frequency bin in the frequency
spectrum.  We can write the likelihood for the ${\rm H}_0$ hypothesis
as:
 \begin{equation}
 p(\tilde{x} | {\rm H}_0)=\frac{1}{B}e^{-\tilde{x}/B}.
 \end{equation}
where $B$ is the mean noise level.  Next, we assume that the mode is
stochastically excited with a known amplitude $A$. The
likelihood for ${\rm H}_1$ is then:
 \begin{equation}
 p(\tilde{x} | {\rm H}_1)=\frac{1}{B+A}e^{-\tilde{x}/(B+A)}.
 \end{equation}
Since the {\it detection significance} is $p=e^{-\tilde{x}/B}$, we use
Eq.~(\ref{bayes}) to give:
 \begin{equation}
 p({\rm H}_{0} | x)= \left(1+\frac{B}{B+A}p^{-A/(B+A)}\right)^{-1}.
 \label{posterior}
 \end{equation}
Figure~\ref{bayes-stat} show the results for two different {\it
detection significances}.  When the {\it detection significance} is
10\,\%, the likelihood ratio can be greater than unity for large
values of the mode amplitude, leading to the acceptance of the null
hypothesis.  This is rather paradoxical, i.e., that large mode
amplitude can lead to the rejection of the alternative hypothesis.  To
resolve the paradox we note that the posterior probability of ${\rm
H}_0$ is in any case never lower than 40\%, or the posterior
probability of ${\rm H}_1$ is never higher than 60\%.  This implies
that both hypotheses are equally likely when the {\it detection
significance} is as low as 10\,\%.  In other words, when we set, a
priori, a large mode amplitude and get a low {\it detection
significance}, the alternative hypothesis is as likely as the null
hypothesis.

The main conclusion to be drawn from this calculation is that the {\it
detection significance} should be set much lower than 10\,\% in order
to avoid misinterpretation of the result.  For example, with a {\it
detection significance} of 1\,\%, the posterior probability for H$_0$
can fall to 10\,\% when the signal-to-noise ratio is above unity.
\citet{Sellke2001} showed that the posterior probability can never
be lower than the lower bound
 \begin{equation}
   p({\rm H}_{0} | x) \geq \left(1-\frac{1}{{\rm e} p \ln p}\right)^{-1}.
 \label{bound}
 \end{equation}
The reader may verify for themselves that this lower bound is
effectively reached for Eq.~(\ref{posterior}).  In the case, when
the amplitude of the mode $A$ is not known, one needs to set, a
priori, value for the likely range of amplitude.  In the case of a
uniform prior, the posterior probability $p({\rm H}_{0} | \tilde{x})$
then does reach a minimum that is higher than the lower bound of
Eq.~(\ref{bound}).

In summary, the significance level should not be used for justifying a
detection (or a non-detection).  Instead we recommend to use the
prescription of \citet{Berger1997}, as given by Eqs.~(\ref{prescrit1})
and (\ref{prescrit2}) and to specify the alternative hypothesis ${\rm
H}_1$.

\vspace{5 mm}

{\bf On the choice of the prior probability} One important question
when applying Bayesian statistics is what value should the prior
probability of the hypothesis ${\rm H}_0$, i.e., $p_0$, take? We
define the prior probability as the probability that the H$_0$
hypothesis is correct.  The probability that the alternative H$_1$
hypothesis is correct can then be defined as $p({\rm H}_1)=1-p_0$.  It
is common to set $p_0=0.5$ so as to avoid prejudicing one hypothesis
over the other.  Given current model prejudices, we know that low-frequency p modes have narrow widths typically between 10 and 30 nHz, \citep{Chaplin05}, which cover a few frequency bins only; and we expect the same to be true for g modes.  In which case, would we
expect the probability that H$_1$ and H$_0$ are true to be the same at
all frequencies? Since Bayesian statistics uses a priori knowledge, it
is possible to use our knowledge of the properties of modes to tell us
which hypothesis is more likely to be true at a given frequency.  We
can also use this knowledge to then guide the regions we search in
frequency (See section 4.5.3).

\subsection{Patterns}
\subsubsection{Rotational splitting}

As shown in Section 2 [see Eq.~(\ref{eq7J})], the mode degeneracy is
lifted by the solar rotation; each $(n,l)$ mode is then split into
$2l+1$ components.  The collapsogramme technique pioneered by
\citet{TA2000} makes use of the pattern created by the rotational
splitting to detect the modes.  It has been extensively used by
\citet{Salabert2007} for detecting modes below 1000 $\mu$Hz, using
resolved-Sun data collected the GONG instrument.  It can also be used
for full-disc instruments producing a single power spectrum: in this
case it is called an {\it overlapogramme}  \citep[For an application see][]{Salabert2008}.  \citet{WJC2002} has
devised a statistical technique based on the detection of an ordered
multiplet (due to rotational splitting) in the power spectrum of
full-disc integrated data.  This lowers the detection level, relative
to the level for just a single peak, depending on the number of peaks
that are found in the multiplet.  The derived limit (under the H$_{0}$
hypothesis) can be translated into the equivalent $\sigma$ level in a
power spectrum for a one-year observing time (where $\sigma$ is the
mean power spectral density in the neighbourhood of the sought-for
modes): 5.9$\sigma$, 4.5$\sigma$, 3.8$\sigma$ for a doublet, triplet
and a quadruplet, respectively.  A similar approach has been used by
\citet{RF2001} for the GOLF data.  They derived levels for detecting
modes of various degrees using Monte-Carlo simulations, which can be
translated for a one-year time series to 5.4$\sigma$ for an $l=1$
doublet, 5.9$\sigma$ for an $l=2$ doublet and 4.3$\sigma$ for an $l=2$
triplet.

\subsubsection{Asymptotic behaviour}

The asymptotic behaviour of g-mode periods may be used to aid
detection [see Eq. (\ref{eq4J}) of Section 2]. Such analyses were pioneered
by Delache in 1983, leading to a claimed detection of g modes
\citep{PDPS83}. This approach is only of relevance to high-order g
modes (i.e., very low-frequency modes below $\sim100\,\mu$Hz) for which the
asymptotic behaviour applies.  Unfortunately, the solar noise
increases towards lower frequencies, and the mode spacing (in
frequency) decreases dramatically. The situation is further
complicated by the effects of rotational splitting \citep[][see also Section 5.2.1]{CFBA95}.


The {\it Exact Fractions Technique} (EFT) was pioneered by
\citet{raay88} and applied by \citet{Palle1998}.  The technique is
rather similar to the {\it echelle diagram} devised by \citet{GG81}.
The main difference is that instead of cutting the spectrum into
several windows of width $\Delta \nu_0$, the frequency axis is first
changed to period, and the spectrum is then cut into several windows
of width $P_0/\sqrt{l(l+1)}$.  The full EFT procedure is described in
more detail in \citet{Palle1998}.

The EFT uses the property that the modes, if they exist, are regularly
spaced in period.  The same property has been used by \citet{RAG2007}.
They compute the periodogram of the frequency power spectrum expressed
in period, i.e., the periodogram of the periodogram. The regular
comb-like pattern expected in the first periodogram then manifests as
another comb-like pattern in the second periodogram. Unlike the
technique devised by \citet{PDPS83} or the EFT, the statistical
understanding is far from trivial, requiring Monte-Carlo simulations
to derive the likelihood of structures appearing above a given
threshold.

\subsubsection{Guided search}

An artificial way of reducing the detection limit is to reduce the
size of the frequency windows over which we wish to search for modes,
e.g., by looking in windows centred around theoretical g-mode
frequencies.  \citet{DD99} provided a simple formula to derive the
number of peaks due to noise that one can find in a power spectrum,
given a list of frequencies and a window containing these frequencies.
Under the H$_{0}$, hypothesis, it is written as follows:
 \begin{equation}
 N=N_{{\rm l}} (1-(1-p_{{\rm det}})^{N_{{\rm w}}}).
 \label{eq2}
 \end{equation}
Here, $N_{\rm l}$ is the number of frequencies guiding the search,
$p_{{\rm det}}$ is the probability level needed for identifying a peak
and $N_{\rm w}$ is the window size in units of frequency bins.  When $N_{{\rm
w}}p_{{\rm det}}$ is much smaller than unity, we can rewrite
Eq~(\ref{eq2}) as:
 \begin{equation}
 N \approx N_{\rm l} N_{{\rm w}} p_{{\rm det}}.
 \label{eq3}
 \end{equation}
This simple formula is quite useful, in that it allows us to realise
that the number of identified peaks will increase with the size of the
window and the number of frequencies guiding the search.  This is the
drawback of such a method: spurious peaks will be detected that are
likely to be wrongly identified as g modes.  Here we should also
remind the reader that theoretical p-mode frequencies showed
systematic errors of the order of several $\mu$Hz, which came from our
inability to model properly the surface of the Sun \citep{JCD90}.
Therefore care should be exercised when using theoretical frequencies
as a guideline for searching for g modes.


The sensitivity of g-mode frequency predictions to solar models is of the order of 1\% (see Section 2.2.3). This error estimate can be used to produce an inverted top-hat prior probability distribution that can be used to guide Bayesian searches for g modes. The centres of each hat were determined by the mode frequencies predicted by the M1 model of \citet{JPGB2000} and the width of each hat was 2\% of the mode frequency. Frequencies where modes are expected to be observed have been given a prior probability that the $\rm H_0$ hypothesis is true ($p_0$ in equation 36) of 0.5 and elsewhere the prior probability is unity. Figure \ref{figure[top hat]} shows such a distribution, where all components that will be visible in Sun-as-a-star data (when $l+m$ is even) have been plotted for $l=1$ and $l=2$ g modes. We have assumed that the different $m$ components are separated by $0.4\,\rm\mu Hz$. However, as we do not know the rotation rate of the solar core this assumption may not be valid (see Section 2.2.4). Notice that despite the large number of g modes that are present at low frequencies the prior is not uniform with frequency and so this approach should limit the number of false detections.

\begin{figure}
\centering
 \includegraphics[width=0.8\textwidth, clip]{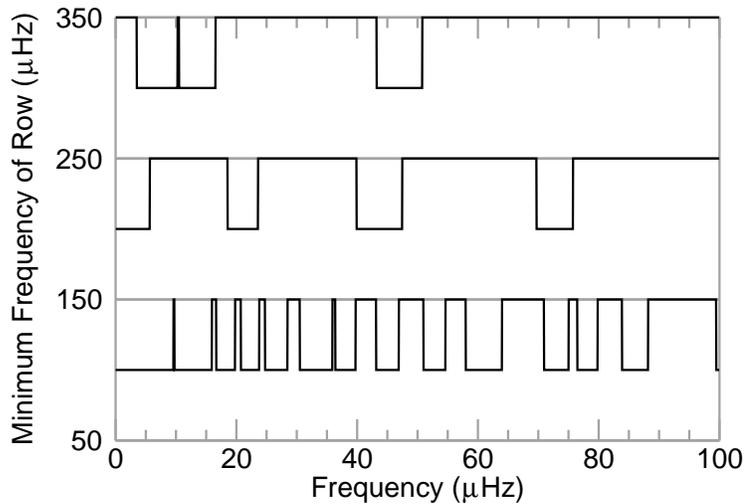}\\
 \caption{An Echelle diagram showing the variation of the of
 the prior probability for g modes. The prior probability, $p_0$, was
 either 1.0, at frequencies where modes are not expected to be
 observed, or 0.5 at frequencies where modes are expected.}\label{figure[top
hat]}
\end{figure}

\subsection{Data combination}

The signal from solar oscillations will be common to contemporaneous
data observed by different instruments. It is, therefore, pertinent to
search spectra constructed from contemporaneous data for statistically
significant prominent concentrations of power, which lie significantly
above the local noise background and that are coincident in frequency
in the different datasets.  The use of different data sets related to
different observables and/or wavelengths could very well be the
solution to the g-mode detection.  There is no doubt that the
combination of more than two signals could considerably lower our
detection limit.  Observables such as radial velocity, intensity
fluctuations, limb displacement and/or brightening are polluted by
different sources of noise, such as supergranulation and active regions,
that produce different signatures.  We can list the possible
combinations as follows:
 \begin{itemize}
 \item use of one instrument
 \item use of more than one instrument
 \end{itemize}
In terms of the analysis techniques we might apply to exploit such
data, autocorrelation falls into the first category.  The
collapsogramme technique used by \citet{TA2000} also belongs to the
first category.  Another technique developed by \citet{RG1999} used a
longer sampling time (e.g., 80\,sec instead of 40\,sec) to create two
overlapping, independent timeseries of the GOLF data.  The first time
series used all the even-sample data points, while the other used the
odd-sample data points.  The combination of the two resulting Fourier
spectra can be used to improve the signal-to-noise ratio of modes with
a lifetime shorter than the observation time \citep{TA2007}.

The Multivariate Spectral Regression Analysis (MSRA) belongs to the
second category \citep{Koop74,TA2000}.  It assumes that the modes are
predominantly coherent over the observing time. The basic assumption
is that low-degree p modes and g modes will have lifetimes that are
significantly longer than the observing time.  The MSRA has been
applied to data from the different SPM channels of the VIRGO data by
\citet{Finsterle2001}.

A derivation of the joint probability, $p$, of the occurrence of
coincident prominent features in two spectra is given in
\citet{AMB2007}. Various statistical tests are derived in
\citet{WJC2002} that determine the probability of observing prominent
structures in a single spectrum.  These tests can be easily adapted so
they can be used to compare two spectra if the probability of
observing a single prominent spike at the same frequency in each of
the spectra can be calculated \citep{AMB2007}. The technique works
best when the signal-to-noise ratio is similar in the different
instruments.  These tests were originally designed to detect p modes,
however, the underlying theory is still applicable when searching for
g modes.

%
%




\section{Applications of detection techniques}

As outlined in the previous section, the building of a detection is made of an assemblage of various elementary bricks.  It is essential that a {\it detection technique} be segmented in different categories.  The key category is {\it statistical testing} because it exemplifies that the potential biases of a detection technique can indeed be understood.

Hereafter, we will try to review the application of detection techniques, and their associated results.

\subsection{Statistical thresholding}

\subsubsection{Radial velocity}

The first reported g-mode detection was made by \citet{PDPS83}.  They
analysed differences in the solar radial velocity recorded at disc
centre, and further out toward the limb. Observations were made at the
Stanford Solar Observatory \citep{Scherrer1979}, and at the Crimean
Astrophysical Observatory \citep{Severnyi1976}.  The length of the
time series was 105 days, but with a duty cycle of only 10\%.  After
computing the power spectrum, they constructed the cumulative power
distribution in the frequency range [45-105] $\mu$Hz.  \citet{PDPS83}
calculated the cumulative distribution of the power spectrum in this
range, and ascribed the strongest peaks as being due to g modes.

The cut-off used to tag prominent peaks was 2.5 $\sigma$ which, given
the assumed $\chi^2$ with 2 d.o.f statistics, means that in the window
considered about 5 $\pm$ 2.5 ``detections'' will have been due solely
to noise (there are about 55 independent frequency bins in the window given the
10\% duty cycle).  Fourteen possible modes were identified, at least
half of which are likely due to noise.  In addition, they did not take
into account that the power spectrum has a $1/\nu$ dependence implying
that the mean varies with frequency; this explains the departure
observed from the expected statistical distribution.  The minimum
amplitude of the detected peaks was 25 cm\,s$^{-1}$.  Since this paper
appeared, no independent confirmation of the results has appeared in
the literature. Indeed, as we shall see we below, quoted upper limits on
g-mode amplitudes, given by analyses of modern datasets, lie some two
orders of magnitude below the claimed \citet{PDPS83} amplitudes.

A similar approach was used by \citet{TA2000} for setting an upper
limit to the amplitude of the g modes, which lay significantly below
the amplitudes claimed by \citet{PDPS83}.  The canonical figure for
H$_0$ was given for a threshold of 10\,\% for a 100$\mu$Hz window,
providing a minimum posterior probability for H$_0$ of 38\% (See
Section 4.2.2).  As shown in Fig.~17, a threshold of 1\,\% would
provide a lower posterior probability for H$_0$ of typically 15 \% for
mode amplitude at a signal-to-noise ratio less than 10 (but greater than
unity).  The 1\,\% threshold would increase by a factor two the upper
limit quoted by \citet{TA2000}.

\citet{AG2002} also used a statistical thresholding approach.  They
used oversampling in an attempt to increase the detectability of
long-lived signals.  They reported the detection of three g-mode
candidates below 290 $\mu$Hz with a significance level of 4\,\%, or a
posterior probability for H$_{0}$ of 25\%.  They could not tag the
candidates as being g modes, and then set an upper limit to the
amplitude of g modes, based upon the H$_{1}$ hypothesis, of 7 mm
s$^{-1}$.

\citet{TC2004} applied thresholding for various kinds of features: for
singlets \citep[as][]{TA2000}, for doublets and for triplets.  The
latter application of the thresholding is more akin to {\it searching
for pattern of g-mode peaks}, which is described below. Several
candidate detections were flagged at a threshold of 10\,\% (implying a
lower bound on the posterior probability of 38\%). But none appeared
in all their test observation periods.

\subsubsection{Solar diameter}

Observations of the solar diameter have also been used in the search
for g modes, e.g., in observations by the SCLERA instrument.  The
lowest amplitude detected corresponded to a radius change of 1 mas for
p modes, and to 0.1 mas for g modes; assuming that these radius
changes were real physical changes, they would have been equivalent to
radial velocities of 50 km s$^{-1}$ and 250 m s$^{-1}$, respectively.
As outlined by \citet{Brown1979}, the observed oscillation signals
were more likely due to brightness effects than to physical radius
changes.  This latter fact was also confirmed by \citet{TA98d} and
\citet{CT99} using space-borne instrumentation.  The amplification is
due to a geometrical effect, which makes optical thickness changes
dominate temperature changes.  The rms amplitudes of the modes
detected by \citet{TA98d} were typically of the order of 8\,ppm,
corresponding to a radius change of 1 mas.  \citet{CT99} showed that
the oscillation signal increases as the distance to the limb
decreases.  This property will be used by the SODISM instrument on the
PICARD mission to try to detect modes at the limb \citep{LD99,
Corbard2008}.

\subsubsection{Solar wind data}

\citet{DT95} used magnetic field and particle data collected by the
Ulysses mission. They used different spectral estimators (based on
multi tapers; see \citet{thomson82}) and different statistical tests
(based on the $F$-test), with a threshold of 5\,\%\footnote{The
$F$-test is used for comparing two random variables having a $\chi^2$
distribution with 2 d.o.f}.  They claimed the detection of several
g-mode frequencies. However, these claimed detections were not
confirmed by \citet{DD99} or by \citet{GHPR98}, who both argued that
the chosen threshold was too weak and gave rise to false detections.
By making use of the work of \citet{Berger1997}, we can now understand
in retrospect that this threshold will have lead to a posterior
probability of H$_{0}$ being true of at least 29\,\% [from
Eq.~(\ref{bound})]: the applied $F$-test threshold was clearly
not low enough.

\subsection{Searching for patterns of g-mode peaks}

\subsubsection{Asymptotic properties of periods}

The first attempt to use the asymptotic properties of the g-mode
frequencies dates back to \citet{CFPHD84a,CFPHD84}.  They
used the asymptotic formula [see Eq.~(\ref{eq4J})], corrected for
rotational splitting, to compute g-mode frequencies as a function of
$P_0$ and $\overline{\Omega}_0$ ($\overline{\Omega}_0$ being the mean
rotation as defined by Eq.~\ref{eq7J}).  They then calculated the
coherence between the calculated g-mode frequencies and the observed
power spectrum, and drew maps of coherence as a function of
($P_0,\overline{\Omega}_0$). No detections were made.  The same
procedure was applied by \citet{CFBA95} also without success.

\citet{CFBA95} concluded that either the g modes were not present or
that the solar noise was too high.  This method has also recently been extended to include also g-mode frequencies excluded by the normal asymptotic equation. The models shown in Table 1 cover a sufficient range of $P_{0}$ rendering an interpolation scheme as function of $P_{0}$ possible. This allows to include g modes with $l=1$ and $l=2$ down to $n=1$.  The results are not yet conclusive, possibly due to the precision of the frequencies and/or the interpolation scheme. Another reason may have been that the range for the rotation in the core was too small. 
One of the major drawbacks of the technique is that the computation of g-mode frequencies needs to be very accurate.


\subsubsection{Rotational splitting}

\citet{TC2004} applied statistical thresholding to search for patterns
of peaks induced by rotational splitting.  They derived, using
Monte-Carlo simulation, pattern rejection thresholds for doublets and
triplets at the 10\,\% level for a 20-$\mu$Hz window (posterior probability greater than 38\,\% ); similar
levels were analytically derived by \citet{WJC2002} for p modes.
Given the appearance, and disappearance, in different observation
periods of the various detected peaks, \citet{TC2004} concluded that
some patterns {\it were due to noise}, thereby supporting the
posterior probability of H$_{0}$ being greater than 38\,\%.  The analysis was further pursued by \citet{Mathur2007} who confirmed detection at a 2\% level for a 10$\mu$Hz window (corresponding to a 4\% level for a 20 $\mu$Hz window, or posterior probability greater than 26\%).  \citet{WJC2002} also
searched for similar patterns, reporting possible detections of p
modes as low as 700 $\mu$Hz. A similar technique was used by
\citet{Salabert2007} to detect low-order p modes, of higher degree, in
resolved-Sun observations made by the GONG network. They detected
modes below 1000 $\mu$Hz, with typical lifetimes of the order of 3
years.

\subsubsection{Periodogram of Fourier space}

\citet{RAG2007} found a very interesting peak in the periodogram of
the periodogram of GOLF data \citep[See also][]{RAG2008}. They claimed
that this peak is due to the superposition of g modes that lie in the
asymptotic region, where the separation in period is approximately
constant. The likelihood that the peak is not due to noise is 99.5$\pm$0.13\%, with a posterior for ${\rm H}_0$ greater than 3.4 \%.  The peak is positioned close to the period predicted by
standard solar models.  The peak can be produced in several ways, all
related to clusters of peaks being equally spaced in period.  
For instance, the power of the g modes could be spread onto several adjacent peaks like the presence of a deep magnetic field or changes in the propagation cavity (a displacement of the tachocline as a consequence of the activity cycle).  Other possibilities were also discussed by \citet{RAG2007}.  They used artificial data to show that, if this
peak is evidence of g modes, the modes could have widths in a
frequency-power spectrum that are commensurate with damping times of
several months, with a possible signal-to-noise ratio not greater than
five in power. If this is the case, and given that the data searched
here are $3071\,\rm d$ in length, the width of these modes would be
resolved in the spectrum. For example, if the mode damping time was
$122\,\rm d$ the power of the modes in the spectrum would be spread
across eight frequency bins.  \citet{Appourchaux2004} showed that it
is possible to detect short-lived modes by smoothing the power
spectrum over several frequency bins.  For example, the detection probability is
higher than 90\,\% for modes spanning at least eight frequency bins with a
signal-to-noise ratio greater than three.  Therefore, we have searched
the contemporaneous, very low-frequency $(\nu<500\,\rm\mu Hz)$
BiSON\footnote{Birmingham Solar Oscillation Network,
\citet{WJC1996}}, GOLF and SOI/MDI data for evidence of prominent peaks
with various widths. We searched for clusters containing 2, 3, 4 and 5
spikes. We allowed a cluster to be spread over twice the width of the
mode and took the widths of the modes to be the number of frequency bins covered
if the mode lifetimes were 1 month, 2 months and 4 months,
respectively (i.e., 32 bins, 16 bins and 8 bins, respectively). However, no statistically significant peaks were found within
$4\,\rm\mu Hz$ of the model g-mode frequencies.

\subsection{Innovative methods using a single set of data}
\subsubsection{Time-distance analysis}

\citet{Duvall2004} devised a method for detecting g modes using
time-distance helioseismology of p mode signals.  The idea is to
detect the small flows induced by the g modes on the travel times of
the p modes.  He used deep-focusing rays to create travel-time maps
for detecting perturbations at about 100\,Mm below the solar surface,
and then applied regular spherical harmonics masks on these maps to
obtain power spectra of these travel times.  While this method has not
yielded any potential detections, plenty of work is need to develop
the technique.

\subsubsection{Cross spectra of GOLF data}

Very recently, \citet{Grec2008} used about 13\,years of GOLF data to
search for g modes using cross spectra.  The idea is to use about 22
sub-series, each of 7-month's duration, to create cross spectra that
are calculated by pair-by-pair. In total, this provides more than 253
cross spectra (i.e., $22 \times 23 / 2$).  The averaged cross spectrum
is then normalised to {\it remove} the low frequency dependence, and
then an autocorrelation of the cross spectrum is calculated.  The
autocorrelation of the cross spectrum shows a maximum at a value that
is tentatively identified as being the g-mode {\it rotational
splitting}.  In addition, they also tried to detect in the 253 cross
spectra the signature of peaks having the same phase.  They compared
their result with computed g-mode frequencies of a solar model, but
found no convincing detections.

\subsection{Use of contemporaneous data}
\subsubsection{A signal at 220.7 $\mu$Hz?}

At the beginning of the SOHO mission, there were several reports of
the detection of a signal close to 220.7 $\mu$Hz, that might possibly
be identified as the $l$=2, $n=-3$ g mode
\citep{Gabriel1999,Finsterle2001}.  Very recently, \citet{AJM2008} and
\citet{RAG2008} reported again on the presence of this signal in data
from the different intensity instruments comprising the VIRGO package
on SOHO.  There is a trace of the signal in GOLF (but not with the same quality), in GONG (but at the level of the noise) and nothing in SOI/MDI.  However, it seems
that the signal has no obvious instrumental or SOHO-related origin.
It is possible that the origin of the signal is related to the noise  created in intensity by convection close to a g-mode frequency, or in fact could be a component of a g-mode.

\subsubsection{Coincidence search}

Statistical tests have been used to search contemporaneous BiSON,
GOLF and SOI/MDI data for low-frequency p modes, g modes and mixed modes
using a frequentist approach.  To take advantage of
the availability of contemporaneous data from different instruments
frequency-amplitude spectra were searched for prominent spikes or
patterns of spikes that were positioned in the same frequency bin or
bins in any two of the three spectra. Details of the statistical
tests can be found in \citet{AMB2007}. Amplitude threshold levels
were calculated for a 1\,\% probability of detecting a prominent
feature by chance at least once in a frequency range of $100\,\rm\mu
Hz$.


\begin{figure*}
\centering
\includegraphics[width=0.9\textwidth, clip]{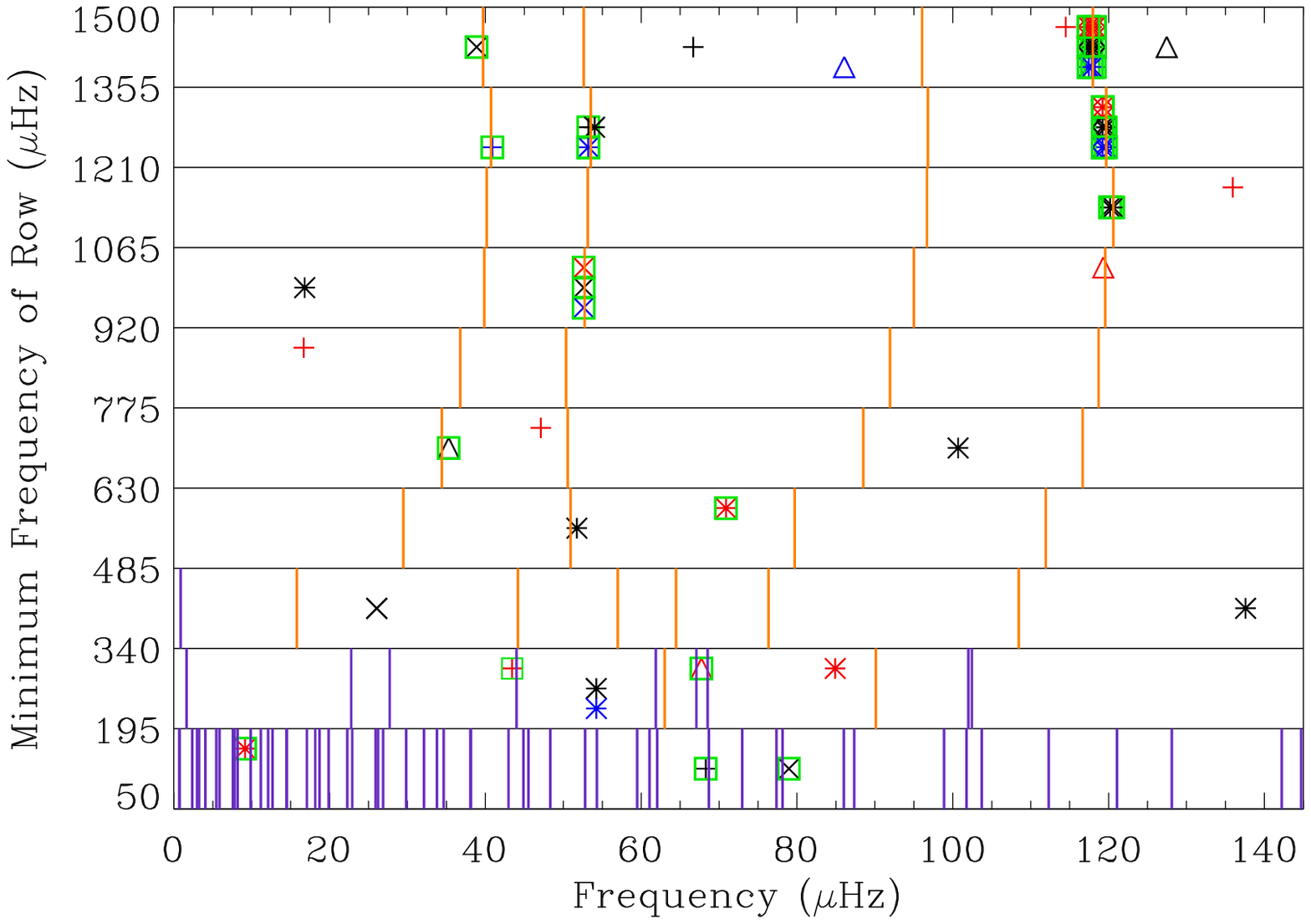}\\
  \vspace{0.3cm}
  \includegraphics[width=0.78\textwidth, clip]{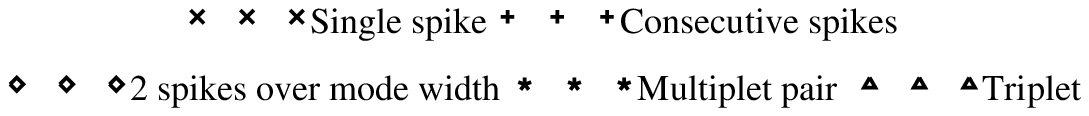}\\
\caption{An Echelle plot, modulo $145\,\rm \mu Hz$, marking
locations in frequency of occurrences uncovered by the test
searches. Locations in frequency where spikes, or patterns of
spikes, were found in the same frequency bin, or bins, in BiSON and GOLF
spectra at levels sufficient to record $p \le 1$\,\% are marked by
the black symbols in the middle of each row. A different symbol has
been used for each test (see figure legend). We have also recorded
prominent spikes or patterns of spikes found by comparing either the
BiSON and SOI/MDI spectra (red symbols at top of each row) or the GOLF
and SOI/MDI spectra (blue symbols at bottom of each row).  The orange
vertical lines mark locations of the frequencies of p modes
predicted by the Saclay seismic model \citep{STC2001}. The vertical
purple lines mark locations of the $m=0$ frequencies of g modes predicted
by the M1 model \citep{JPGB2000}.}
\label{figure[results echelle]}
\end{figure*}


Figure \ref{figure[results echelle]} shows visually, in the form of
an echelle diagram, the locations in frequency where prominent
spikes or patterns of spikes were observed. Also plotted on Figure
\ref{figure[results echelle]} are the mode frequencies predicted by
solar models. Detections were only considered as possible mode
candidates if they were positioned within $1\,\rm\mu Hz$ of a
predicted mode frequency and such candidates are highlighted by a 
green square in Figure \ref{figure[results echelle]}.

In the p-mode frequency range all of the candidates correspond to
previously claimed detections of p modes \citep[See][]{TT1998,
LB2000a,RAG2001,WJC2002,RAG2004,AMB2007}. Some of the detections in the
g-mode range lie close to the predicted frequencies. However, it
should be noted that, since the number of g modes that lie in the
frequency range searched is large, the chances of mistaking a false
detection for a g mode candidate is large. Therefore, to improve
confidence in the detections we required that each mode candidate
passed more than one of the statistical tests, thus significantly
reducing the number of false detections. This extra requirement
discounts all of the detections in the g-mode range and some of the
detections in the p-mode range.

It is apparent from Figure \ref{figure[results echelle]} that the
analysis has uncovered several occurrences of $p\le1$\% that lie
well away from the predicted mode frequencies. The number of these
detections exceeds the number expected given the range in
frequencies searched and the threshold probability. This possibly
indicates that the statistical distribution in the noise is slightly
different from the Gaussian distribution, which was assumed when
deriving the statistical tests. Alternatively it could be due to the
misleading significance levels assigned when adopting a frequentist
approach (see Section 4.4.2).


\section{Discussion and Conclusion}

Figure~\ref{fig:amp_dog_kumar_kevin} shows the comparison of the best
observational measurements with theoretical amplitudes of the g modes.
The theoretical g-mode amplitudes range from 10$^{-2}$\,mm\,s$^{-1}$ to
a few mm\,s$^{-1}$.  The most optimistic theoretical mode amplitudes
could be collectively detectable, after the collection of more than 10
years of observations of the solar radial velocity.  Unfortunately, the detection of individual peaks is very far
from being feasible since the only possibility would be to have the most optimistic amplitude larger by 50\%, or twice as much in power \citep{KB2009}.  Since the noise
reduction scales with the observing time $T$ like $\log(T)/\sqrt{T}$,
another 80 years or so of data would be required to reduce the actual
detection limit by a factor two.  The limit does not scale like
$1/\sqrt{T}$ because the probability limit is kept constant in a given
detection window while the number of frequency bins in that window increases
like $T$ \citep{TA99}. At the time of writing, there is indeed a consensus
amongst the authors of this review that {\it there is currently no undisputed detection of solar g  modes.}


Despite this state of affairs, the search is not over yet.  What can
be done to turn our fortunes around? First of all, some of the
detection techniques presented in this review paper that rely on the
theoretical knowledge presented in Sections 2 and 3. Several are yet
to exploit fully the theoretical information: this calls for the
application of full Bayesian statistical inference on the data (yet to
be done). 

The observation of signals from the the solar limb will
hopefully offer another way of improving the potential visibility of
the g modes.  These observations will be carried out by PICARD, which
is due to be launched in 2009. 

Because the solar atmosphere is
relatively noisy, reductions of noise levels can potentially be
achieved by making multiple observations that are sensitive to
perturbations at different heights in the atmosphere. Here, one would
seek to rely on changes in the coherence of the noise with height.  The GOLF-NG instrument recently put in operation uses this principle for measuring solar radial velocities in 8 different heights of the solar atmosphere \citep{STC2006,Salabert2008b}.  

Development of methods based on time-distance helioseismology offer promises not only for detecting directly the g modes, but also for probing the solar core by making observations from two widely separated point of view.  The technique is called stereoscopic seismology, and may provide the structure and the dynamics of the core with a completely different approach with the Solar Orbiter mission \citep{AG2003,Gizon2006}.

Finally, we could try to detect the perturbation of the gravitational
potential created by the g modes.  For instance, the LISA\footnote{The
ESA/NASA gravitational wave laser interferometric space antenna}
mission may provide a lower detection limit than the current classical
helioseismic techniques.  \citet{Polnarev2009} deduced a lower
detection limit for LISA that is about a factor five lower in
amplitude than the actual GOLF limit for modes of $m=\pm2$ (see
Figure~\ref{fig:amp_dog_kumar_kevin}).  This limit would be low enough
to detect g modes if they had amplitudes as high as those predicted by
\citet{KB2009}.  The variations in the potential can be also
detected by using laser ranging, e.g., the ASTROD mission
\citep{Ni2007, Appourchaux2009}.  As shown by \citet{RB2008}, the capability of ASTROD
would be sufficient to allow the detection of g modes even if their
amplitudes were to be as low as those predicted by \citet{Kumar96}.

We conclude with an optimistic comment. It is superficially
unfortunate that we may not yet have detected g modes in the Sun.  On the
brighter side, on which we always prefer to be, it shows that we have
before us a greater challenge which will yield greater satisfaction
when we overcome it.  We are all now much more prepared to continue
the search.

\acknowledgement The whole Phoebus group would like to thank the
long-time support provided by the European Space Agency, in particular we thank Clare Bingham, Cecilia Nillson, Mylene Riemens, Birgit Schroeder for secretarial support, and Martin Huber, Peter Wenzel and Bernard Foing for political and financial support; and the support provided by the International Space Science Institute, in particular we thank Brigitte Schutte, Saliba Saliba, Vittorio Manno and Roger-Maurice Bonnet for supporting our program.  SOHO is a mission of international collaboration between ESA and NASA.  The Phoebus group would like also to thank Boris Dintrans for setting the CVS server at the Observatoire Midi-Pyr\'en\'ees which allows for the parallel writing of the review.

\appendix

\section{Nonlinear amplitude limitation}
\label{app:nonlinear}

\subsection{Illustrative case of one-dimensional acoustic wave}

Consider a toy acoustic mode in a uniform medium in which Eulerian 
perturbations, denoted by a prime, to pressure, $p$, density, $\rho$, and 
temperature, $T$, are related by the perturbed equation of state at first order:
\begin{equation}
p^\prime=c^2\rho^\prime+\widehat\alpha T^\prime\,.
\end{equation}
Acoustic modes
confined between two values, 0 and 1, of a spatial variable $x$
are slowly damped by viscosity and excited by a thermal process 
which we model crudely, and approximately, by
\begin{equation}
\frac{\partial T^\prime}{\partial t}=-\widehat\beta u\,,
\end{equation}
where $u$ is velocity. The following
equation then governs the wave motion:
\begin{equation}
\frac{\partial^2\rho^\prime}{\partial t^2}
-c^2\frac{\partial^2\rho^\prime}{\partial x^2}
-\frac{\partial}{\partial t}\left(\lambda\rho^\prime+\nu\frac{\partial^2\rho^\prime}{\partial x^2}\right)
=\frac{\partial}{\partial x}\left(\rho u\frac{\partial u}{\partial x}\right)\,,
\label{eq:wavemotion}
\end{equation}
where $c^2$, $\rho$, $\lambda=\widehat\alpha\widehat\beta/c\rho$ and 
the kinematic 
viscosity $\nu$ are all constants. We have also introduced the device 
of replacing $\partial/\partial x$ by $c^{-1}\partial/\partial t$ in 
the thermal term, which is now $\lambda\partial\rho^\prime/\partial t$,
to ensure that it excites both forward and backward propagating components
of the mode. A more physically realistic model not requiring that
artificial device for the excitation could easily have been adopted, but the
governing equation would have been unnecessarily more complicated.

The Eulerian density perturbation may be represented as a sum of normal modes:
\begin{equation}
\rho^\prime=\sum_{n}\frac{1}{2}\left( A_n  \,{\rm e}^{-{\rm i}\omega_nt}
                              +A^*_n\,{\rm e}^{ {\rm i}\omega_nt}
                      \right)\cos k_nx
\label{eq:lineigen}
\end{equation}
with $n=0$ corresponding to the parent mode and $n=1,2$ to the
daughter modes, where the complex amplitude $A_n(t)$ is slowly varying
and the asterisk denotes complex conjugate. The wavenumber $k_n$ is
determined by the boundary conditions, and the real frequency
$\omega_n$ satisfies $\omega_n=ck_n$, so that the amplitude of a
linearized unforced mode grows exponentially with time at rate
$\sigma_n=(\lambda-\nu k^2_n)/2$, whose magnitude we consider to
be small compared with $\omega_n$.  It is now straightforward to show
that when the weak nonlinear interactions between the parent and the two
low-amplitude daughters are taken into account as a
small perturbation, by substituting the linear eigenfunctions of
Eq.~(\ref{eq:lineigen}) into Eq.~(\ref{eq:wavemotion}),
keeping only second-order terms, and integrating over $x$, the
resulting equations for the amplitudes are given approximately by
\begin{equation}
\frac{\rd A_0}{\rd t}-\sigma_0A_0
=\frac{4{\rm i}\Lambda\omega_0}{I_0}A_1A_2\,{\rm e}^{{\rm i}\Delta\omega t}
\end{equation}
and
\begin{equation}
\frac{\rd A_{1,2}}{\rd t}-\sigma_{1,2}A_{1,2}
=\frac{{\rm i}\Lambda\omega^2_0}{I_{1,2}\omega_{1,2}}A_0A^*_{2,1}\,{\rm e}^{-{\rm i}\Delta\omega t}
\label{eq:pdynamics}
\end{equation}
in which $I_n=\langle\rho^\prime_n\rho^{\prime*}_n\rangle$ is a measure 
of the inertia of mode, the angular brackets denoting integration over
$x$, and $\Lambda=\left[\cos\Delta kx\right]^1_0/4\Delta k$ is a mode
coupling constant which is an integral of terms trilinear in
eigenfunctions; also -$\sigma_{1,2}$ are the intrinsic
damping rates of the daughters, and are presumed to be positive,
$\Delta\omega=\omega_0-\omega_1-\omega_2$ is the frequency mismatch, and
$\Delta k=k_0-k_1-k_2$ depends on the chosen
boundary conditions. These equations are very similar to those
derived by \citet[see][]{WD82} for stellar g modes.

Equation~(\ref{eq:pdynamics}) describes the pulchral\footnote{derived from Appius Claudius Pulcher, Roman politician of the 1st century BC
who had several daughters and grand-daughters all named Claudia.} dynamics: if it is 
accepted that the daughters dissipate much more rapidly than the parent 
can grow, then the variation of $A_0$ may at first be neglected; the 
equations are linear in $A_{n\ne0}$, and admit solutions 
$A_n=Q_n\,\exp(-{\rm i}\Delta\omega t/2)$ with $Q_n\propto\exp(st)$ in
which $s\ge0$ if $|A_0|^2\ge A^2_{\rm c}$, where 
\begin{equation} 
A^2_{\rm c}=\frac{I_1I_2\omega_1\omega_2}{\Lambda^2\omega^4_0}
\left\{\sigma_1\sigma_2+\frac{1}{4}
\left[1+\left(\frac{\sigma_2-\sigma_1}{\sigma_2+\sigma_1}\right)^2\right]
                             \left(\Delta\omega\right)^2\right\}\,.
\label{eq:ac2}
\end{equation} 
As we discuss later, $A_{\rm c}$ estimates the limiting amplitude
of the parent.
Note that for a given coupling constant $\Lambda$ and close resonance
($\Delta\omega$ very small), the amplitude above which the parent excites
her daughters is approximately proportional to the (harmonic) mean of
their damping rates, and inversely proportional to the coupling
constant.  It is independent of the intrinsic growth rate of the 
parent (of course), provided that the growth rate is small compared with the
damping rates -$\sigma_n$. A state of steady amplitudes exists when
$|A_0|=A_{\rm c}$. In that case, other things being equal, the 
amplitude of the main mode would decrease if the viscosity were to be
reduced, a result which at first sight might seem counterintuitive.

To proceed further we need to apply these results to solar g modes,
which requires considering the g-mode frequency spectrum in spherical symmetry. Not
surprisingly, for a nontrivial coupling constant one needs
$m_1+m_2=m_0$, and also $l_2-l_1=l_0$ or $l_0-2$, where $m$ and $l$
are azimuthal order and degree.  These selection rules come from the integral of spherical harmonics that appears 
as a factor of the coupling coefficient \citep[see][for details]{WD82}
\begin{equation}
Z = \int Y_0^\star \, Y_1 \, Y_2 \, \sin \theta {\rm d}\theta {\rm d}\phi 
\end{equation}
where $Y_{0,1,2}$ denotes the spherical harmonic associated to the modes $(0,1,2)$, 
and $(\theta,\phi)$ are spherical polar angles.
If the coupling constant is not
to be small, one needs the orders $n_1$, $n_2$ not to be very 
different, so that the product of the daughter eigenfunctions varies
on a scale similar to that of their parent. These conditions, together
with the resonance condition $|\Delta\omega|\!<\!<\!\omega_0$, are most
likely to be satisfied when $l_1$ and $l_2$ are large, requiring
$l_2\simeq l_1$ and consequently 
$\omega_2\simeq\omega_1\simeq\frac{1}{2}\omega_0$, because when $l$ is
large the frequency distribution is dense, as can be seen from the
approximate asymptotic eigenfrequency equation derivable from
Eq.~(\ref{wave}):
\begin{equation}
\pi^{-1}\int\left(\frac{N^2}{\omega^2}-1\right)^{1/2}\rd\ln r
\simeq\frac{n-\frac{1}{2}}{l+\frac{1}{2}}\,,
\end{equation}
the integral being between two radii at which $N=\omega$.
From that relation it follows that the frequency difference 
$\delta\omega$ between two modes of consecutive order satisfies
$\delta\omega\simeq\mu\omega l^{-1}$ with $\mu$ of order unity.
Also the damping rates of the daughters satisfy $-\sigma_2\simeq-\sigma_1$,
which, for convenience, we here denote simply by $\eta$.

\subsection{Effect of frequency mismatches}

The scenario proposed by \citet{WD83} is that an `equilibrium'
state of steady oscillations is reached (ignoring temporal
variation of the background state of the Sun) in which twin
daughters of high degree extract energy nonlinearly from 
the parent and subsequently dissipate it at the same rate
that the parent extracts energy linearly from the background
state. Following a similar study by \citet{WersingerEtal80},
Dziembowski also demanded that the equilibrium be stable,
which requires $\Delta\omega>2\eta$. Then, not knowing
the structure of the Sun precisely enough to calculate
the resonances, he considered instead the probable
amplitude of the parent, making certain assumptions
about the probability distribution of frequencies: the
probability of being within $\Delta\omega$ of resonance is
proportional to $\Delta\omega/\delta\omega$, and since
$\delta\omega\simeq l^{-1}$ one might expect to find 
amplitude limitation to be most likely amongst daughters with
greatest $l$. But also one must recognise that damping rates
increase with $l$, according to $\eta\simeq\eta_0l^2$ (for a damping dominated
by radiative losses), rendering the energy-extraction process at high $l$ less effective 
and thereby permitting more freedom for the parent to grow.
The outcome is a compromise. Dziembowski estimated the
probability $P$ that the parent has its amplitude limited to $A$ to be
\begin{equation}
P=1-\exp \left( -\frac{\pi A^2}{16\mu\eta}\sum_i\Lambda^2_i \right)\,,
\end{equation}
the summation being over all pairs $i$ of twins (having coupling
constants $\Lambda_i$). Of course, the formula for the coupling
constant, which depends on the precise way in which one chooses
to define $A$, is much more complicated than that for the simple 
acoustic interaction discussed at the beginning of the section.

\bibliographystyle{aa}
\bibliography{thierrya}
\end{document}